\documentclass[%
 reprint,
 amsmath,amssymb,
 aps,
floatfix
]{revtex4-2}

\usepackage{epsfig,amssymb,amsmath,latexsym, upgreek}
\usepackage{natbib,hyperref}
\usepackage{graphicx}
\usepackage{dcolumn}
\usepackage{bm}
\usepackage{color}
\usepackage{import}
\usepackage{xr} 

\newcommand{\tF}{t_\text{F}}
\newcommand{\mfh}[1]{\textcolor{magenta}{#1}}

\newcommand{\xCOM}{x_\text{COM}}
\newcommand{\yCOM}{y_\text{COM}}
\newcommand{\xII}{x_\text{I}}
\newcommand{\yII}{y_\text{I}}
\newcommand{\xF}{x_\text{F}}
\newcommand{\yF}{y_\text{F}}
\newcommand{\vBar}{\bar{v}}
\newcommand{\xBar}{\bar{y}}
\newcommand{\tOne}{t_\text{adv}}
\newcommand{\tTwo}{t_\text{reform}}

\newcommand{\OmegaL}{\Omega_\text{L}}
\newcommand{\OmegaR}{\Omega_\text{R}}

\newcommand{\rhoC}{\rho_\text{c}}
\newcommand{\JT}{\mathcal{J}_\text{T}}
\newcommand{\epsNoise}{\epsilon_\text{n}}

\begin{document}

\preprint{APS/123-QED}

\title{Spatiotemporal control of structure and dynamics in a polar active fluid}

\author{Saptorshi Ghosh$^1$, Chaitanya Joshi$^2$, Aparna Baskaran$^1$}
 \email{aparna@brandeis.edu}
\author{Michael F. Hagan$^1$}%
\email{hagan@brandeis.edu}
 
\affiliation{%
$^1$Martin Fisher School of Physics, Brandeis University, Waltham, Massachusetts 02453, USA \\%
$^2$Department of Physics and Astronomy, Tufts University, Medford, Massachusetts 02155, USA
}%

\begin{abstract}
We apply optimal control theory to a model of a polar active fluid (the Toner-Tu model), with the objective of driving the system into particular emergent dynamical behaviors or programming switching between states on demand. We use the effective self-propulsion speed as the control parameter (i.e. the means of external actuation). We identify control protocols that achieve outcomes such as relocating asters to targeted positions, forcing propagating solitary waves to reorient to a particular direction, and switching between stationary asters and propagating fronts. We analyze the solutions to identify generic principles for controlling polar active fluids. Our findings have implications for achieving spatiotemporal control of active polar systems in experiments, particularly in vitro cytoskeletal systems. Additionally, this research paves the way for leveraging optimal control methods to engineer the structure and dynamics of active fluids more broadly.
\end{abstract}

\maketitle


\section{\label{sec:intro}Introduction}

Active systems are a diverse class of non-equilibrium assemblies composed of anisotropic components that transform stored or ambient energy into motion. Idealized realizations that have contributed to the development and refinement of this conceptual framework include bacterial suspensions \cite{Liu2012, Czirok2000, Pierce2018}, minimal systems of purified cytoskeletal proteins \cite{Needleman2017, Sarfati2022, Ndlec1997, Surrey2001, Gardel2008, Gardel2010, Dogterom2019, SoareseSilva2011}, synthetic self-propelled colloids \cite{Wagner2022, Wang2015, GomezSolano2017}, swarming 
bacterial cells \cite{Hallatschek2023, Liu2021, Hamby2018, Ni2020, Lauga2009}, and model tissues and cell sheets \cite{GonzalezRodriguez2012, Henkes2020, Fodor2018, Nichol2009} \cite{Zhang2020, Doostmohammadi2016, Saw2017, Trepat2009, Trepat2018, Duclos2014, BlanchMercader2018, Garcia2015, Ghibaudo2008}. 
In these systems, the interplay between internal activity and the interactions among the active agents results in a wide range of emergent collective behaviors \cite{Vicsek2012, Chate2020, Doostmohammadi2018, Bechinger2016, Gompper2020, Marchetti2013}.  These behaviors emerge spontaneously without requiring a central control mechanism. However, there is typically no means to select which behavior emerges or to switch between states, which significantly limits the functionality of active materials. 


Recent experimental advances have put the objective of control within our reach. Experiments have demonstrated that light can be used as a control field to assemble self-limited functional structures in active colloids \cite{Aubret2018, Palacci2013} and to exert spatiotemporal control of motility induced phase-separation (MIPS) \cite{Arlt2018, Frangipane2018}. By shining sequences of light that vary in space and time on active materials constructed with light-activated motor proteins, researchers can control the average speed of active flows \cite{Lemma2023, Zarei2023} and steer defects in active nematics \cite{Zhang2021}, and  drive the  formation and movement of asters in isotropic suspensions \cite{Ross2019}.

Theoretical progress toward functionalizing active matter has taken two paths. The first has been to impose spatiotemporal activity patterns and observe their effects on the system dynamics \cite{Shankar2022, Zhang2021, Nasiri2023, Knezevic2022}, thereby identifying easily accessible target states. The second is to use the framework of optimal control \cite{Brunton2022, Dullerud2013} and optimal transport \cite{Villani2009} to identify spatiotemporal activity patterns that will drive the system to a pre-chosen target dynamics \cite{Norton2020, Shankar2022, Sinigaglia2023}. 

The work described in this article belongs in this second class.  We study the optimal control theory of the classic active matter theory, a dry 2D active polar fluid, first considered by Toner and Tu \cite{Toner1995, Toner1998}. We treat the convective speed of the active particles as the control parameter and identify control solutions that drive the system to targeted steady states, including forcing an aster to move to a given spatial location, causing propagating stripes to reorient along a particular direction, and driving a propagating stripe to convert into a stationary aster.  In the previous work most closely related to this one, Norton et al. \cite{Norton2020} obtained a control solution to switch a confined active nematic between two symmetric attractors (changing handedness of a circular flow state). Here, we show that an optimal control framework can solve  much more diverse  problems, including switching among attractors with very different broken symmetry patterns and driving the system into states which are not stable attractors at a given set of parameters. Further, we analyze the identified optimal solutions to uncover generic control principles that are broadly applicable to active polar fluids.

This paper is laid out as follows. In section~\ref{sec:model}, we review the hydrodynamic theory and describe the key features of the steady states that arise in the absence of control. Section~\ref{sec:optimalControl} describes the method for implementing optimal control theory.  In section~\ref{sec:results}, we report the results of the control solutions for driving the system to particular steady states or switching between them. Then, we analyze the control solutions in terms of the dynamical equations of motion to identify the essential mechanisms that drive the system into the desired behavior, with the goal of identifying generic principles. Finally section~\ref{sec:discussion} concludes with a discussion on testing these results in experiments.


\begin{figure*}
\centering
  \includegraphics[width=\textwidth]{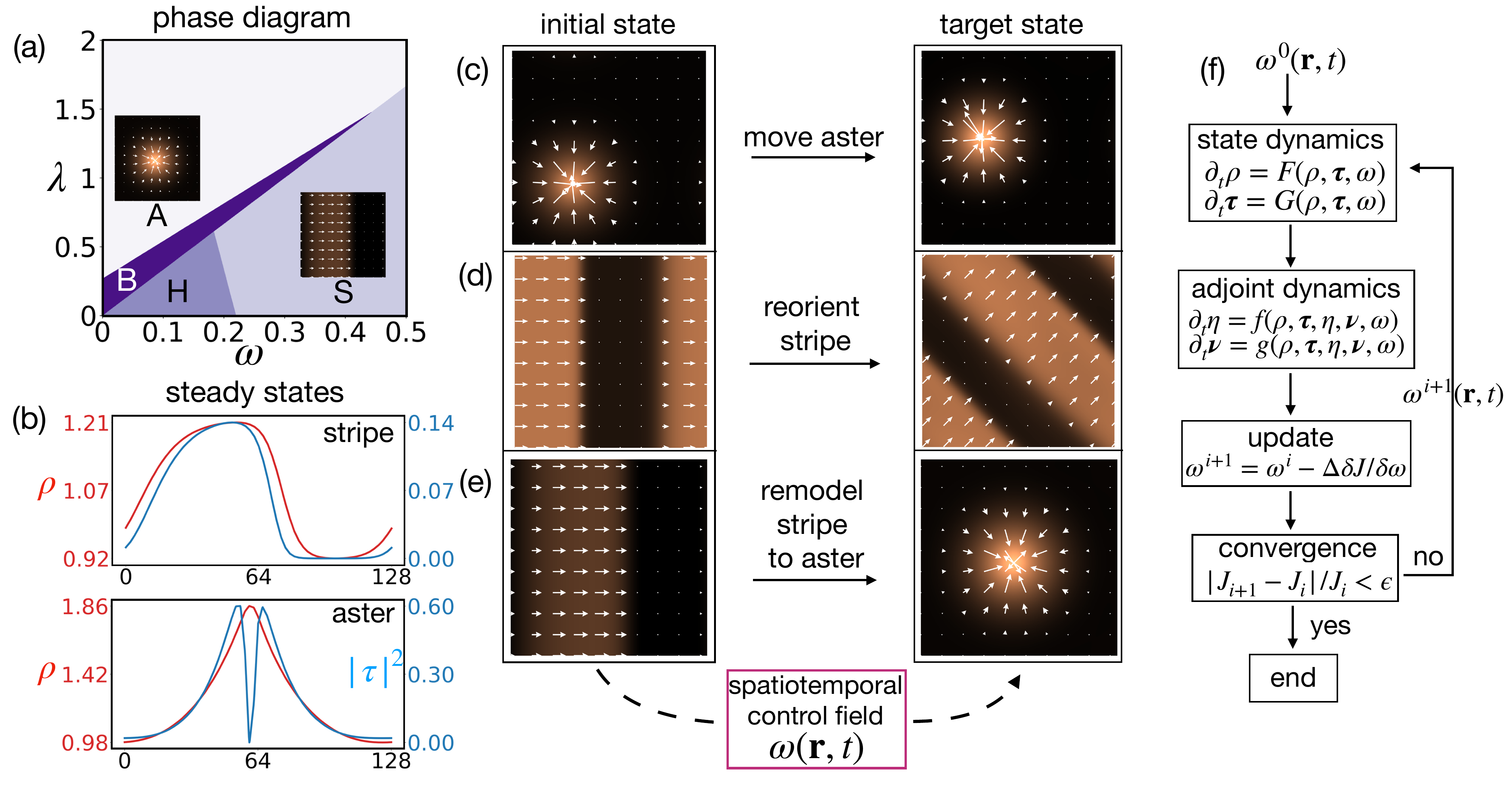}
  \caption{ Phenomenology of the bulk active polar fluid in the absence of control: \textbf{(a)} Phase diagram and representative snapshots of the inhomeogenous steady states as a function of model parameters $\lambda$ and $\omega$, with the mean density set to  $\rho_{0} = 1.07$. The propagating stripes and asters are phase-separated domains of high polar order in a background of a low density disordered state. The intermediate state labeled B (which is not relevant for the present work) corresponds to a state of transient blobs of polar order coexisting with a disordered background.\textbf{(b)} Linear scans of the density $\rho$ and magnitude of polarization field $\boldsymbol{\uptau}$ along the direction perpendicular to the interface of the phase-separated domains. In the  stripes, the polar order is homogeneous in the domain while the aster is a domain of high splay with a defect with topological charge $+1$ at its center.  \textbf{(c-e)}   Illustration of optimal control goals considered in this study. \textbf{(f)} Schematic of the method that we use to solve the optimal control problem, the direct adjoint looping (DAL) method.}
  \label{fig:fig1}
\end{figure*}

\section{Model}
\label{sec:model}

 We consider a macroscopic description of an active polar fluid in 2D, in terms of a conserved density field $\rho(\boldsymbol{r}, t)$ and the polarization field $\boldsymbol\uptau(\boldsymbol r, t) = \rho(\boldsymbol{r}, t)\boldsymbol P(\boldsymbol{r}, t)$, a vector characterizing orientational order in the fluid. As noted by Toner and Tu \cite{Toner1995}, the order parameter is also a velocity that convects mass in an active fluid. While a number of distinct dynamical equations for these hydrodynamic quantities have been considered in the literature \cite{Lee2001,Husain2017,Geyer2018,Worlitzer2021}, the particular model we study is of the form 
 \begin{eqnarray}
    \partial_{t}\rho = -\boldsymbol{\nabla}\cdot({\omega}\boldsymbol{\uptau} - D\boldsymbol{\nabla}\rho) \label{rho1}
\end{eqnarray}
\begin{eqnarray}
    \partial_{t}\boldsymbol{\uptau} + \lambda_{1}\boldsymbol{\uptau}.\boldsymbol{\nabla}\boldsymbol{\uptau}= -\nu(a_{2}(\rho) + a_{4}(\rho)|\boldsymbol{\uptau}|^2)\boldsymbol{\uptau} - \boldsymbol{\nabla}({\omega}\rho) \nonumber \\
    + K\boldsymbol{\nabla}^{2}\boldsymbol{\uptau} + {\lambda_{2}}\boldsymbol{\uptau}_{\alpha}\boldsymbol{\nabla}\boldsymbol{\uptau}_{\alpha} + \lambda_{3}\boldsymbol{\uptau}\boldsymbol{\nabla}.\boldsymbol{\uptau} \label{tau1}.
\end{eqnarray}
We briefly discuss the physics it captures and the emergent phenomenology that results from it. Extensive studies can be found   in prior work on this model \cite{Mishra2010,Gopinath2012,Caussin2014,Reinken2018,Ngamsaad2018}.  

The density dynamics Eq.~\ref{rho1} is an advection-diffusion equation where the advective velocity is proportional to the orientational order parameter $\boldsymbol P(\boldsymbol{r}, t)$.  The dynamics of the orientational order has three features: (i) It has a self-convection term with coefficent $\lambda_{1}$, encoding the absence of Galilean invariance in the `dry' theory \cite{Toner1995}. (ii) It has terms consistent with Model A dynamics that drive the system downhill on a free energy landscape where $F=$ $\int_{\mathbf{r}}\frac{a_2(\rho)}{2}|\mathbf{\boldsymbol{\uptau}}|^{2} + \frac{a_4(\rho)}{4}|\boldsymbol{\uptau}|^{4} + \frac{K}{2}(\partial_{\alpha}\uptau_{\beta})(\partial_{\alpha}\uptau_{\beta}) + \frac{\lambda}{2}|\boldsymbol{\uptau}|^{2}\nabla\cdot\boldsymbol{\uptau} $, encoding the fact that the flocking occurs due to spontaneous symmetry breaking. (iii) It contains a hydrostatic pressure term of the form $P\equiv \omega \rho - \frac{\lambda_{2}}{2} |\boldsymbol{\uptau}|^2$, where $\lambda_2$ encodes the tendency for elongated self-propelled particles to splay due to recollision events \cite{Bertin2006,Gopinath2012,Peshkov2012}.



In this study of spatiotemporal control, we choose $a_2(\rho) = (1 - \rho/\rhoC)$ and $a_4(\rho) = (1 + \rho/\rhoC)/\rho^{2}$,   yielding a continuous mean-field transition from an isotropic $\boldsymbol{\uptau}$ to a homogeneous, polar or a swarming state ($|\boldsymbol{\uptau}|> 0$) at the critical density $\rho = \rhoC$. For simplicity, we choose $D = K$. 
Without loss of generality we set the critical density $\rhoC = 1$. We further reduce the number of independent parameters and set $\lambda_{1} = \lambda_{2} = \lambda_{3}$. We set the unit of time as $\uptau_0=\nu^{-1}$, the relaxation time scale of the orientation field, and the unit of length as $l_0=(D/\nu)^{\frac{1}{2}}$.The simplified dimensionless equations are:   
\begin{align}
    \partial_{t}\rho = -\boldsymbol{\nabla}\cdot({\omega}\boldsymbol{\uptau} - \boldsymbol{\nabla}\rho)
    \label{eq:density}
\end{align}
\begin{align}
    \partial_{t}\boldsymbol{\uptau} = -(a_{2}(\rho) + a_{4}(\rho)|\boldsymbol{\uptau}|^2)\boldsymbol{\uptau} - \boldsymbol{\nabla}({\omega}\rho) + \boldsymbol{\nabla}^{2}\boldsymbol{\uptau} +\nonumber\\ {\lambda}\left(\boldsymbol{\uptau}_{\alpha}\boldsymbol{\nabla}\boldsymbol{\uptau}_{\alpha} + \boldsymbol{\uptau}\boldsymbol{\nabla}.\boldsymbol{\uptau} - \boldsymbol{\uptau}.\boldsymbol{\nabla}\boldsymbol{\uptau}\right).
    \label{eq:uptau}
\end{align}
They involve three parameters: (1) the mean density $\rho_{0}$, which is set by the initial condition; (2) the dimensionless convective velocity $\omega$; and (3) $\lambda$, which is controlled by the strength of interparticle interactions.


The phenomenology of this model is described in \cite{Gopinath2012} and summarized in Fig. ~\ref{fig:fig1}a. For the purposes of this work, we note that the dynamics of this system admits two inhomogeneous steady states: (i) propagating stripes composed of ordered swarms moving through a disordered background, which are referred to as polar drops elsewhere in the literature \cite{Solon2013,Caussin2014}, when $\omega/\lambda \gg 1$,  and (ii) a stationary high density aster, again in an isotropic background, when $\lambda / \omega \gg 1$.  Both of these states arise close to the threshold density for orientational ordering (which we set to $\rhoC=1$), and correspond to the system phase separating into a dense ordered phase and a dilute disordered phase. In this study, we fix the homogeneous density at $\rho=1.07$ and consider the problem of controlling these inhomogeneous steady states using spatiotemporal patterning of the convective strength $\omega$, referred to as the activity or control parameter in the rest of this paper.

This model has the virtue of mathematical simplicity and a minimal number of parameters, thus enabling physical insight from the optimal control solutions. At the same time, the states we seek to control are directly realizable in experiments (see section~\ref{sec:discussion}). Further, in the SI \cite{SIref}, we show that this theory is linearly controllable on short length scales. Thus it is ideal for investigating applications of optimal control in active systems.



\section{OPTIMAL CONTROL}
\label{sec:optimalControl}

\begin{figure*}
  \centering
  \includegraphics[width=\textwidth]{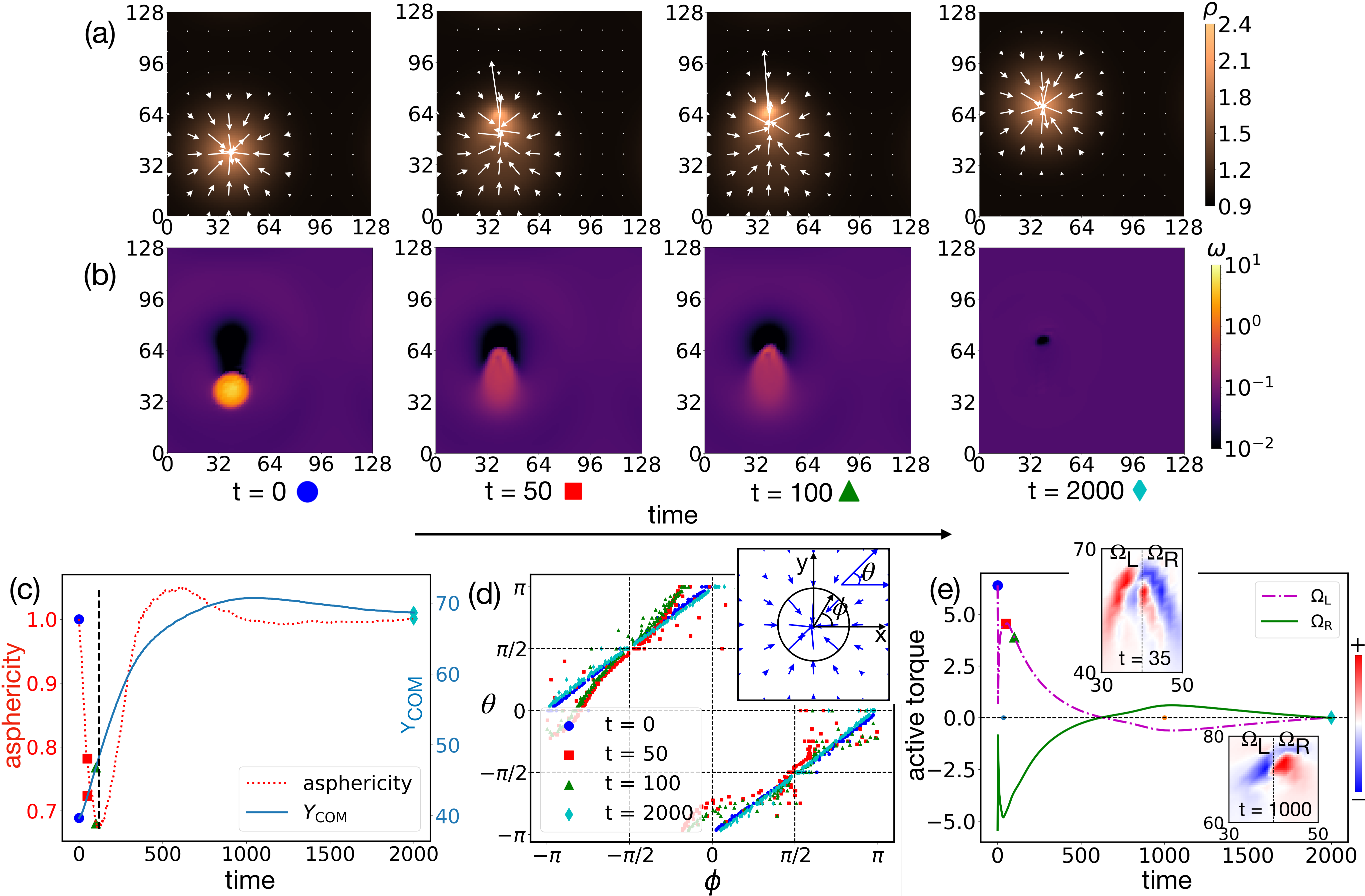}
  \caption{\textbf{Advecting an aster.}  The control solution for moving an aster from ($\xII = 60, \yII = 40$)  at $t=0$ to ($\xF = 60, \yF = 70$) at time $\tF=2000$. All length and timescales are presented in dimensionless units, which are defined in section~\ref{sec:model}.  \textbf{(a,b)} Snapshots of (a) the density $\rho$ (color map) and polarization $\boldsymbol{\uptau}$ (arrows) profiles and (b) the control solution activity field $\omega^2$ (color map). At times $t = 50, 100$ the upper quadrant of the aster unwinds and the aster becomes prolate  while the aster core maintains the  $+1$ defect. At $t=2000$ the aster is reformed at the target location. \textbf{(c)} Analysis of the aster shape: The left y-axis shows the asphericity of the aster (defined as the ratio of eigenvalues of the shape tensor, see section~\ref{sec:asterAdvection}) and the right y-axis shows the y-coordinate of the center of mass of the aster as a function of time. \textbf{(d)} The aster profile, as characterized by the polarization direction $\theta$ as a function of the azimuthal angle $\phi$ around the aster center (defined as the position at which the density is maximum, which coincides with the defect core, where $\uptau=0$). The measurement is taken at radius $r=20$ from the core.    \textbf{(e)} The active torque, $\boldsymbol{\nabla}\omega\times\boldsymbol{\uptau}$ , integrated over the left, $\OmegaL$, and right, $\OmegaR$, subdomains of the aster (see section~\ref{sec:asterAdvection}) as a function of time, showing the driving forces for unwinding and closing of the aster in each subdomain due to activity gradients. 
  The objective function parameters are $\{A, B, C, D, K\} = \{0.1, 1.0, 2.0, 2.0, 0\}$ and the simulation box size is $128 \times 128$. The baseline control value for all results reported here is $\omega_0=0.2236$. 
  A video of this trajectory is in supplemental Movie S1  \cite{SIref}.
\label{fig:asterAdvection}
 }
\end{figure*}

\begin{figure*}
  \includegraphics[width=\textwidth]{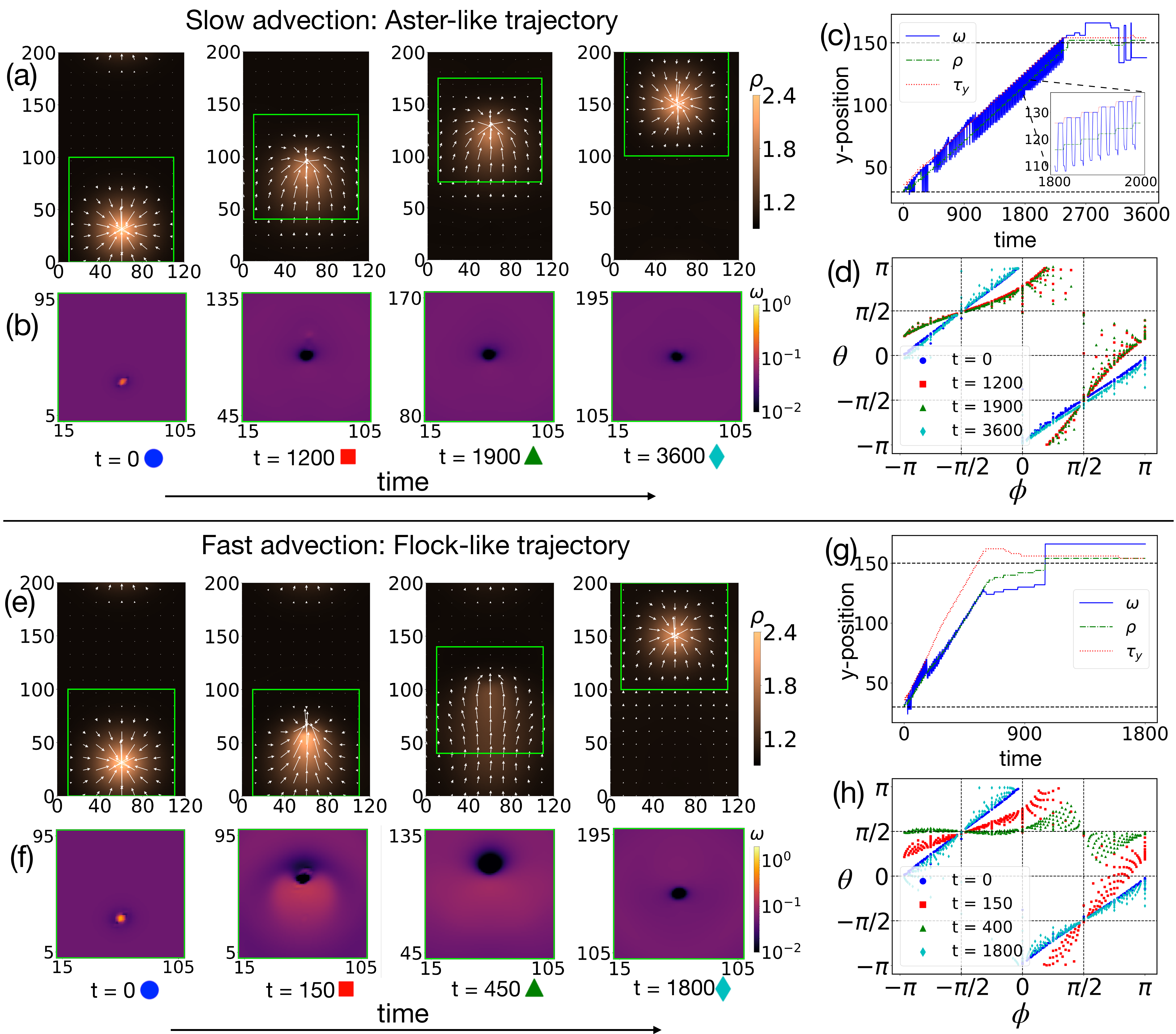}
  \caption{\textbf{Prescribing the path and speed of aster advection.}   The control problem is formulated in two stages:  In advection, the aster moves at a mean speed of $\vBar$ for $\tOne$ time units; in reformation, the aster reacquires its steady-state profile over a timescale $\tTwo$. The figure shows two examples. In both cases, the control task is to move the aster by 120 length units in the $\hat{y}$ direction. \textbf{(a-d)} Example 1: Slower advection, with $\vBar=1/20$, $\tOne=2400$, $\tTwo=1200$. \textbf{(e-h)} Example 2: Faster advection, with $\vBar=1/5$, $\tOne=600$, $\tTwo=1200$. \textbf{(a,e)}  Snapshots of the density (color map) and polarization (arrows) profiles for Examples 1 (a) and 2 (e). For both examples, snapshots are shown for the initial state  $t=0$, two intermediate times during the advection phase, and the final point $t=3600$. \textbf{(b,f)} Corresponding snapshots for the activity field (color map). 
\textbf{(c,g)} Tracking the progress of the aster and the control solution. The plot shows the y-components of the position corresponding to the aster core (density maximum, $\rho$, green curve); activity maximum ($\omega$, blue curve); and the minimum torque ($\uptau_y$, red curve).  \textbf{(d,h)} The polarization direction $\theta$ as a function of the azimuthal angle $\phi$, measured at a distance $r=20$ from the aster core, at indicated times. 
The objective function parameters for both examples  are $\{A, B, C, D, K\} = \{0.1, 1.0, 2.0, 2.0, 0\}$ and the simulation box size is $120 \times 200$. 
Videos corresponding to Examples 1 and 2 are provided in supplemental Movie S2  and Movie S3 \cite{SIref}.
  \label{fig:asterTrajectory}
}
\end{figure*}

\begin{figure*}
  \centering
  \includegraphics[width=\textwidth]{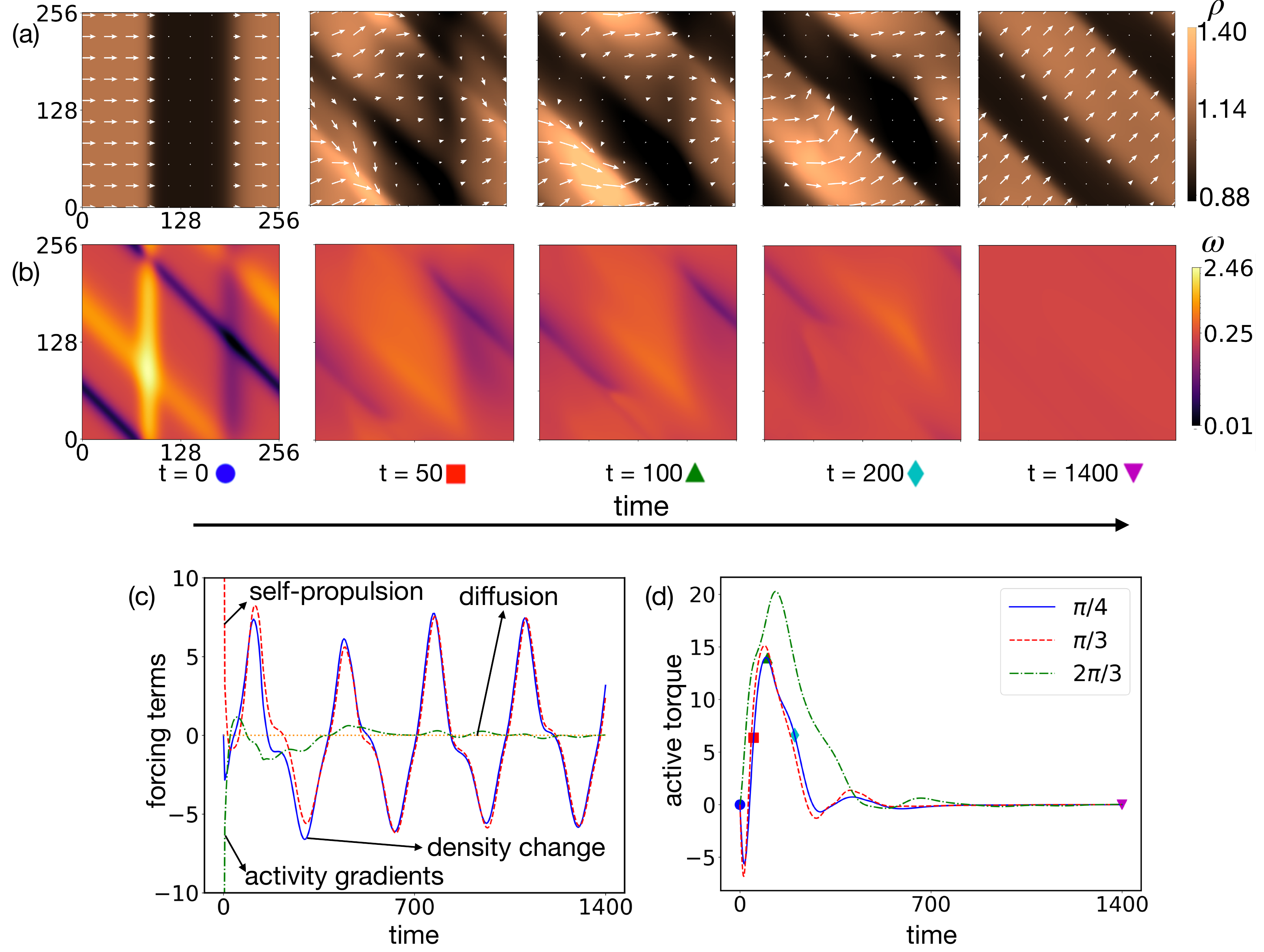}
  \caption{   \textbf{Changing the propagation direction of stripes.} \textbf{(a,b)} Snapshots of (a) the density  (color map) and polarization (arrows) profiles and (b) the activity field $\omega$ (color map). The system is initialized in an unperturbed stripe steady-state traveling in the $+\hat{x}$ direction, with parameters $\{\omega, \lambda\} = \{0.4, 0.0\}$. The control solution begins at $t=0$ and the system state is shown at indicated times. \textbf{(c)}  Contribution of each term of the density dynamics, ~\eqref{eq:density}, evaluated by integration across the defined sub-domain: $\Omega_{\delta}: 0 < x \le 50$ and $0 < y \le 100$ (see section \ref{sec:stripe}). The terms are: $\int_{{\Omega_\delta}} \partial t \rho$ (\textit{density change}), which is driven by $- \omega \nabla\cdot\boldsymbol{\uptau}$ (\textit{self-propulsion}),   $-\boldsymbol{\uptau}\cdot\nabla\omega$ (density flow due to \textit{activity gradients}) and $\nabla^{2}\rho$ (\textit{diffusion} due to density gradients). \textbf{(d)} The active torque $\boldsymbol{\nabla}\omega \times \boldsymbol{\uptau}$ integrated over the entire domain with time for different target orientations. 
The objective function parameters  are $\{A, B, C, D, K\} = \{0.1, 1.0, 7.0, 7.0, 0\}$, and the simulation box size is $256 \times 256$.
Supplemental videos S4 and S6 respectively show this trajectory and an independent run in which the stripe is forced to re-oriented by $90^{\circ}$ \cite{SIref}.  
  \label{fig:stripe}}
\end{figure*}


\begin{figure}
  \centering
  \includegraphics[width=0.5\textwidth]{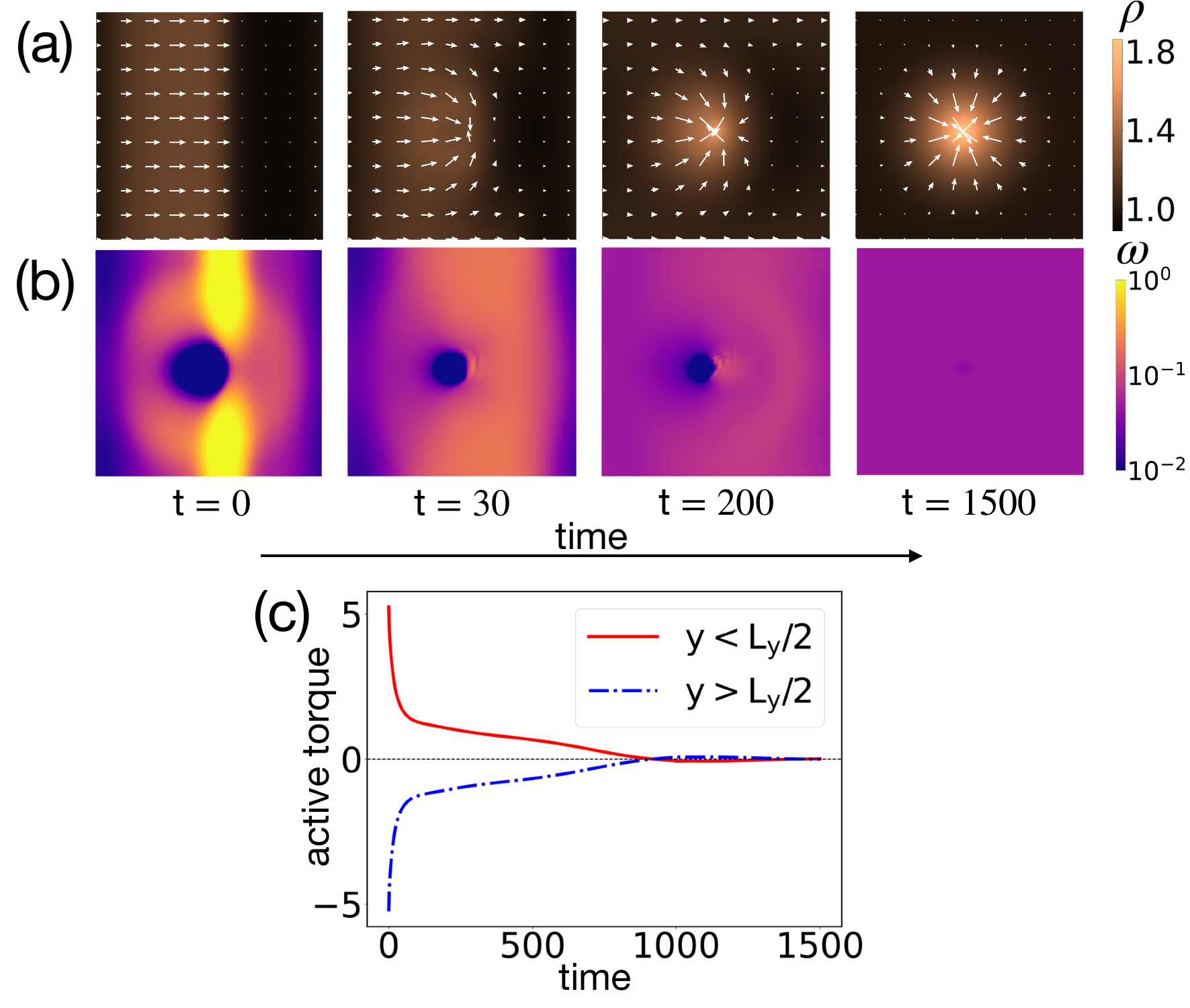}
  \caption{\textbf{Remodelling a propagating stripe to a stationary aster.} The system is initialized in an unperturbed stripe steady-state traveling in the $+\hat{x}$ direction, with parameters  $\{\omega, \lambda\} = \{0.4, 0.6\}$ seeded at $(x_0 = 60, y_0 = 60)$.   The target state is obtained from a simulation of an aster steady state with parameters $\{\omega, \lambda\} = \{0.04, 0.6\}$. 
\textbf{(a,b)} Snapshots of (a) the density  (color map) and polarization (arrows) profiles  and (b) the activity field (color map). \textbf{(c)} The active torque $\nabla\omega \times \boldsymbol{\uptau}$ integrated over time for each of the two subdomains $0 < x \le L_\text{x}$,  $0 < y \le L_\text{y}/2$ and  $L_\text{y}/2 < y \le L_\text{y}$. The objective function parameters  are $\{A, B, C, D, K\} = \{0.1, 1.0, 5.5, 5.5, 0\}$, and the simulation box size is $128 \times 128$.
A video of this trajectory is in supplemental Movie S5 \cite{SIref}.
\label{fig:stripeToAster}
}
\end{figure}


Spatiotemporal control of the system involves identifying an activity field $\omega(\mathbf{r},t)^{2}$  in an interval $t \in [0, \tF]$ such that the state of the system evolves from some initial condition {$\rho(\mathbf{r}, 0)$, $\boldsymbol\uptau(\mathbf{r}, 0)$} to a chosen target state {$\rho^{*}(\mathbf{r},t)$, $\boldsymbol\uptau^{*}(\mathbf{r}, t)$} within the control window $[0, \tF]$.  We define this in terms of $\omega^2$ to constrain solutions to positive activity. In the optimal control framework, we identify such a solution by minimizing the scalar objective function $\mathcal{J}$ defined as
\begin{align}
    \mathcal{J} = \int_{o}^{\tF} dt\int_{\Omega}d\mathbf{r} & \bigg[ \frac{A}{2}\left(\omega^{2} - \omega_{0}^2\right)^{2} + \frac{B}{2}\boldsymbol{\nabla}\omega^{2}\cdot\boldsymbol{\nabla}\omega^{2} \nonumber \\ + \frac{K}{2}(d\omega^{2}/dt)^{2}  
    & + \frac{C}{2}\left(\rho - \rho^{*}\right)^{2} +\frac{D}{2}\left(\boldsymbol{\uptau} - \boldsymbol{\uptau^{*}}\right)^{2} \bigg]
    \label{eq:objective}
\end{align}
subject to the constraints that the dynamical fields {$\rho(\mathbf{r}, t)$, $\boldsymbol\uptau(\mathbf{r}, t)$} obey Eq.(\ref{eq:density}) and Eq.(\ref{eq:uptau}) at every time point in the control window. The term $(\omega^2 - \omega_{0}^2)^{2}$ penalizes deviations of the magnitude of the control variable $\omega$ from some baseline $\omega_{0}$, and thereby discourages control solutions to be away from some chosen value of activity. 
The terms $\frac{B}{2}\boldsymbol {\nabla}\omega^{2}\cdot\boldsymbol{\nabla}\omega^{2}$ and $\frac{K}{2}(d\omega^{2}/dt)^{2}$ promote smoothness of the activity field in space and time. To simplify the presentation of results, we set $K=0$ and thus do not penalize time-variations. 

The terms $\left(\rho - \rho^{*}\right)^{2}$ and $\left(\boldsymbol{\uptau} - \boldsymbol\uptau^{*}\right)^{2}$ measure the deviations from the target state, $\rho^{*}$ and $\boldsymbol{\uptau^{*}}$.
We constrain our
search of optimal state trajectories to those that obey the
system dynamics by introducing Lagrange multipliers, $\eta$ and $\boldsymbol{\nu}$, which are adjoint variables for $\rho$ and $\boldsymbol{\uptau}$. These dynamical constraints are enforced in the optimization by adding them to the cost function as 
\begin{align}
    \mathcal{L} = \mathcal{J} + \int_{0}^{\tF}dt\int_{\Omega}d\boldsymbol{r}[ \eta \left(\partial_{t}\rho + ..\right) + \boldsymbol\nu\cdot\left(\partial_{t}\boldsymbol{\uptau} + ..\right)].
\end{align}

The necessary condition for optimality is $\nabla \mathcal{L} = 0$ \cite{Kirk2004, Lenhart2007}, so  ${\delta \mathcal{L}}/{\delta \eta}$, ${\delta \mathcal{L}}/{\delta \boldsymbol{\nu}}$, ${\delta \mathcal{L}}/{\delta \rho}$, ${\delta \mathcal{L}}/{\delta \boldsymbol{\uptau}}$, ${\delta \mathcal{L}}/{\delta \omega}$, ${\delta \mathcal{L}}/{\delta \rho(\tF)}$, ${\delta \mathcal{L}}/{\delta \boldsymbol{\tau}(\tF)}= 0$. The first two conditions return  equations (\ref{eq:density})-(\ref{eq:uptau}) governing 
$\rho$ and $\boldsymbol{\uptau}$. The following two conditions yield the dynamical equations for the adjoint variables $\eta$ and $\boldsymbol{\nu}$, 
\begin{align}
    \partial_{t}\eta = & C\left(\rho - \rho^{*}\right) - \boldsymbol\nabla^{2}\eta - \omega^{2}\boldsymbol{\nabla}\cdot\boldsymbol{\nu}\nonumber \\
     & + (\delta_{\rho}a_{2}(\rho) + \delta_{\rho}a_{4}(\rho)|\boldsymbol{\uptau}|^{2})(\boldsymbol{\nu}.\boldsymbol{\uptau}).
     \label{eq:eta}
\end{align}
\begin{align}
    \partial_{t}\boldsymbol{\nu} = & D\left(\boldsymbol{\uptau} - \boldsymbol{\uptau^{*}}\right) + \left(a_{2}(\rho) + a_{4}(\rho)|\boldsymbol{\uptau}|^{2}\right)\boldsymbol{\nu}  + 2a_{4}(\rho)\boldsymbol{\uptau}(\boldsymbol{\nu}\cdot\boldsymbol{\uptau}) \nonumber \\
    & - \lambda\left[-\boldsymbol{\uptau}\boldsymbol{\nabla}\cdot\boldsymbol{\nu} + 2\boldsymbol{\nu}\boldsymbol{\nabla}\cdot\boldsymbol{\uptau} -\boldsymbol{\nabla}(\boldsymbol{\nu}\cdot\boldsymbol{\uptau}) \right. \nonumber \\
    & \left. + \boldsymbol{\uptau}\cdot\boldsymbol{\nabla}\boldsymbol{\nu} - \boldsymbol{\nu_{\alpha}}\boldsymbol{\nabla}\boldsymbol{\uptau_{\alpha}}\right]  -\boldsymbol{\nabla}^{2}\boldsymbol{\nu} -\omega^{2}\boldsymbol{\nabla}\eta.
    \label{eq:nu}
\end{align}
with boundary conditions at $\tF$ : $\{ \eta, \boldsymbol{\nu} \}(\boldsymbol{r}, \tF) = 0$, and periodic boundary conditions on the domain. ${\delta \mathcal{J}}/{\delta \omega} = 0$ yields an equation to update the control input as,
\begin{align}
      2 A\omega\left(\omega^{2} - \omega_{0}^2\right) - 2 B\omega\boldsymbol{\nabla}^{2}\omega^{2} - 2 K \omega (d^{2}\omega/dt^{2}) \nonumber \\
     - 2\omega\boldsymbol{\uptau}\cdot\boldsymbol{\nabla}\eta  - 2\omega\boldsymbol{\nu}\cdot\boldsymbol{\nabla}\rho = 0.
\label{eq:dOmega}
\end{align}

We use the direct-adjoint-looping
(DAL) method \cite{Kerswell2014} to minimize the cost function under the constraint that the dynamics
satisfies Eqs.~\eqref{eq:density} and \eqref{eq:uptau},  to yield the optimal schedule of activity in space and time (see SI sections IV and V for more details). 
Specifically, we construct an initial condition by performing a simulation with unperturbed dynamics (Eqs.~\eqref{eq:density} and \eqref{eq:uptau}) until reaching steady-state, at a parameter set that leads to a desired initial behavior. We construct a target configuration in the same manner, using a different parameter set that leads to the desired target behavior. We also specify a time duration $\tF$ over which the control protocol will be employed, and an initial trial control protocol. 
We then perform a series of DAL
iterations; in each iteration the system and the adjoint dynamics are integrated from the initial condition for time $\tF$ under the current control protocol, and the cost function (Eq.~\eqref{eq:objective}) is computed from the
resulting trajectory. The adjoint equations are integrated backwards in time to propagate the residuals. After each backward run,  the control protocol is updated via gradient descent, $\omega^{i+1} = \omega^{i} - \Delta {\delta \mathcal{J}}/{\delta \omega} $, to minimize the cost function. We employ
Armijo backtracking \cite{Borzi2011} to adaptively choose the step-size for gradient descent and to ensure convergence of the DAL algorithm. We have implemented this calculation in the open-source Python finite element method solver FEniCS \cite{Dupont2003}.

\section{RESULTS}
\label{sec:results}


Using the optimal control framework described in section~\ref{sec:optimalControl} and the hydrodynamic equations  Eqs.~\eqref{eq:density} and \eqref{eq:uptau}, we have computed spatiotemporal control solutions that steer the system state toward the target configuration, for each of the target behaviors shown in Fig.~\ref{fig:fig1} (c-e). In this section, we describe these calculations, and physical insights that can be learned by studying the computed control solutions.

\subsection{Aster advection} 
\label{sec:asterAdvection}

First, starting in a parameter regime where a stationary aster is stable ($\omega = 0.05$ and $\lambda = 0.8$), we seek to advect an aster to a new location. The control problem specifies the initial and final states of the system as well as the elapsed time; that is, the spatial dependence of the density and polarization fields at every point in space at times $t=0$ and $t=\tF=2000$.

Fig.~\ref{fig:asterAdvection} summarizes the results of this computation. Fig.~\ref{fig:asterAdvection}a,b respectively show the time evolution of the system configuration and the applied control field that drives the transformation. At early times, the applied control is strongest at the aster core while it is lowest in front of the aster along the direction we seek to move it.  The aster then elongates to assume a comet-like shape, with a denser, polar-ordered head (see snapshots at $t= 50, 100$), as it advects toward the target location. Note that the activity is largest to the rear of the aster in this region. Thus, the control solution pushes (rather than pulls) the aster. 

To quantify how the position and profile of the aster change over the course of advection, we measure its center-of-mass position $(\xCOM, \yCOM)$ and asphericity. Here, we track the y-coordinate of the center of mass, which is calculated as $\yCOM = \Sigma_{{\text{i}\text{j}}:\rho_{\text{i}\text{j}} > \rho_{0}}\text{j}\rho_{\text{i}\text{j}}/\Sigma_{\text{i}\text{j}:\rho_{\text{i}\text{j}} > \rho_{0}}\rho_{\text{i}\text{j}}$, and the asphericity is given by the ratio of eigenvalues of shape tensor: $\mathcal{I}_{\alpha \beta} = \sum_{\text{ij}:\rho_{\text{i}\text{j}} > \rho_{0}} \rho_{\text{ij}}\left(||\mathbf{r}_{\text{ij}}||^{2}\delta_{\alpha\beta} - r_{\alpha}^{\text{ij}}r_{\beta}^{\text{ij}}\right)$, where Latin indices denote grid points and Greek indices correspond to Cartesian coordinates, and $r_{\text{ij}}$ is the distance of the $\text{ij}^{th}$ grid point from the center of mass.  As shown in Fig.~\ref{fig:asterAdvection}c, the control window naturally partitions into two stages. During the initial stage ($0 < t \lesssim 100$) the aster rapidly changes shape into the comet-like configuration,  as seen by the decrease in its asphericity, while simultaneously undergoing advection, moving towards the target point. Then, over the remaining long time window ($100 \lesssim t < \tF)$ the aster reforms slowly, the asphericity increases back to 1, and it moves the remaining small distance to the target position. 

For further description of the aster profile during these stages, we present the  angle of the polarization field $\theta$ as a function of the azimuthal angle $\phi$ around the aster core at four time points in Fig.~\ref{fig:asterAdvection}d. At the initial time ($t=0$) the system is radially symmetric with polarization vectors pointing toward the aster center, while by $t=50$ and $t=100$ the symmetry of $\theta$ in the top $[0,\pi]$ and bottom  $[-\pi,0]$ quadrants is broken, with more polarization at the bottom points toward the advection direction, and the front-end remains aster-like with a radial configuration. The orientation returns to the aster configuration by $t=\tF$.


Further, we can understand the control solution physically by noting that the dynamics of $\boldsymbol{\uptau}$ is such that gradients in the control field $\omega$ create a torque on the orientation field, i.e., $\partial_{t}\boldsymbol{\uptau} \sim   -\rho\boldsymbol{\nabla}\omega$ or equivalently $\partial_{t}\theta \sim \boldsymbol{\nabla}\omega\times\boldsymbol{\uptau}$. 
%
%
We then calculate the integral of the torque ($ \uptau_{\pm} =  \int_{\Omega'} \partial_t \theta$) over two domains,   $\Omega_{\text{L}}: [x \in \{x_0 - 10,  x_0\}, y \in \{0,  L_{y}\}]$ on the left and and $\Omega_{\text{R}}: [x \in \{x_0,  x_0  + 10\}, y \in \{0,  L_{y}\}]$ to the right, with $x_{0} = 40$ in our case (Fig.~\ref{fig:asterAdvection}e).  When the aster unwinds and advects, the region to the left has a positive torque (countercockwise rotation) and the region on the right experiences negative torque (clockwise rotation), which correspond to the partial unwinding of the aster. This is also illustrated by the snapshot shown for $t = 35$, where the dashed line ($x = 40$) represents the axis along the aster's motion. During the subsequent reformation phase, as the aster regains circular symmetry, the profile winds back such that the left/right subdomains experience clockwise/counterclockwise torque respectively. The snapshot shown at $t = 1000$ illustrates this behavior. Eventually, the system relaxes sufficiently close to an unperturbed aster configuration that the net torque becomes zero.

\subsubsection{\textit{Specifying the trajectory of aster advection}}
A limitation of the approach described thus far is that convergence of the control solution becomes unreliable when trying to advect the aster over distances significantly greater than its size ($\Delta \xBar \approx 30$). This is because the gradients in the objective function are extremely shallow for the initial stages of the trajectory when the target state is far from the initial state. 
While techniques to find global minima can potentially overcome this problem, an alternative approach is to change the objective function to ensure sufficient gradients at all stages. A simple example of the latter strategy is to prescribe the entire trajectory of the aster. This approach has the added benefit of controlling the translocation speed, but has the potential drawback of arriving at a suboptimal solution (either slower translocation or higher control cost), since the problem is more constrained.

We applied the latter strategy to the problem of translocating an aster a distance of $\Delta \xBar=120$. We formulated the control problem to translocate the aster at a constant speed $\vBar$ for a time $\tOne$, follow by a time $\tTwo$ for reformation of the aster. We find that specifying the path in this way allows specifying a target distance that is arbitrarily far without any difficulties in achieving convergence of the control solution.



Figs.~\ref{fig:asterTrajectory}a,b show the system configurations and corresponding control solution $\omega$ for an example with $\vBar=1/20$,  $\tOne=2400$, and $\tTwo=1200$. At $t=0$, the activity is maximum at the core of the aster, but unlike the previous setup where only initial and final state of the aster are specified, the order of magnitude remains same throughout the advection phase of aster.  During the first phase of the solution (constant advection), the aster undergoes partial dissolution and, as intended, a roughly steady rate of translation toward the target (see snapshots at $t=1500, 2500$). However, because we specified the trajectory at discrete intervals spaced by $\delta \xBar=1$, the optimal control field oscillates with a period of about $\delta t \approx \delta \xBar / \vBar = 20$.  This behavior is evident in Fig.~\ref{fig:asterTrajectory}c, which shows the positions of the maxima of  $\rho$ and $\omega$ and the minimum of the y-component of the torque, $\uptau_y$, as a function of time. The maximum in the control solution $\omega$ exhibits strong oscillations of $~20$ length units between the front and rear of the aster (while the activity remains low at the aster core, see Fig.~\ref{fig:asterTrajectory}b), whereas the density maximum moves at a nearly continuous speed toward the target.  The minimum $\uptau_y$ tracks polarization toward the $-\hat{y}$ direction and it consistently coincides with the high activity point at the front of the aster. Taken together, these observations show that the control solution pushes the aster from the rear, while exerting torques at the front that maintain aster-like polarization.  This is captured in Fig.~\ref{fig:asterTrajectory}d which shows  $\theta$ as a function of the azimuthal angle $\phi$ at two intermediate times during the advection phase, t= 1200 and 1900. Finally, during the second (reformation) phase, the aster re-acquires its radially symmetric steady state density and polarization profile. The dynamical interplay among these forces can be seen in the supplemental video \cite{SIref}.


Since we are specifying the path of the aster, we can investigate how the control solution depends on the chosen advection rate. Fig.~\ref{fig:asterTrajectory}e-h shows analogous results for a trajectory in which the advection phase is shortened to $\tOne=600$, forcing a higher translation speed $\vBar=1/5$. The higher speed leads to a qualitatively different type of trajectory; the aster unwinds into a flock during the advection phase, and then reforms during the second phase.  Here the control solution takes a bean-shaped spatial profile, which initially pulsates periodically to unwind the leading edge of the aster and push the aster toward the target position. At early times (by $t=150$) the rear of the aster adopts a flock-like state with polarization primarily pointing in the $\hat{y}-$direction; by $t=400$ most of the aster becomes flock-like. The extent of polarization along $\hat{y}$ is particularly clear from the plot of $\theta(\phi)$ at $t= 400$ (Fig.~\ref{fig:asterTrajectory}h, green triangles).  

These two control problems indicate that the specification of the cost function can result in very different activity profiles and intermediate states in the optimal solution. We can introduce other metrics and constraints to obtain solutions that are potentially more readily implementable in experiments. 

\subsection{Changing the direction of propagating stripes.}
\label{sec:stripe}
Next, starting at a parameter set for which stripes are stable, we obtain an activity profile to change the stripe propagation direction, with initial direction along $+\hat{x}$ and a target direction diagonally oriented along $45^\circ$. Note that we obtain similar results for any target orientation, including reversing the stripe direction by $180^\circ$.  Fig.~\ref{fig:stripe} a,b show the system configurations and corresponding control solutions at several time points. At $t=0$, the applied activity is strongest at the leading edge of the stripe, and decays over the width of the leading boundary layer (i.e., the region where the polarization changes from isotropic to uniform). These gradients in activity lead to both melting (reduction of the magnitude of polarization) and turning (reorientation of polarization toward the target direction). At the next two time-points ($t=100, 200$) the activity has decreased in magnitude, but continues to turn the polarization.  By $t=500$ the activity is nearly uniform in space and approaching its steady-state value of 0.4. However, some curvature remains near the leading edge of the stripe. The stripe has completely reformed by the last time point ($t=1400$). 
Notably, the timescale for dissolving and reorienting the stripes at this periodic box size $256\times 256$ (SI Movie S4 \cite{SIref}) was about $500$ in our dimensionless units, which is 2 orders of magnitude lower than obtaining stripes from a random homogenous initial condition in the absence of control.

While an intuitive route to reorienting a strip would be to melt the stripe to an isotropic domain and then have it reform in the new direction, this is not the optimal solution given by the control theory.  Starting with the density equation, Eq.\eqref{eq:density}, we investigate the primary forces influencing density evolution during the stripe reorientation process. For this we choose a subdomain $0 < x \le 50, 0 < y \le 100$ within the simulation box of size $256 \times 256$. 
 We integrate each of the three terms in Eq.\eqref{eq:density} over the specified subdomain as a function of time (Fig.~\ref{fig:stripe}c): $- \omega \nabla\cdot\boldsymbol{\uptau}$, which governs the convection of active particles at convection speed $\omega$;  $-\boldsymbol{\uptau}\cdot\nabla\omega$, which determines the local density dynamics due to gradients in activity; and $\nabla^{2}\rho$, which determines the diffusion due to density gradients. We find that, at all times, the dominant contribution arises from self-propulsion, $- \omega \nabla\cdot\boldsymbol{\uptau}$; contributions from gradients in activity and density have negligible contributions. Thus, we conclude that activity gradients are not the driving force for the density dynamics, but rather lead to the torques that reorient the polarization field, as described in our analysis in \ref{sec:asterAdvection}. To quantify the effect of active torque in this case, we illustrate in Fig ~\ref{fig:stripe}c that as the difference in orientation between the initial state and target state increases, the active torque also increases, and as the system settles into the target orientation, the active torque goes to 0.

\subsection{Remodeling stripe to aster}
So far, we have considered cases where the initial and target states are both steady states of the uncontrolled forward dynamics of our system at specified parameters. To demonstrate the power of the control theory,  we start in a parameter regime in which propagating stripes are stable, and obtain an activity profile that drives the system into a stationary aster (not a steady state at these parameters). For the initial condition, we run to steady-state under parameters that lead to propagating stripes, $\lambda = 0.6$ and $\omega=0.4$. 
To obtain a configuration to specify the target state, we perform an independent simulation in which we obtain a stationary aster steady-state with $\lambda = 0.6$ and $\omega=0.04$. Fig.~\ref{fig:stripeToAster} shows the trajectory and corresponding control solution. We see that at early times the applied activity is strongest at the top- and bottom-edges of the leading boundary layer of the stripe, which bends the polarization vectors toward the core of the target aster. The magnitude of the activity field decreases quickly in time, but the spatial profile remains similar, thus continuing to steer polarization toward the core, and decreasing the net momentum in the $+\hat{\boldsymbol x}$ direction. Due to the coupling between $\rho$ and $\boldsymbol{\uptau}$ (see  Eq.~\eqref{eq:density}), the resulting gradients and polarization lead to convection of density toward the core. By $t=1500$, there is a density maximum at the core and the system has achieved a radially symmetric state, which leads to a balance of propulsion forces and thus a stationary state.


\begin{figure*}[hbt]
  \centering
  \includegraphics[width=\textwidth]
  {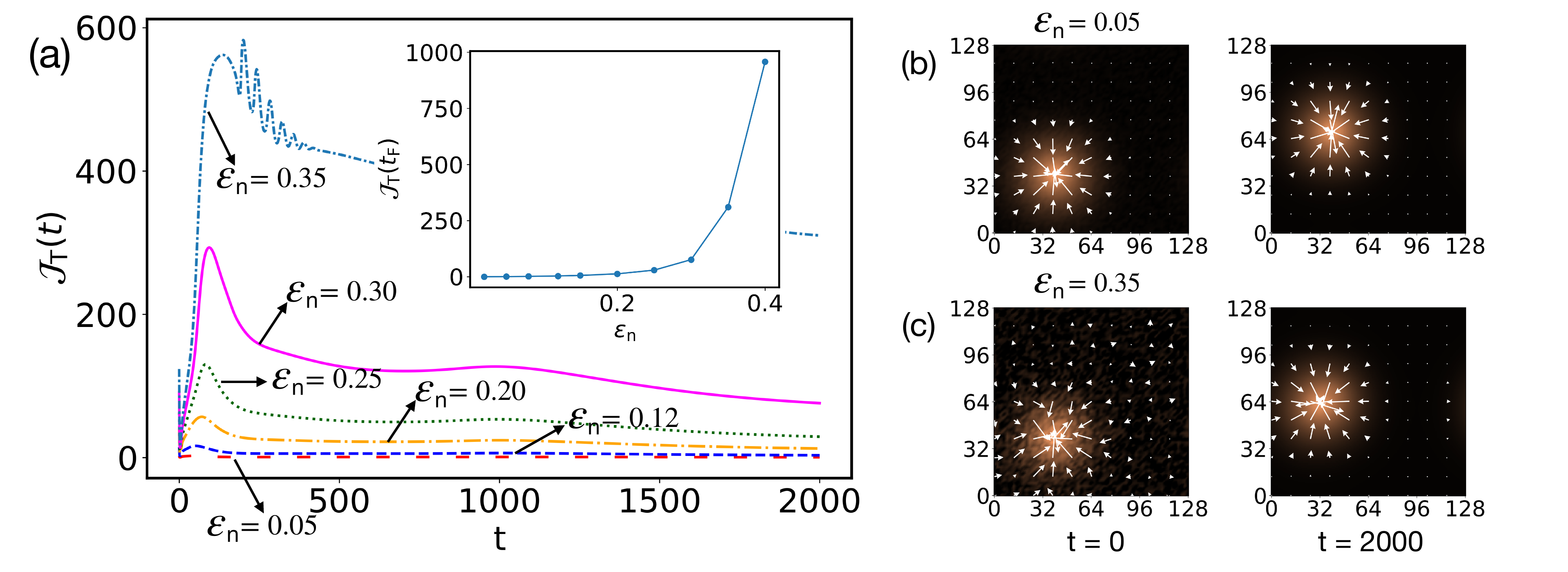}
  \caption{\textbf{Robustness of the control solution to adding noise to the initial condition.} \textbf{(a)} The plot shows the deviation of the system from the target state as a function of time, $\JT(t) = \frac{1}{2} \left( (\rho(t) - \rho^{*})^{2} + (\boldsymbol{\uptau(t) - \uptau^{*})^{2}} \right)$, when adding Gaussian noise to the initial condition with indicated magnitude $\epsNoise$, and integrating the dynamics using the control protocol computed in the absence of noise ($\epsNoise=0$). The inset shows the deviation of the final state $\JT(\tF)$ as a function of noise magnitude $\epsNoise$. \textbf{(b,c)} The initial and final states for (b) small ($\epsNoise=0.05$) and (c) large ($\epsNoise=0.35$) noise. We see robust aster structures even when the cost function is large in the end state for 35\% noise.
\label{fig:noise}
}
\end{figure*} 

\subsection{Robustness of optimal control solution to noise.} Since models are never completely accurate and noise is inevitable in any experimental system, we investigated the robustness of our control solution to errors or noise. Because our implementation uses deterministic PDEs, we tested the effects of noise by perturbing the initial condition for the aster translocation problem presented in Fig.~\ref{fig:asterAdvection}. Specifically, we added Gaussian noise with a relative magnitude $\epsNoise$ to the values $\rho, \uptau_x,$ and $\uptau_y$ at each pixel, and then integrated the dynamics with the control protocol computed in the absence of noise. Fig.~\ref{fig:noise} shows the performance of the control solution. We plot the deviation from the target state,  $\JT(t) = \frac{1}{2} \left( (\rho(t) - \rho^{*})^{2} + (\boldsymbol{\uptau(t) - \uptau^{*})^{2}} \right)$, and the inset shows the deviation of the final state from the target $\JT(\tF)$, as a function of $\epsNoise$. We see that noise has a relatively small effect on the performance up to a magnitude of about $\epsNoise=20\%$, after which deviations in the objective function rise dramatically. However, even with $\epsNoise=35\%$ and a relatively large value of $\mathcal{J}\approx 350$ at $\tF$, the final state is remarkably close to the target in all qualitative aspects (see Fig.~\ref{fig:noise}c), with a well-formed aster close to the target position. Thus, we conclude that the optimal control solution is robust to noise, at least in the initial condition.

\section{Discussion and conclusions}
\label{sec:discussion}

We demonstrate an optimal control theory framework that can prescribe activity patterns to guide an active material into desired emergent behaviors, focusing on an active polar fluid as a model system. The capabilities include programmed switching among dynamical attractors with very different dynamics and distinct broken symmetry patterns, and reprogramming the dynamics of existing attractor states. As an example of the former, we identify a spatiotemporal activity pattern that converts a propagating stripes state into a stationary aster. As an example of the latter, we show that a stationary aster can be programmed to self-advect to a new target configuration, either via an arbitrary trajectory or along a prescribed path. Similarly, propagating stripes can be forced to reorient in arbitrary directions. Depending on the choice of terms and weights in the objective function, the spatiotemporal variations of the control inputs can be regulated to limit experimental cost or ensure smooth trajectories.

Further, we show that the optimal control solutions are robust to noise. In particular, perturbing the initial condition by up to 20\% leads to minimal quantitative deviations from the target behavior, and the solution remains qualitatively accurate for significantly larger perturbations. Also, we note that additional strategies can be employed for experimental systems where larger noise sources or systematic errors are unavoidable. This includes integrating closed-loop control components. For example, one can observe the current state of the system at regular intervals along a trajectory, and recompute the optimal control solution using the current state as the initial condition. Alternatively, one can add linear feedback terms that analyze deviations from the pre-computed optimal control trajectory \cite{Bechhoefer2021}.

In addition to directly applying the computed activity protocols, examination of their forms provides both fundamental and practical insights into controlling active materials. In particular, we show how the spatial gradients in the applied activity field lead to localized torques which rotate polarization directions, leading to the programmed reformulation of the pattern of interest (e.g. aster or stripe). Unsurprisingly, the form of the trajectory is different depending on the task being encoded for --- changing the broken symmetry state of the system (e.g. stripe-to-aster, Fig.~\ref{fig:stripeToAster}) requires very different spatial arrangements of active torques then advection (Figs.~\ref{fig:asterAdvection} and \ref{fig:asterTrajectory}) or reorientation (Fig.~\ref{fig:stripe}). Notably however, the applied activity field and corresponding trajectory also depend strongly on the time allowed for the transformation. In the example of advecting the aster over a distance many times its size (Fig.~\ref{fig:asterTrajectory}), the aster mostly retains its form throughout the course of the trajectory when moving at moderate speed, but when forced to complete the journey $5\times$ faster, the applied activity field reshapes the aster into a localized flock or swarm, which reformulates into an aster upon reaching the target position. 

The states we seek to control are realizable in experiments. Propagating concentration waves of aligned self-propelled particles have been observed in dense actin motility assays \cite{Schaller2010, Huber2018, Sciortino2021, Suzuki2017} and in self-chemotactic bacterial systems\cite{Saragosti2011, Pohl2014, Bhattacharjee2022, Alert2022, Narla2021, Cremer2019, Bhattacharjee2021, Brenner1998, Fei2020} . Asters are ubiquitous in cell biology in processes such as the formation of the mitotic spindle, oogenesis, and plant cell cytokinesis. \cite{Mayer1999, Schaffner2006, Loughlin2010, ShirasuHiza2004, Heidemann1975, Heidemann1978, Schmit1983, Kumagai2001}. They can be reliably obtained in \textit{in vitro} suspensions of cytoskeletal filaments and motor proteins\cite{ShirasuHiza2004, Ndlec1997, Najma2024, Wollrab2019, Foster2015, Colin2018, Luo2013, Stam2017, Berezney2022, Thoresen2011, Koster2016, Glaser2016, Nedelec2003, Ross2019}.  In such systems, activity can be controlled in space and time by constructing active materials with optogenetic molecular motors, and using a digital light projector to shine a programmed spatiotemporal sequence of light on the sample  \cite{Lemma2023, Zarei2023,Zhang2021,Ross2019}. Thus, our results can be directly tested.

The optimal control framework presented here is highly generalizable, and can be readily applied to any system provided there is a means to externally actuate the system and there is a reasonably accurate continuum model. Importantly, the control variable need not be limited to the activity field, since the objective function can be extended to include any property of the material that can be actuated. With the recent success of automated PDE learning tools in discovering continuum models for active systems (e.g. \cite{Joshi2022, Dunkel2023, Golden2023}), applications need not be limited to systems with accurate models already available. Furthermore, since model discovery tools work better when provided with a variety of data, including from non-steady-state observations, we anticipate that combining model discovery tools with optimal control could be a powerful approach to both discover more accurate models and enhance the reliability of the control solutions.


\begin{acknowledgments}
This work was supported by the Department of Energy (DOE) DE-SC0022291. Preliminary data and analysis were supported by the National Science Foundation (NSF) DMR-1855914 and the Brandeis Center for Bioinspired Soft Materials, an NSF MRSEC (DMR-2011846). Computing resources were provided by the NSF XSEDE allocation TG-MCB090163 (Stampede and Expanse) and the Brandeis HPCC which is partially supported by the NSF through DMR-MRSEC 2011846 and OAC-1920147. AB acknowledges support from NSF-2202353. 
\end{acknowledgments}

%

%
%
%
%
%
%
%

\newpage

\onecolumngrid
\appendix

\section{ Moment of Inertia}

The moment of inertia matrix for the aster is given by
\begin{equation}
\mathcal{I}_{ij} = \sum_{k} m_{k}\left(||r_{k}||^{2}\delta_{ij} - x_{i}^{k}x_{j}^{k}\right)
\label{eq:inertia}
\end{equation}
with $m_k$ a mask to restrict the calculation to the vicinity of the aster:
\begin{align}
m_k = \Theta(\rho(x,y) - \rho_0)
\label{eq:mk}
\end{align}
where $\Theta$ is the Heaviside step function and $\rho$ is the density field. That is, $m_k$ is the density field value at all points with density above the mean value; otherwise $m_k=0$. 
 The aster's center of mass is dynamically tracked, and $x_{i}^{k}$ represents the distance of the $k$-th point from the $i$-th axis, which passes through aster's center of mass. 

Upon diagonalizing this moment of inertia matrix, we obtain the eigenvalues $\lambda_1$ and $\lambda_2$, which  correspond to the principal moments of inertia. We then define the \textit{asphericity} as the ratio $\frac{\lambda_2}{\lambda_1}$, which quantifies the extent of non-sphericity (asphericity) of the moving object.

A ratio closer to 1 indicates a more spherical shape, while significantly deviating values suggest an elongated or flattened form.

\section{ Orientation field}
We further quantify the motion of aster by measuring the orientation of polarization vectors relative to the polar axis. The reference point of the polar axis is determined by the point where density field has the maximum value. Our region of interest is a circular region of radius $\sim 10$ (in non-dimensional units), which is roughly the radius of an aster. We mask out the remaining space. This is depicted in the inset to Figure 2 in the main text, where the circle approximates the aster.

\section{Analyzing the optimal control solution: calculating the torque}
\label{sec:torque}
We start with the full hydrodynamic equation for polarization: 
\begin{align}
    \partial_{t}\boldsymbol{\uptau} = -(a_{2}(\rho) + a_{4}(\rho)|\boldsymbol{\uptau}|^2)\boldsymbol{\uptau} - \boldsymbol{\nabla}({\omega}\rho) + \boldsymbol{\nabla}^{2}\boldsymbol{\uptau} +\nonumber\\ {\lambda}\left(\boldsymbol{\uptau}_{\alpha}\boldsymbol{\nabla}\boldsymbol{\uptau}_{\alpha} + \boldsymbol{\uptau}\boldsymbol{\nabla}.\boldsymbol{\uptau} - \boldsymbol{\uptau}.\boldsymbol{\nabla}\boldsymbol{\uptau}\right).
    \label{eq:uptau}
\end{align}
During the initial stages, the applied activity profile exhibits large spatial gradients, with the predominant contribution to polarization dynamics largely stemming from the term $\nabla(\omega\rho)$: 
\begin{align}
    \partial_{t}\boldsymbol{\uptau} =   -\rho\boldsymbol{\nabla}\omega
\label{eq:torque}
\end{align}
We express $ \uptau = |\uptau|[ cos\theta \hat{x} + sin\theta \hat{y}]$, and re-write equation \eqref{eq:torque} as:
\begin{align}
    |\boldsymbol{\uptau}|\partial_{t}\left[\cos\theta\boldsymbol{\hat{x}} + \sin\theta\boldsymbol{\hat{y}}\right] =   -\rho\left[\partial_{x}\omega \boldsymbol{\hat{x}} + \partial_{y}\omega \boldsymbol{\hat{y}}\right]
\end{align}
Further simplification yields:
\begin{align}
    |\boldsymbol{\uptau}|\partial_{t}\theta\left[-\sin\theta\boldsymbol{\hat{x}} + \cos\theta\boldsymbol{\hat{y}}\right] =   -\rho\left[\partial_{x}\omega \boldsymbol{\hat{x}} + \partial_{y}\omega \boldsymbol{\hat{y}}\right]
\end{align}
\begin{align}
    |\boldsymbol{\uptau}|\partial_{t}\theta =   \rho\left[\partial_{x}\omega \sin\theta - \partial_{y}\omega \cos\theta\right].
\end{align}
Finally, this formulation results in:
\begin{align}
\partial_{t}\theta \sim \boldsymbol{\nabla}\omega\times\boldsymbol{\hat{\uptau}}/|\boldsymbol{\uptau}|
\label{eq:torqueTheta},
\end{align}
where the expression \eqref{eq:torqueTheta} elucidates the temporal evolution of $\theta$ influenced by the cross product of the gradient of activity $\boldsymbol{\nabla}\omega$ and the unit vector $\boldsymbol{\hat{\uptau}}$ in the direction of the torque $\boldsymbol{\uptau}$, normalized by the magnitude of $\boldsymbol{\uptau}$.

\section{Direct-adjoint-looping (DAL) method}
\label{sec:DAL}
 We use direct-adjoint-looping (DAL), an iterative optimization method \cite{Kerswell2014} to solve for optimal schedule of activity in space and time that accomplishes our control goals. We start by writing the Lagrangian $\mathcal{L}$ of optimization, Eq. \eqref{eq:objective}, where $\boldsymbol{\nu}$ and $\eta$ act as Lagrange multipliers or adjoint variables that constrain the dynamics to follow Eqs.~\eqref{eq:density} and \eqref{eq:uptau}.

 We construct an initial condition by performing a simulation with unperturbed dynamics (Eqs.~\eqref{eq:density} and \eqref{eq:uptau}) until reaching steady-state, at a parameter set that leads to a desired initial behavior. We construct a target configuration in the same manner, using a different parameter set that leads to the desired target behavior. We also specify a time duration $\tF$ over which the control protocol will be employed, and an initial trial control protocol $\omega^{0}(\mathbf{r}, t)$. 
We then perform a series of DAL
iterations, with each iteration involving the following steps:
\begin{itemize}
    \item \textbf{Step 1:} The equations of motion, ~\eqref{eq:density} and \eqref{eq:uptau}, are integrated forward in time from $t = 0$ to $t = \tF$ with the current protocol of spatiotemporal activity $\omega^{i}(\mathbf{r}, t)$ (where $i$ is the current iteration) and fixed initial conditions, $\rho(\mathbf{r}, 0)$ and $\mathbf{\uptau}(\mathbf{r}, 0)$. 
    \item  \textbf{Step 2:} The adjoint equations, ~\eqref{eq:eta} and ~\eqref{eq:nu}, are integrated backward in time from $t = \tF$ to $t = 0$ with the initial condition, $\eta(\mathbf{r}, \tF) = 0$ and $\boldsymbol{\nu}(\mathbf{r}, \tF) = 0$.
    \item \textbf{Step 3:} The control protocol is updated via gradient descent, $\omega^{i+1} = \omega^{i} - \Delta {\delta \mathcal{J}}/{\delta \omega} $, to minimize the cost function. 
\end{itemize}
Iterations are continued until the gradient $\delta \mathcal{J}/{\delta \omega}$ falls below a user-defined tolerance. We employ Armijo backtracking \cite{Borzi2011} to adaptively choose the step-size $\Delta$ for gradient descent and to ensure convergence of the DAL algorithm.

\section{Derivation of Adjoint Equations}

We begin with the equation of motion for the state variables:
\begin{align}
    \partial_{t}\rho = -\boldsymbol{\nabla}\cdot({\omega^{2}}\boldsymbol{\uptau} - \boldsymbol{\nabla}\rho)
    \label{eq:density}
\end{align}
\begin{align}
    \partial_{t}\boldsymbol{\uptau} = -(a_{2}(\rho) + a_{4}(\rho)|\boldsymbol{\uptau}|^2)\boldsymbol{\uptau} - \boldsymbol{\nabla}({\omega^{2}}\rho) + \boldsymbol{\nabla}^{2}\boldsymbol{\uptau} +\nonumber\\ {\lambda}\left(\boldsymbol{\uptau}_{\alpha}\boldsymbol{\nabla}\boldsymbol{\uptau}_{\alpha} + \boldsymbol{\uptau}\boldsymbol{\nabla}.\boldsymbol{\uptau} - \boldsymbol{\uptau}.\boldsymbol{\nabla}\boldsymbol{\uptau}\right).
    \label{eq:uptau}
\end{align}
We write the full objective function, which is the sum of terminal state and running state penalties that we aim to minimize, as follows:
 $$ \mathcal{J} = \frac{1}{2}\int_{\Omega}d\mathbf{r}\, E\left(\rho_{\tF} - \rho^{*}\right)^{2} + F\left(\boldsymbol{\uptau}_{\tF} - \boldsymbol{\uptau}^{*}\right)^{2} + \int_{o}^{\tF}dt \int_{\Omega}d\mathbf{r}\mathcal{H}, $$
where, 
\begin{align}
    \mathcal{H} =  \bigg[ \frac{A}{2}\left(\omega^{2} - \omega_{0}^{2}\right)^{2} + \frac{B}{2}\boldsymbol{\nabla}\omega^{2}\cdot\boldsymbol{\nabla}\omega^{2} + \frac{K}{2}(d\omega^{2}/dt)^{2}  
    & + \frac{C}{2}\left(\rho - \rho^{*}\right)^{2} +\frac{D}{2}\left(\boldsymbol{\uptau} - \boldsymbol{\uptau^{*}}\right)^{2} \bigg].
    \label{eq:objective}
\end{align}
We then introduce Lagrange multipliers $\eta$ and $\boldsymbol{\nu}$ that constrain the dynamics to the equations of motion for density and polarization respectively, and write the Lagrangian as:
\begin{align}
    \mathcal{L} = \mathcal{J} +  \int_{0}^{\tF} dt\int_{\Omega}d\mathbf{r}  
     \eta\left[\partial_{t}\rho + \boldsymbol{\nabla}\cdot(\omega^{2}\boldsymbol{\uptau} - \boldsymbol{\nabla}\rho)\right]
    + \boldsymbol{\nu}\cdot\left[\partial_{t}\boldsymbol{\uptau} + (a_{2} + a_{4}|\boldsymbol{\uptau}|^2)\boldsymbol{\uptau} + \boldsymbol{\nabla}(\omega^{2}\rho) - \boldsymbol{\nabla}^{2}\boldsymbol{\uptau} \right. 
    &\left. - \lambda\left(\uptau_{\alpha}\boldsymbol{\nabla}\uptau_{\alpha} + \boldsymbol{\uptau}\boldsymbol{\nabla}\cdot\boldsymbol{\uptau} - \boldsymbol{\uptau}\cdot\boldsymbol{\nabla}\boldsymbol{\uptau}\right)\right].
\end{align}

The necessary condition for optimality is $\nabla \mathcal{L} = 0$ \cite{Kirk2004, Lenhart2007}, so  ${\delta \mathcal{L}}/{\delta \eta}$, ${\delta \mathcal{L}}/{\delta \boldsymbol{\nu}}$, ${\delta \mathcal{L}}/{\delta \rho}$, ${\delta \mathcal{L}}/{\delta \boldsymbol{\uptau}}$, ${\delta \mathcal{L}}/{\delta \rho(\tF)}$, ${\delta \mathcal{L}}/{\delta \boldsymbol{\uptau}(\tF)}$, and ${\delta \mathcal{L}}/{\delta \omega} = 0$. 
The first two conditions of optimality yield back the dynamics equations. The next two conditions, ${\delta \mathcal{L}}/{\delta \rho}$, ${\delta \mathcal{L}}/{\delta \boldsymbol{\uptau}}$, yield dynamics equations for the adjoint variables $\eta$ and $\boldsymbol{\nu}$ as:
\begin{align}
    \partial_{t}\eta = & C\left(\rho - \rho^{*}\right) - \boldsymbol\nabla^{2}\eta - \omega\boldsymbol{\nabla}\cdot\boldsymbol{\nu}\nonumber \\
     & + (\delta_{\rho}a_{2}(\rho) + \delta_{\rho}a_{4}(\rho)|\boldsymbol{\uptau}|^{2})(\boldsymbol{\nu}.\boldsymbol{\uptau}).
     \label{eq:eta}
\end{align}
\begin{align}
    \partial_{t}\boldsymbol{\nu} = & D\left(\boldsymbol{\uptau} - \boldsymbol{\uptau^{*}}\right) + \left(a_{2}(\rho) + a_{4}(\rho)|\boldsymbol{\uptau}|^{2}\right)\boldsymbol{\nu}  + 2a_{4}(\rho)\boldsymbol{\uptau}(\boldsymbol{\nu}\cdot\boldsymbol{\uptau}) \nonumber \\
    & - \lambda\left[-\boldsymbol{\uptau}\boldsymbol{\nabla}\cdot\boldsymbol{\nu} + 2\boldsymbol{\nu}\boldsymbol{\nabla}\cdot\boldsymbol{\uptau} -\boldsymbol{\nabla}(\boldsymbol{\nu}\cdot\boldsymbol{\uptau}) \right. \nonumber \\
    & \left. + \boldsymbol{\uptau}\cdot\boldsymbol{\nabla}\boldsymbol{\nu} - \boldsymbol{\nu_{\alpha}}\boldsymbol{\nabla}\boldsymbol{\uptau_{\alpha}}\right]  -\boldsymbol{\nabla}^{2}\boldsymbol{\nu} -\omega\boldsymbol{\nabla}\eta.
    \label{eq:nu}
\end{align}
with a boundary condition at time $\tF$ set by ${\delta \mathcal{L}}/{\delta \rho(\tF)}$, ${\delta \mathcal{L}}/{\delta \boldsymbol{\uptau}(\tF)}$ as $\eta(\tF) = -E(\rho(\tF) - \rho^{*}) = 0$ and $\boldsymbol{\nu}(
\tF) = -F(\boldsymbol{\uptau}(\tF) - \boldsymbol{\uptau}^{*})$. For our computations, we choose $E, F=0$ for simplicity since we obtained adequate convergence from the time-integrated penalty. 

Finally, the condition ${\delta \mathcal{L}}/{\delta \omega}$ yields: 
\begin{align}
     {\delta \mathcal{L}}/{\delta \omega} = 2 A \omega\left(\omega^{2} - \omega_{0}^2\right) – 2 B \omega\boldsymbol{\nabla}^{2}\omega^{2} - 2 K \omega (d^{2}\omega/dt^{2}) 
     - 2\omega\boldsymbol{\uptau}\cdot\boldsymbol{\nabla}\eta  - 2\omega\boldsymbol{\nu}\cdot\boldsymbol{\nabla}\rho.
\label{eq:dOmega}
\end{align}
which is used to update the control variable $\omega$ during gradient descent.

\onecolumngrid

\section{Controllability}
\label{sec:controllability}
Let us consider the problem we want to solve in control theory. If we define $\mathbf{X}=\left(\begin{array}{c} \rho \\\uptau_{\text x}\\ \uptau_{\text y} \end{array}\right)$, we seek to solve the set of nonlinear partial differential equations $\frac{\partial \mathbf{X}}{\partial t}=H\left[\mathbf{X},\omega \right]$ for the control solution $\omega(\mathbf{r},t)$, subject to a given initial condition $X(\mathbf{r},0)$, and a boundary condition in time $X(\mathbf{r},t_{\text{F}})$, which is the target state with $\{0,t_{\text F}\}$ as the control window. A particular dynamical system is considered controllable if we can demonstrate the existence of a solution to the above problem.  When the dynamics is nonlinear, demonstrations of controllability have been limited to a few simple systems where the nonlinearities have special properties.  What we do instead is consider the controllability of Eqs. \ref{eq:density} -\ref{eq:uptau}  when linearized about the unstable fixed point of a homogeneous polar state.  Demonstrating controllability of the homogeneous fixed point tells us that at short enough length scales, we will be able to drive the system to desired values of the dynamical fields, which can be thought of as different fixed points in the continuous space of fixed points associated with translational symmetry and broken rotational symmetry characteristic of our system. 

Linearizing our theory using  $\rho = \rho_{0} + \delta \rho(\boldsymbol{r}, t)$ and $\boldsymbol{\uptau} = \boldsymbol{\uptau_0} + \delta \uptau_{{\|}}(\boldsymbol{r},t)\hat{x} + \delta \uptau_{\perp}(r,t)\hat{y}$, and introducing the Fourier transform, $\Tilde{x}(\boldsymbol{q}, t) = \int d\boldsymbol{r}e^{i\boldsymbol{q\cdot r}}x(\boldsymbol{r}, t)$, we obtain

\onecolumngrid
$$
\partial_t\left(\begin{array}{c}
\delta \Tilde{\rho} \\
\delta \Tilde{\uptau}_{\|} \\
\delta \Tilde{\uptau}_{\perp}
\end{array}\right)=\left(\begin{array}{ccc}
-q^2 & i \omega q_{\|} & i \omega q_{\perp} \\
\alpha_1\left(\rho_0\right)+i \omega q_{\|} & -\left(\alpha_2+i \lambda \uptau_0 q_{\|}+q^2\right) & -i \lambda \uptau_0 q_{\perp} \\
i \omega q_{\perp} & -i \lambda \uptau_0 q_{\perp} & i \lambda \uptau_0 q_{\|}-q^2
\end{array}\right)\left(\begin{array}{c}
\delta \Tilde{\rho} \\
\delta \Tilde{\uptau}_{\|} \\
\delta \Tilde{\uptau}_{\perp}
\end{array}\right)+\left(\begin{array}{c}
i \tau_0 q_{\|} \\
i \rho_0 q_{\|} \\
i \rho_0 q_{\perp}
\end{array}\right) \delta \Tilde{\omega}
$$
where $q_{{\|},\perp}$  denote the wavevectors along and orthogonal to direction of polarization, $\alpha_1 (\rho)=-\uptau_{0}\left( \frac{\delta a_2}{\delta \rho}-\frac{a_2}{a_4}\frac{\delta a_4}{\delta \rho}\right)$, and $\alpha_2(\rho)=-2a_2$. Note that $\alpha_{1,2}>0$ for all $\rho_0>\rho_c$.

Let us now introduce the notation
$$ X_{\text q} = 
\left(\begin{array}{c}
\delta \Tilde{\rho} \\
\delta \Tilde{\uptau}_{\|} \\
\delta \Tilde{\uptau}_{\perp}
\end{array}\right),
$$
$\omega_{\text q}=\delta \Tilde{\omega}(\mathbf{q},t)$, and
$$ B_{\text q} = 
\left(\begin{array}{c}
i \uptau_0 q_{\|} \\
i \rho_0 q_{\|} \\
i \rho_0 q_{\perp}
\end{array}\right).
$$
Our linearized theory is then of the form, 
$$
\partial_{t} X_{\text q}(t)=A_{\text q} X_{\text q}(t)+B_{\text q} \omega_{\text q}(t)
$$
One can readily establish that this linear system has a solution to the boundary value control problem when  the controllability matrix
$$
C=\left[\begin{array}{lllll}
B_{\text q} & A_{\text q} B_{\text q} & A_{\text q}^{2} B_{\text q} 
\end{array}\right]
$$
is of rank 3 \cite{Brunton2022}. 

Computing the column vectors of $C$, we obtain 
\begin{equation*}
A_{\text q}B_{\text q} = 
\begin{bmatrix}
-\omega \rho_0 q^2  - i \uptau_0 q_{\parallel}q^2 \\
i \tau_0 q_{\parallel} \left(\alpha_1(\rho_0) + i \omega q_{\parallel}\right) - i \rho_0 q_{\parallel} \left(\alpha_2(\rho_0)+q^2\right) 
+ \lambda \rho_0 \tau_0 q^2 \\
-i \rho_0 q_{\bot} q^2  - \tau_0 \omega q_{\parallel} q_{\bot} \\
\end{bmatrix}
\end{equation*}
%
    \begin{equation*}
        A^{2}B = 
        \begin{bmatrix}
        i \rho_0 q_{\parallel} \left(\lambda \uptau_0 \omega q^2-i \omega q_{\parallel} \left(\alpha_2 +2q^2\right) \right) \\
        + i \uptau_0 q_{\parallel} \left(i \omega q_{\parallel}\alpha_1-\omega^2 q^2-q^2 \right) \\
        
        + 2\omega \rho_0 q_{\bot}^2 q^2 
        \\[10pt]   
        i \rho_0 q_{\parallel} \left(\left(\alpha_2 + i \lambda \uptau_0 q_{\parallel} +q^2 \right)^2  - \lambda^2 \uptau_0^2 q_{\bot}^2\right) \\
        + i \rho_0 q_{\bot} \left(i \lambda \uptau_0 q_{\bot} \left(\alpha_2  + 2 q^2\right)  \right) -\omega\rho_0 q^2 \left(\alpha_1+i\omega q_{\parallel}\right)\\
        - i \tau_0 q_{\parallel} \left(\left(\alpha_1 + i \omega q_{\parallel}\right) \left(\alpha_2 + i \lambda \uptau_0 q_{\parallel} - 2q^2\right) - \lambda \uptau_0 \omega q_{\bot}^2\right)
        \\[10pt]
        i\rho_0 q_{\parallel} \left(i \lambda \uptau_0 q_{\bot} \left(\alpha_2  + 2 q^2\right)  - \omega^2 q_{\parallel} q_{\bot}\right) \\        
        + i \uptau_0 q_{\parallel} \left(-i \lambda \uptau_0 q_{\bot}\alpha_1  -2 i \omega q_{\bot} q^2\right) \\
        + i \rho_0 q_{\bot} \left(-\lambda^2 \uptau_0^2 q^2 + q^2\left(q^2-2i \lambda \uptau_0 q_{\parallel} \right) - \omega^2 q_{\bot}^2\right)
        \end{bmatrix}
    \end{equation*}
%
To examine the rank of the matrix $C$ in a physically informative way, let us consider 3 special cases. First, let us consider the case of spatial gradients orthogonal to the direction of order. Setting $q_{\parallel}=0$ and truncating the matrix elements to quadratic order in $q$ we obtain\begin{equation*}
    C = \left(\begin{array}{ccc}
0&-\omega\rho_0 q_{\perp}^2 &0 \\
0& \begin{array}{c}
     \lambda \rho_{0}\uptau_{0}q_{\perp}^{2} 
\end{array} & \begin{array}{c}
      -\rho_0 \lambda\uptau_{0}\alpha_2q_{\perp}^2 - \omega\rho_{0}\alpha_{1}q_{\perp}^2 \\ -\lambda\uptau_{0}\omega q_{\perp}^2
\end{array}  \\
i \rho_{0} q_{\perp} & 0 & 0
\end{array}\right)
\end{equation*}
As is apparent from the form of the matrix, the three column vectors are indeed linearly independent and hence the linearized theory of an active polar fluid is  controllable in the presence of gradients in the direction perpendicular to that of the spontaneously broken symmetry.  Next, let us consider two cases that show the limits on the controllability of the linear theory. If we consider the long wavelength limit of the controllability matrix $C$, we see that, when truncated to lowest order in wavevector 
\begin{equation*}
   C= \left(\begin{array}{ccc}
i \uptau_0 q_{\parallel}&0 &0 \\
i \rho_0 q_{\parallel}& i \uptau_0 q_{\parallel}\alpha_1+i \rho_0 q_{\parallel}\alpha_2 &i \rho_0 q_{\parallel}\alpha_{2}^{2}-i\uptau_0 q_{\parallel}\alpha_{1} \alpha_{2}  \\
i \rho_{0} q_{\perp} & 0 & 0
\end{array}\right)
\end{equation*}
which clearly is not of rank 3. Thus, the system is not controllable on the longest length scales. To identify the length scale up to which the system is controllable, let us compare the relevant terms in the second column. We will need to retain terms to quadratic order in the gradients when $\lambda\rho_{0}\uptau_{0}q_{\perp}^{2}\sim (\rho_{0}\alpha_{2}+\uptau_{0}\alpha_{1})q_{\parallel}$. Setting aside the direction of spatial homegeneity we get an estimate of the length scale up to which the system is controllable as $\ell_\text{max}=\frac{\lambda\rho_{0}\uptau_{0}}{(\rho_{0}\alpha_{2}+\uptau_{0}\alpha_{1})}$.  Given that all the terms on the right hand side scale with the mean density $\rho_{0}$, the length scale up to which the linear system is controllable is set by the strength of the nonlinearities $\lambda$. Recall that our system is non-dimensionalized using the diffusive length scale $(D/\nu)^{1/2}$ and $\lambda$ has the units of a diffusion coefficient. So, our system is controllable on length scales that are comparable to the diffusive length scale. 

Finally, note that when we restrict attention to spatial gradients that lie purely along the direction of broken symmetry, the controllability matrix becomes 
\begin{equation*}
    C = \left(\begin{array}{ccc}
i \uptau_0 q_{\parallel}&-i\omega_0 q_{\parallel}^2 &\rho_0 \omega_0 \alpha_2 q_{\parallel}^2- \uptau_0\omega_0 \alpha_1 q_{\parallel}^2\\
i \rho_0 q_{\parallel}& \begin{array}{c}
     i \uptau_0 q_{\parallel}\alpha_1-i \rho_0 q_{\parallel}\alpha_2 \\
     -\uptau_{0}\omega q_{\parallel}^{2} + \lambda\rho_{0}\uptau_{0}q_{\parallel}^{2}
\end{array} & \begin{array}{c}
    i \rho_0 q_{\parallel}\alpha_{2}^{2}-i\uptau_0 q_{\parallel}\alpha_{1} \alpha_{2} \\
     -\omega\rho_{0}q_{\parallel}^{2}\alpha_{1} + \alpha_{1}\lambda\uptau_{0}^{2} q_{\parallel}^{2} \\
     -i \alpha_{2}\uptau_{0}\omega q_{\parallel}^{2}    
\end{array}  \\
0 & 0 & 0
\end{array}\right)
\end{equation*}
and the system is clearly not controllable. Thus, we see that spatial gradients orthogonal to the direction of the local polar order are critical to obtaining control solutions for an active polar fluid. 

\onecolumngrid
\section{Residues: deviations from target state as a function of time}
\label{sec:residues}
\begin{figure*}[hbt]
  \centering
  \includegraphics[width=\textwidth]{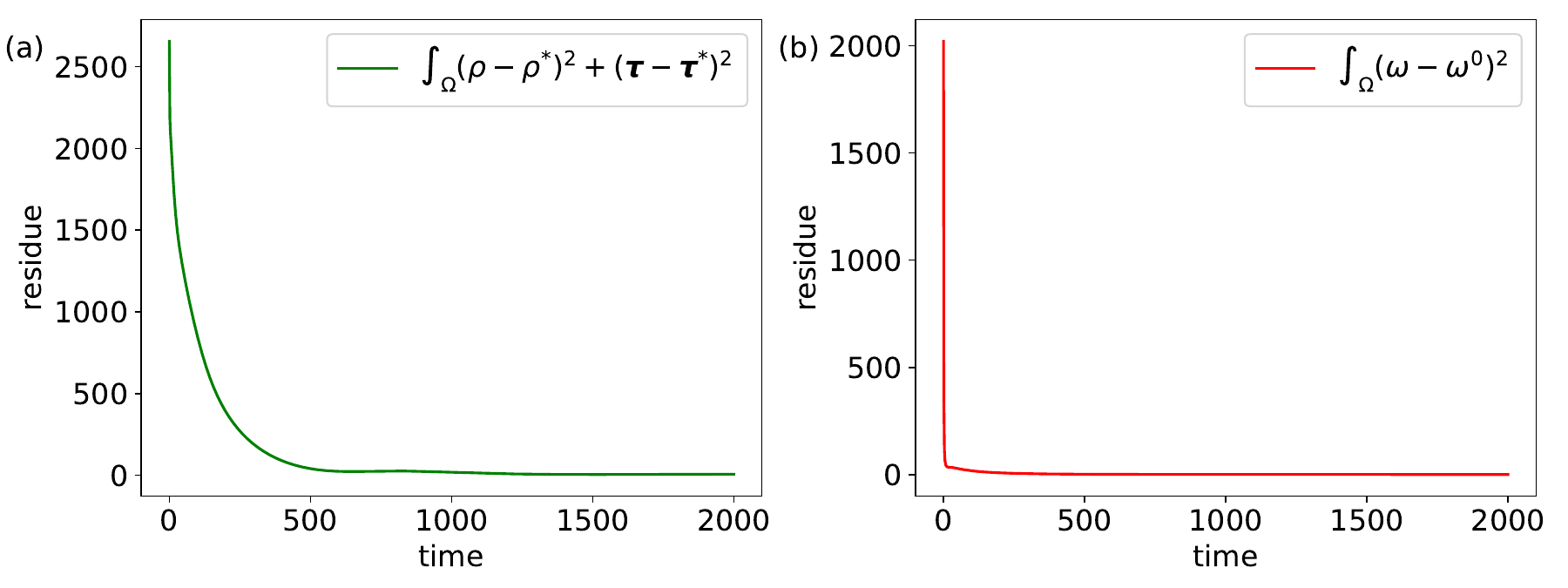}
  \caption{   \textbf{Residues as a function of time for aster advection with only initial and target state specified (Fig. 2 main text).} (a) State residue: deviation of the system state from the target.   (b) Control residue: the deviation of the control variable from its baseline value.
  \label{fig:res1}}
\end{figure*}

\begin{figure*}[hbt]
  \centering
  \includegraphics[width=\textwidth]{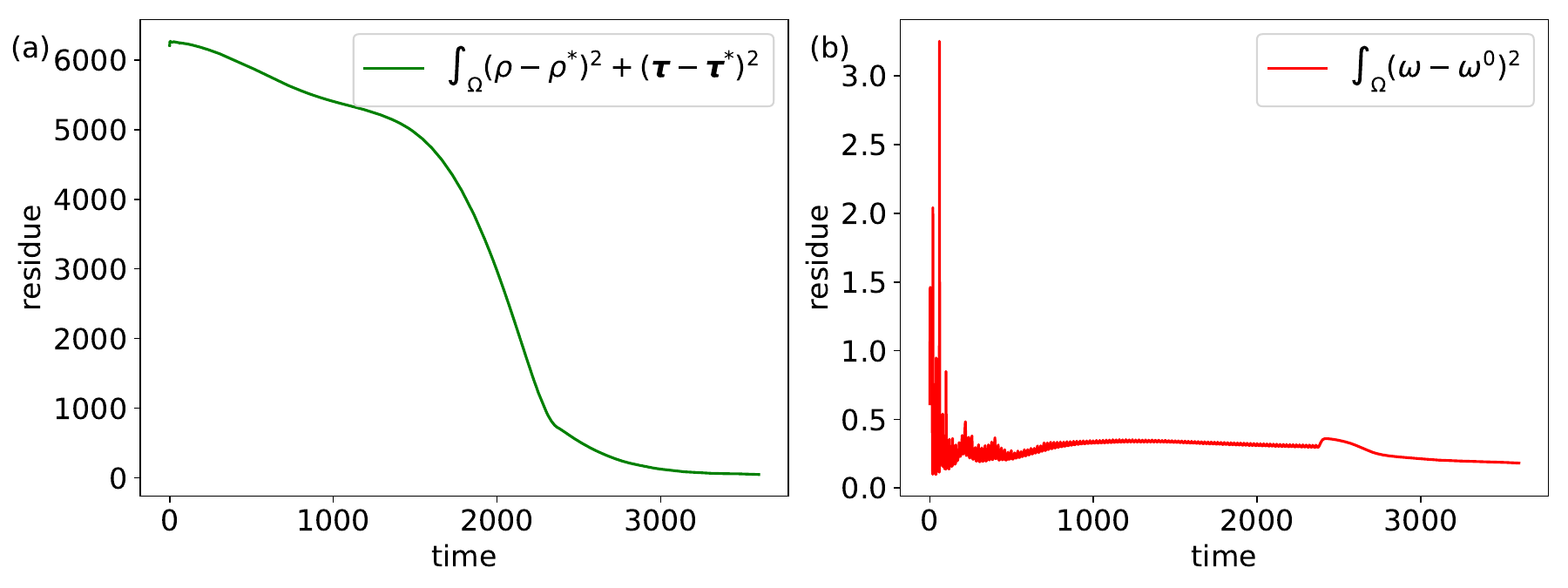}
  \caption{    \textbf{Residues for aster advection with path specified, slow advection (Aster-like trajectory, Fig. 3a main text).} (a) State residue: deviation of the system state from the target.   (b) Control residue: the deviation of the control variable from its baseline value.
  \label{fig:res2}}
\end{figure*}

\begin{figure*}[hbt]
  \centering
  \includegraphics[width=\textwidth]{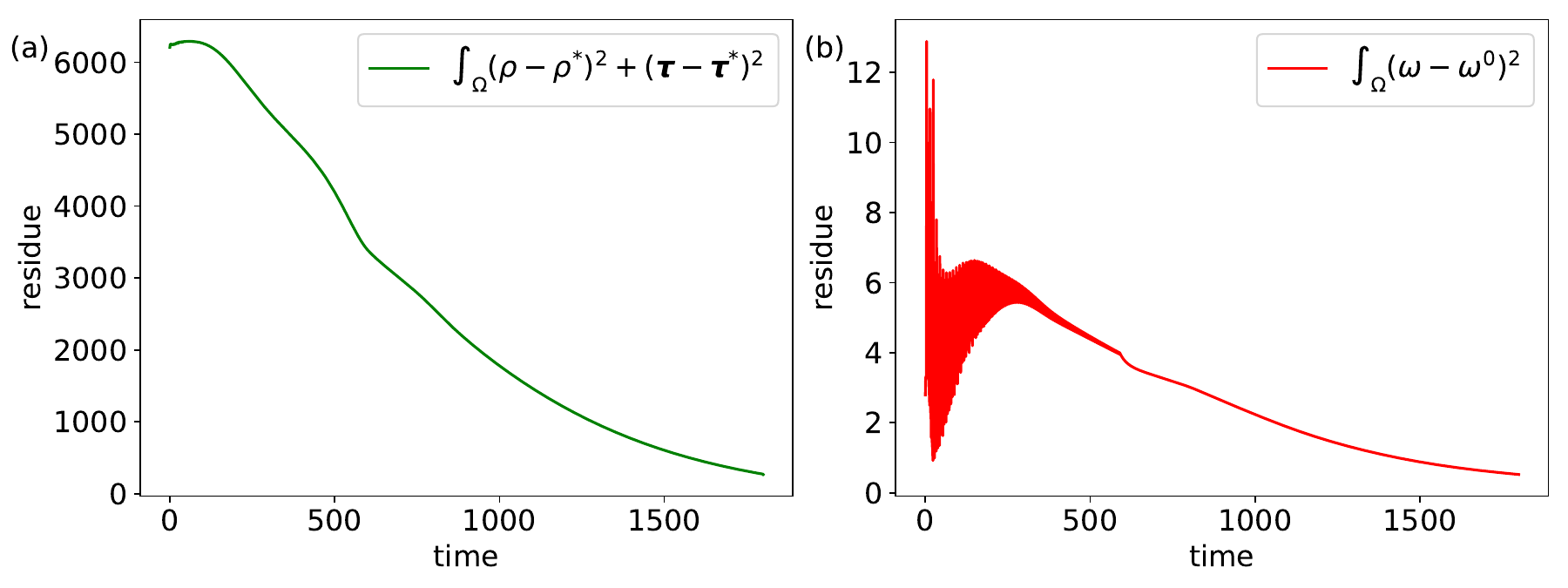}
  \caption{   \textbf{Residues for aster advection with path specified, fast advection (Flock-like trajectory, Fig. 3b main text).}   (a) State residue: deviation of the system state from the target.   (b) Control residue: the deviation of the control variable from its baseline value.
  \label{fig:res3}}
\end{figure*}

\begin{figure*}[hbt]
  \centering
  \includegraphics[width=\textwidth]{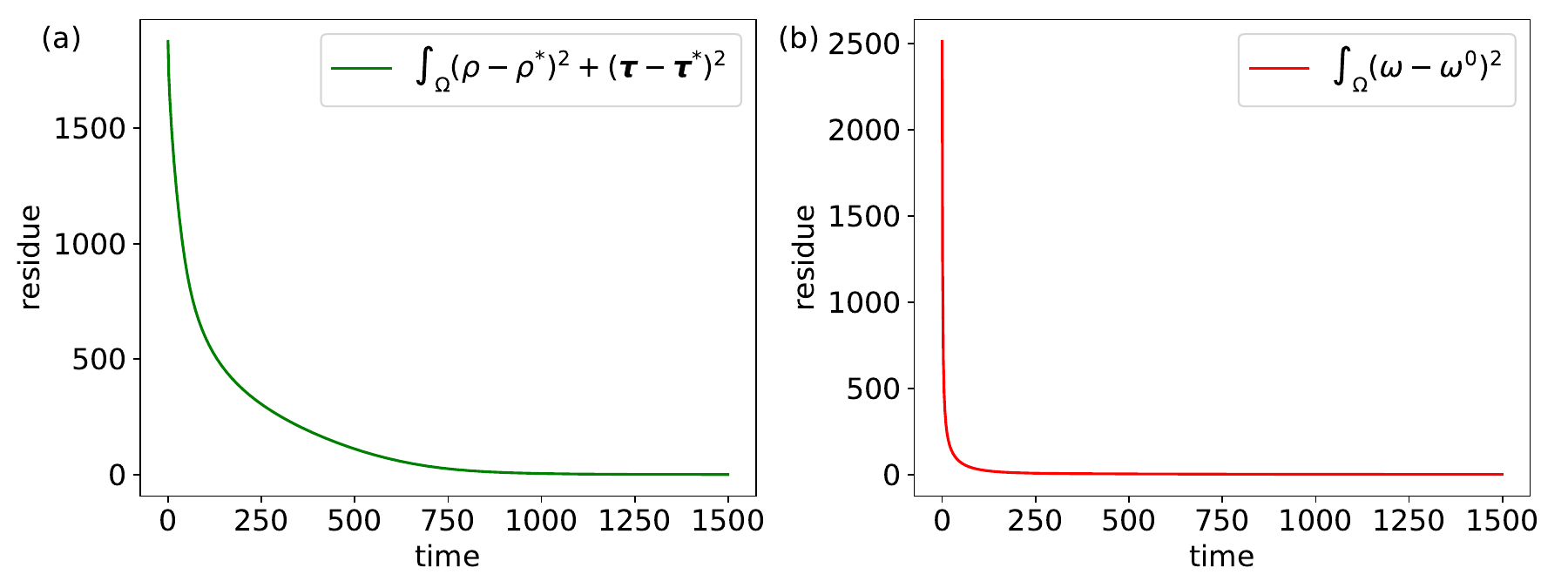}
  \caption{   \textbf{Residues for example in which a propagating stripe is remodelled into a stationary aster (Fig. 5 main text).} (a) State residue: deviation of the system state from the target.   (b) Control residue: the deviation of the control variable from its baseline value.
  \label{fig:res4}}
\end{figure*}

In this section, we assess how quickly the solutions approach their target by plotting the residues, meaning the deviations from the target state, $\left(\rho - \rho^{*}\right)^{2} + \left(\boldsymbol{\uptau} - \boldsymbol\uptau^{*}\right)^{2}$, and the deviation of the control variable from its baseline value, $\omega-\omega_0$. SI Figs.~\ref{fig:res1} -- \ref{fig:res4} show these residues  as a function of time for the examples of: aster advection  (Fig. 2 main text), aster advection with the trajectory specified (Figs.3a, 3b main text), and remodelling a stripe into an aster (Fig. 5 main text).

\clearpage

\section{Movie Descriptions}

\begin{itemize}
    \item \textbf{Movie S1}: Aster translocation and reformation in the case where only the final state of the aster is specified (Fig. 2 main text). The left panel depicts density $\rho$ (color map) and polarization $\boldsymbol{\uptau}$ (arrows) profiles. The right panel shows the activity field $\omega$, with a logarithmic scale colorbar.
    \item \textbf{Movie S2}: Aster translocation and reformation where the trajectory of aster is specified, with slow advection (Aster-like trajectory, Fig. 3a main text). The left panel depicts  density $\rho$ (color map) and polarization $\boldsymbol{\uptau}$ (arrows) profiles. The right panel shows the activity field $\omega$, with a logarithmic scale colorbar.
    \item \textbf{Movie S3}: Aster translocation and reformation where the trajectory of aster is specified, with fast advection (Flock-like trajectory, Fig. 3b main text).. The left panel depicts density $\rho$ (color map) and polarization $\boldsymbol{\uptau}$ (arrows) profiles. The right panel shows the activity field $\omega$, with a logarithmic scale colorbar.
    \item \textbf{Movie S4}: Reorienting the direction of propagation of a stripe by $45^{0}$ (Fig. 4 main text).  The left panel depicts density $\rho$ (color map) and polarization $\boldsymbol{\uptau}$ (arrows) profiles. The right panel shows the activity field $\omega$, with a logarithmic scale colorbar.
    \item \textbf{Movie S5}: Remodelling a propagating stripe to a stationary aster (Fig. 5 main text).  The left panel depicts density $\rho$ (color map) and polarization $\boldsymbol{\uptau}$ (arrows) profiles. The right panel shows the activity field $\omega$, with a logarithmic scale colorbar. 
    \item \textbf{Movie S6}: Reorienting the direction of propagation of a stripe by $90^{0}$. The left panel depicts density $\rho$ (color map) and polarization $\boldsymbol{\uptau}$ (arrows) profiles. The right panel shows the activity field $\omega$, with a logarithmic scale colorbar.
\end{itemize}


\begin{thebibliography}{110}%
\makeatletter
\providecommand \@ifxundefined [1]{%
 \@ifx{#1\undefined}
}%
\providecommand \@ifnum [1]{%
 \ifnum #1\expandafter \@firstoftwo
 \else \expandafter \@secondoftwo
 \fi
}%
\providecommand \@ifx [1]{%
 \ifx #1\expandafter \@firstoftwo
 \else \expandafter \@secondoftwo
 \fi
}%
\providecommand \natexlab [1]{#1}%
\providecommand \enquote  [1]{``#1''}%
\providecommand \bibnamefont  [1]{#1}%
\providecommand \bibfnamefont [1]{#1}%
\providecommand \citenamefont [1]{#1}%
\providecommand \href@noop [0]{\@secondoftwo}%
\providecommand \href [0]{\begingroup \@sanitize@url \@href}%
\providecommand \@href[1]{\@@startlink{#1}\@@href}%
\providecommand \@@href[1]{\endgroup#1\@@endlink}%
\providecommand \@sanitize@url [0]{\catcode `\\12\catcode `\$12\catcode
  `\&12\catcode `\#12\catcode `\^12\catcode `\_12\catcode `\%12\relax}%
\providecommand \@@startlink[1]{}%
\providecommand \@@endlink[0]{}%
\providecommand \url  [0]{\begingroup\@sanitize@url \@url }%
\providecommand \@url [1]{\endgroup\@href {#1}{\urlprefix }}%
\providecommand \urlprefix  [0]{URL }%
\providecommand \Eprint [0]{\href }%
\providecommand \doibase [0]{http://dx.doi.org/}%
\providecommand \selectlanguage [0]{\@gobble}%
\providecommand \bibinfo  [0]{\@secondoftwo}%
\providecommand \bibfield  [0]{\@secondoftwo}%
\providecommand \translation [1]{[#1]}%
\providecommand \BibitemOpen [0]{}%
\providecommand \bibitemStop [0]{}%
\providecommand \bibitemNoStop [0]{.\EOS\space}%
\providecommand \EOS [0]{\spacefactor3000\relax}%
\providecommand \BibitemShut  [1]{\csname bibitem#1\endcsname}%
\let\auto@bib@innerbib\@empty
\bibitem [{\citenamefont {Liu}\ and\ \citenamefont {Lin}(2012)}]{Liu2012}%
  \BibitemOpen
  \bibfield  {author} {\bibinfo {author} {\bibfnamefont {K.-A.}\ \bibnamefont
  {Liu}}\ and\ \bibinfo {author} {\bibfnamefont {I.}~\bibnamefont {Lin}},\
  }\href@noop {} {\bibfield  {journal} {\bibinfo  {journal} {Physical Review
  E}\ }\textbf {\bibinfo {volume} {86}},\ \bibinfo {pages} {011924} (\bibinfo
  {year} {2012})}\BibitemShut {NoStop}%
\bibitem [{\citenamefont {Czir{\'o}k}\ and\ \citenamefont
  {Vicsek}(2000)}]{Czirok2000}%
  \BibitemOpen
  \bibfield  {author} {\bibinfo {author} {\bibfnamefont {A.}~\bibnamefont
  {Czir{\'o}k}}\ and\ \bibinfo {author} {\bibfnamefont {T.}~\bibnamefont
  {Vicsek}},\ }\href@noop {} {\bibfield  {journal} {\bibinfo  {journal}
  {Physica A: Statistical Mechanics and its Applications}\ }\textbf {\bibinfo
  {volume} {281}},\ \bibinfo {pages} {17} (\bibinfo {year} {2000})}\BibitemShut
  {NoStop}%
\bibitem [{\citenamefont {Pierce}\ \emph {et~al.}(2018)\citenamefont {Pierce},
  \citenamefont {Wijesinghe}, \citenamefont {Mumper}, \citenamefont {Lower},
  \citenamefont {Lower},\ and\ \citenamefont {Sooryakumar}}]{Pierce2018}%
  \BibitemOpen
  \bibfield  {author} {\bibinfo {author} {\bibfnamefont {C.}~\bibnamefont
  {Pierce}}, \bibinfo {author} {\bibfnamefont {H.}~\bibnamefont {Wijesinghe}},
  \bibinfo {author} {\bibfnamefont {E.}~\bibnamefont {Mumper}}, \bibinfo
  {author} {\bibfnamefont {B.}~\bibnamefont {Lower}}, \bibinfo {author}
  {\bibfnamefont {S.}~\bibnamefont {Lower}}, \ and\ \bibinfo {author}
  {\bibfnamefont {R.}~\bibnamefont {Sooryakumar}},\ }\href@noop {} {\bibfield
  {journal} {\bibinfo  {journal} {Physical Review Letters}\ }\textbf {\bibinfo
  {volume} {121}},\ \bibinfo {pages} {188001} (\bibinfo {year}
  {2018})}\BibitemShut {NoStop}%
\bibitem [{\citenamefont {Needleman}\ and\ \citenamefont
  {Dogic}(2017)}]{Needleman2017}%
  \BibitemOpen
  \bibfield  {author} {\bibinfo {author} {\bibfnamefont {D.}~\bibnamefont
  {Needleman}}\ and\ \bibinfo {author} {\bibfnamefont {Z.}~\bibnamefont
  {Dogic}},\ }\href@noop {} {\bibfield  {journal} {\bibinfo  {journal} {Nature
  reviews materials}\ }\textbf {\bibinfo {volume} {2}},\ \bibinfo {pages} {1}
  (\bibinfo {year} {2017})}\BibitemShut {NoStop}%
\bibitem [{\citenamefont {Sarfati}\ \emph {et~al.}(2022)\citenamefont
  {Sarfati}, \citenamefont {Maitra}, \citenamefont {Voituriez}, \citenamefont
  {Galas},\ and\ \citenamefont {Estevez-Torres}}]{Sarfati2022}%
  \BibitemOpen
  \bibfield  {author} {\bibinfo {author} {\bibfnamefont {G.}~\bibnamefont
  {Sarfati}}, \bibinfo {author} {\bibfnamefont {A.}~\bibnamefont {Maitra}},
  \bibinfo {author} {\bibfnamefont {R.}~\bibnamefont {Voituriez}}, \bibinfo
  {author} {\bibfnamefont {J.-C.}\ \bibnamefont {Galas}}, \ and\ \bibinfo
  {author} {\bibfnamefont {A.}~\bibnamefont {Estevez-Torres}},\ }\href@noop {}
  {\bibfield  {journal} {\bibinfo  {journal} {Soft Matter}\ }\textbf {\bibinfo
  {volume} {18}},\ \bibinfo {pages} {3793} (\bibinfo {year}
  {2022})}\BibitemShut {NoStop}%
\bibitem [{\citenamefont {Ndlec}\ \emph {et~al.}(1997)\citenamefont {Ndlec},
  \citenamefont {Surrey}, \citenamefont {Maggs},\ and\ \citenamefont
  {Leibler}}]{Ndlec1997}%
  \BibitemOpen
  \bibfield  {author} {\bibinfo {author} {\bibfnamefont {F.}~\bibnamefont
  {Ndlec}}, \bibinfo {author} {\bibfnamefont {T.}~\bibnamefont {Surrey}},
  \bibinfo {author} {\bibfnamefont {A.~C.}\ \bibnamefont {Maggs}}, \ and\
  \bibinfo {author} {\bibfnamefont {S.}~\bibnamefont {Leibler}},\ }\href@noop
  {} {\bibfield  {journal} {\bibinfo  {journal} {Nature}\ }\textbf {\bibinfo
  {volume} {389}},\ \bibinfo {pages} {305} (\bibinfo {year}
  {1997})}\BibitemShut {NoStop}%
\bibitem [{\citenamefont {Surrey}\ \emph {et~al.}(2001)\citenamefont {Surrey},
  \citenamefont {N{\'e}d{\'e}lec}, \citenamefont {Leibler},\ and\ \citenamefont
  {Karsenti}}]{Surrey2001}%
  \BibitemOpen
  \bibfield  {author} {\bibinfo {author} {\bibfnamefont {T.}~\bibnamefont
  {Surrey}}, \bibinfo {author} {\bibfnamefont {F.}~\bibnamefont
  {N{\'e}d{\'e}lec}}, \bibinfo {author} {\bibfnamefont {S.}~\bibnamefont
  {Leibler}}, \ and\ \bibinfo {author} {\bibfnamefont {E.}~\bibnamefont
  {Karsenti}},\ }\href@noop {} {\bibfield  {journal} {\bibinfo  {journal}
  {Science}\ }\textbf {\bibinfo {volume} {292}},\ \bibinfo {pages} {1167}
  (\bibinfo {year} {2001})}\BibitemShut {NoStop}%
\bibitem [{\citenamefont {Gardel}\ \emph {et~al.}(2008)\citenamefont {Gardel},
  \citenamefont {Kasza}, \citenamefont {Brangwynne}, \citenamefont {Liu},\ and\
  \citenamefont {Weitz}}]{Gardel2008}%
  \BibitemOpen
  \bibfield  {author} {\bibinfo {author} {\bibfnamefont {M.~L.}\ \bibnamefont
  {Gardel}}, \bibinfo {author} {\bibfnamefont {K.~E.}\ \bibnamefont {Kasza}},
  \bibinfo {author} {\bibfnamefont {C.~P.}\ \bibnamefont {Brangwynne}},
  \bibinfo {author} {\bibfnamefont {J.}~\bibnamefont {Liu}}, \ and\ \bibinfo
  {author} {\bibfnamefont {D.~A.}\ \bibnamefont {Weitz}},\ }\href@noop {}
  {\bibfield  {journal} {\bibinfo  {journal} {Methods in cell biology}\
  }\textbf {\bibinfo {volume} {89}},\ \bibinfo {pages} {487} (\bibinfo {year}
  {2008})}\BibitemShut {NoStop}%
\bibitem [{\citenamefont {Gardel}\ \emph {et~al.}(2010)\citenamefont {Gardel},
  \citenamefont {Schneider}, \citenamefont {Aratyn-Schaus},\ and\ \citenamefont
  {Waterman}}]{Gardel2010}%
  \BibitemOpen
  \bibfield  {author} {\bibinfo {author} {\bibfnamefont {M.~L.}\ \bibnamefont
  {Gardel}}, \bibinfo {author} {\bibfnamefont {I.~C.}\ \bibnamefont
  {Schneider}}, \bibinfo {author} {\bibfnamefont {Y.}~\bibnamefont
  {Aratyn-Schaus}}, \ and\ \bibinfo {author} {\bibfnamefont {C.~M.}\
  \bibnamefont {Waterman}},\ }\href@noop {} {\bibfield  {journal} {\bibinfo
  {journal} {Annual review of cell and developmental biology}\ }\textbf
  {\bibinfo {volume} {26}},\ \bibinfo {pages} {315} (\bibinfo {year}
  {2010})}\BibitemShut {NoStop}%
\bibitem [{\citenamefont {Dogterom}\ and\ \citenamefont
  {Koenderink}(2019)}]{Dogterom2019}%
  \BibitemOpen
  \bibfield  {author} {\bibinfo {author} {\bibfnamefont {M.}~\bibnamefont
  {Dogterom}}\ and\ \bibinfo {author} {\bibfnamefont {G.~H.}\ \bibnamefont
  {Koenderink}},\ }\href@noop {} {\bibfield  {journal} {\bibinfo  {journal}
  {Nature reviews Molecular cell biology}\ }\textbf {\bibinfo {volume} {20}},\
  \bibinfo {pages} {38} (\bibinfo {year} {2019})}\BibitemShut {NoStop}%
\bibitem [{\citenamefont {Soares~e Silva}\ \emph {et~al.}(2011)\citenamefont
  {Soares~e Silva}, \citenamefont {Depken}, \citenamefont {Stuhrmann},
  \citenamefont {Korsten}, \citenamefont {MacKintosh},\ and\ \citenamefont
  {Koenderink}}]{SoareseSilva2011}%
  \BibitemOpen
  \bibfield  {author} {\bibinfo {author} {\bibfnamefont {M.}~\bibnamefont
  {Soares~e Silva}}, \bibinfo {author} {\bibfnamefont {M.}~\bibnamefont
  {Depken}}, \bibinfo {author} {\bibfnamefont {B.}~\bibnamefont {Stuhrmann}},
  \bibinfo {author} {\bibfnamefont {M.}~\bibnamefont {Korsten}}, \bibinfo
  {author} {\bibfnamefont {F.~C.}\ \bibnamefont {MacKintosh}}, \ and\ \bibinfo
  {author} {\bibfnamefont {G.~H.}\ \bibnamefont {Koenderink}},\ }\href@noop {}
  {\bibfield  {journal} {\bibinfo  {journal} {Proceedings of the National
  Academy of Sciences}\ }\textbf {\bibinfo {volume} {108}},\ \bibinfo {pages}
  {9408} (\bibinfo {year} {2011})}\BibitemShut {NoStop}%
\bibitem [{\citenamefont {Wagner}\ \emph {et~al.}(2022)\citenamefont {Wagner},
  \citenamefont {Norton}, \citenamefont {Park},\ and\ \citenamefont
  {Grover}}]{Wagner2022}%
  \BibitemOpen
  \bibfield  {author} {\bibinfo {author} {\bibfnamefont {C.~G.}\ \bibnamefont
  {Wagner}}, \bibinfo {author} {\bibfnamefont {M.~M.}\ \bibnamefont {Norton}},
  \bibinfo {author} {\bibfnamefont {J.~S.}\ \bibnamefont {Park}}, \ and\
  \bibinfo {author} {\bibfnamefont {P.}~\bibnamefont {Grover}},\ }\href
  {\doibase 10.1103/PhysRevLett.128.028003} {\bibfield  {journal} {\bibinfo
  {journal} {Physical Review Letters}\ }\textbf {\bibinfo {volume} {128}},\
  \bibinfo {pages} {028003} (\bibinfo {year} {2022})}\BibitemShut {NoStop}%
\bibitem [{\citenamefont {Wang}\ \emph {et~al.}(2015)\citenamefont {Wang},
  \citenamefont {Duan}, \citenamefont {Ahmed}, \citenamefont {Sen},\ and\
  \citenamefont {Mallouk}}]{Wang2015}%
  \BibitemOpen
  \bibfield  {author} {\bibinfo {author} {\bibfnamefont {W.}~\bibnamefont
  {Wang}}, \bibinfo {author} {\bibfnamefont {W.}~\bibnamefont {Duan}}, \bibinfo
  {author} {\bibfnamefont {S.}~\bibnamefont {Ahmed}}, \bibinfo {author}
  {\bibfnamefont {A.}~\bibnamefont {Sen}}, \ and\ \bibinfo {author}
  {\bibfnamefont {T.~E.}\ \bibnamefont {Mallouk}},\ }\href@noop {} {\bibfield
  {journal} {\bibinfo  {journal} {Accounts of chemical research}\ }\textbf
  {\bibinfo {volume} {48}},\ \bibinfo {pages} {1938} (\bibinfo {year}
  {2015})}\BibitemShut {NoStop}%
\bibitem [{\citenamefont {Gomez-Solano}\ \emph {et~al.}(2017)\citenamefont
  {Gomez-Solano}, \citenamefont {Samin}, \citenamefont {Lozano}, \citenamefont
  {Ruedas-Batuecas}, \citenamefont {van Roij},\ and\ \citenamefont
  {Bechinger}}]{GomezSolano2017}%
  \BibitemOpen
  \bibfield  {author} {\bibinfo {author} {\bibfnamefont {J.~R.}\ \bibnamefont
  {Gomez-Solano}}, \bibinfo {author} {\bibfnamefont {S.}~\bibnamefont {Samin}},
  \bibinfo {author} {\bibfnamefont {C.}~\bibnamefont {Lozano}}, \bibinfo
  {author} {\bibfnamefont {P.}~\bibnamefont {Ruedas-Batuecas}}, \bibinfo
  {author} {\bibfnamefont {R.}~\bibnamefont {van Roij}}, \ and\ \bibinfo
  {author} {\bibfnamefont {C.}~\bibnamefont {Bechinger}},\ }\href@noop {}
  {\bibfield  {journal} {\bibinfo  {journal} {Scientific reports}\ }\textbf
  {\bibinfo {volume} {7}},\ \bibinfo {pages} {14891} (\bibinfo {year}
  {2017})}\BibitemShut {NoStop}%
\bibitem [{\citenamefont {Hallatschek}\ \emph {et~al.}(2023)\citenamefont
  {Hallatschek}, \citenamefont {Datta}, \citenamefont {Drescher}, \citenamefont
  {Dunkel}, \citenamefont {Elgeti}, \citenamefont {Waclaw},\ and\ \citenamefont
  {Wingreen}}]{Hallatschek2023}%
  \BibitemOpen
  \bibfield  {author} {\bibinfo {author} {\bibfnamefont {O.}~\bibnamefont
  {Hallatschek}}, \bibinfo {author} {\bibfnamefont {S.~S.}\ \bibnamefont
  {Datta}}, \bibinfo {author} {\bibfnamefont {K.}~\bibnamefont {Drescher}},
  \bibinfo {author} {\bibfnamefont {J.}~\bibnamefont {Dunkel}}, \bibinfo
  {author} {\bibfnamefont {J.}~\bibnamefont {Elgeti}}, \bibinfo {author}
  {\bibfnamefont {B.}~\bibnamefont {Waclaw}}, \ and\ \bibinfo {author}
  {\bibfnamefont {N.~S.}\ \bibnamefont {Wingreen}},\ }\href@noop {} {\bibfield
  {journal} {\bibinfo  {journal} {Nature Reviews Physics}\ ,\ \bibinfo {pages}
  {1}} (\bibinfo {year} {2023})}\BibitemShut {NoStop}%
\bibitem [{\citenamefont {Liu}\ \emph {et~al.}(2021)\citenamefont {Liu},
  \citenamefont {Shankar}, \citenamefont {Marchetti},\ and\ \citenamefont
  {Wu}}]{Liu2021}%
  \BibitemOpen
  \bibfield  {author} {\bibinfo {author} {\bibfnamefont {S.}~\bibnamefont
  {Liu}}, \bibinfo {author} {\bibfnamefont {S.}~\bibnamefont {Shankar}},
  \bibinfo {author} {\bibfnamefont {M.~C.}\ \bibnamefont {Marchetti}}, \ and\
  \bibinfo {author} {\bibfnamefont {Y.}~\bibnamefont {Wu}},\ }\href@noop {}
  {\bibfield  {journal} {\bibinfo  {journal} {Nature}\ }\textbf {\bibinfo
  {volume} {590}},\ \bibinfo {pages} {80} (\bibinfo {year} {2021})}\BibitemShut
  {NoStop}%
\bibitem [{\citenamefont {Hamby}\ \emph {et~al.}(2018)\citenamefont {Hamby},
  \citenamefont {Vig}, \citenamefont {Safonova},\ and\ \citenamefont
  {Wolgemuth}}]{Hamby2018}%
  \BibitemOpen
  \bibfield  {author} {\bibinfo {author} {\bibfnamefont {A.~E.}\ \bibnamefont
  {Hamby}}, \bibinfo {author} {\bibfnamefont {D.~K.}\ \bibnamefont {Vig}},
  \bibinfo {author} {\bibfnamefont {S.}~\bibnamefont {Safonova}}, \ and\
  \bibinfo {author} {\bibfnamefont {C.~W.}\ \bibnamefont {Wolgemuth}},\
  }\href@noop {} {\bibfield  {journal} {\bibinfo  {journal} {Science advances}\
  }\textbf {\bibinfo {volume} {4}},\ \bibinfo {pages} {eaau0125} (\bibinfo
  {year} {2018})}\BibitemShut {NoStop}%
\bibitem [{\citenamefont {Ni}\ \emph {et~al.}(2020)\citenamefont {Ni},
  \citenamefont {Colin}, \citenamefont {Link}, \citenamefont {Endres},\ and\
  \citenamefont {Sourjik}}]{Ni2020}%
  \BibitemOpen
  \bibfield  {author} {\bibinfo {author} {\bibfnamefont {B.}~\bibnamefont
  {Ni}}, \bibinfo {author} {\bibfnamefont {R.}~\bibnamefont {Colin}}, \bibinfo
  {author} {\bibfnamefont {H.}~\bibnamefont {Link}}, \bibinfo {author}
  {\bibfnamefont {R.~G.}\ \bibnamefont {Endres}}, \ and\ \bibinfo {author}
  {\bibfnamefont {V.}~\bibnamefont {Sourjik}},\ }\href@noop {} {\bibfield
  {journal} {\bibinfo  {journal} {Proceedings of the National Academy of
  Sciences}\ }\textbf {\bibinfo {volume} {117}},\ \bibinfo {pages} {595}
  (\bibinfo {year} {2020})}\BibitemShut {NoStop}%
\bibitem [{\citenamefont {Lauga}\ and\ \citenamefont
  {Powers}(2009)}]{Lauga2009}%
  \BibitemOpen
  \bibfield  {author} {\bibinfo {author} {\bibfnamefont {E.}~\bibnamefont
  {Lauga}}\ and\ \bibinfo {author} {\bibfnamefont {T.~R.}\ \bibnamefont
  {Powers}},\ }\href@noop {} {\bibfield  {journal} {\bibinfo  {journal}
  {Reports on progress in physics}\ }\textbf {\bibinfo {volume} {72}},\
  \bibinfo {pages} {096601} (\bibinfo {year} {2009})}\BibitemShut {NoStop}%
\bibitem [{\citenamefont {Gonzalez-Rodriguez}\ \emph
  {et~al.}(2012)\citenamefont {Gonzalez-Rodriguez}, \citenamefont {Guevorkian},
  \citenamefont {Douezan},\ and\ \citenamefont
  {Brochard-Wyart}}]{GonzalezRodriguez2012}%
  \BibitemOpen
  \bibfield  {author} {\bibinfo {author} {\bibfnamefont {D.}~\bibnamefont
  {Gonzalez-Rodriguez}}, \bibinfo {author} {\bibfnamefont {K.}~\bibnamefont
  {Guevorkian}}, \bibinfo {author} {\bibfnamefont {S.}~\bibnamefont {Douezan}},
  \ and\ \bibinfo {author} {\bibfnamefont {F.}~\bibnamefont {Brochard-Wyart}},\
  }\href@noop {} {\bibfield  {journal} {\bibinfo  {journal} {Science}\ }\textbf
  {\bibinfo {volume} {338}},\ \bibinfo {pages} {910} (\bibinfo {year}
  {2012})}\BibitemShut {NoStop}%
\bibitem [{\citenamefont {Henkes}\ \emph {et~al.}(2020)\citenamefont {Henkes},
  \citenamefont {Kostanjevec}, \citenamefont {Collinson}, \citenamefont
  {Sknepnek},\ and\ \citenamefont {Bertin}}]{Henkes2020}%
  \BibitemOpen
  \bibfield  {author} {\bibinfo {author} {\bibfnamefont {S.}~\bibnamefont
  {Henkes}}, \bibinfo {author} {\bibfnamefont {K.}~\bibnamefont {Kostanjevec}},
  \bibinfo {author} {\bibfnamefont {J.~M.}\ \bibnamefont {Collinson}}, \bibinfo
  {author} {\bibfnamefont {R.}~\bibnamefont {Sknepnek}}, \ and\ \bibinfo
  {author} {\bibfnamefont {E.}~\bibnamefont {Bertin}},\ }\href@noop {}
  {\bibfield  {journal} {\bibinfo  {journal} {Nature communications}\ }\textbf
  {\bibinfo {volume} {11}},\ \bibinfo {pages} {1405} (\bibinfo {year}
  {2020})}\BibitemShut {NoStop}%
\bibitem [{\citenamefont {Fodor}\ and\ \citenamefont
  {Marchetti}(2018)}]{Fodor2018}%
  \BibitemOpen
  \bibfield  {author} {\bibinfo {author} {\bibfnamefont {{\'E}.}~\bibnamefont
  {Fodor}}\ and\ \bibinfo {author} {\bibfnamefont {M.~C.}\ \bibnamefont
  {Marchetti}},\ }\href@noop {} {\bibfield  {journal} {\bibinfo  {journal}
  {Physica A: Statistical Mechanics and its Applications}\ }\textbf {\bibinfo
  {volume} {504}},\ \bibinfo {pages} {106} (\bibinfo {year}
  {2018})}\BibitemShut {NoStop}%
\bibitem [{\citenamefont {Nichol}\ and\ \citenamefont
  {Khademhosseini}(2009)}]{Nichol2009}%
  \BibitemOpen
  \bibfield  {author} {\bibinfo {author} {\bibfnamefont {J.~W.}\ \bibnamefont
  {Nichol}}\ and\ \bibinfo {author} {\bibfnamefont {A.}~\bibnamefont
  {Khademhosseini}},\ }\href@noop {} {\bibfield  {journal} {\bibinfo  {journal}
  {Soft matter}\ }\textbf {\bibinfo {volume} {5}},\ \bibinfo {pages} {1312}
  (\bibinfo {year} {2009})}\BibitemShut {NoStop}%
\bibitem [{\citenamefont {Zhang}\ \emph {et~al.}(2020)\citenamefont {Zhang},
  \citenamefont {Mueller}, \citenamefont {Doostmohammadi},\ and\ \citenamefont
  {Yeomans}}]{Zhang2020}%
  \BibitemOpen
  \bibfield  {author} {\bibinfo {author} {\bibfnamefont {G.}~\bibnamefont
  {Zhang}}, \bibinfo {author} {\bibfnamefont {R.}~\bibnamefont {Mueller}},
  \bibinfo {author} {\bibfnamefont {A.}~\bibnamefont {Doostmohammadi}}, \ and\
  \bibinfo {author} {\bibfnamefont {J.~M.}\ \bibnamefont {Yeomans}},\
  }\href@noop {} {\bibfield  {journal} {\bibinfo  {journal} {Journal of the
  Royal Society Interface}\ }\textbf {\bibinfo {volume} {17}},\ \bibinfo
  {pages} {20200312} (\bibinfo {year} {2020})}\BibitemShut {NoStop}%
\bibitem [{\citenamefont {Doostmohammadi}\ \emph {et~al.}(2016)\citenamefont
  {Doostmohammadi}, \citenamefont {Thampi},\ and\ \citenamefont
  {Yeomans}}]{Doostmohammadi2016}%
  \BibitemOpen
  \bibfield  {author} {\bibinfo {author} {\bibfnamefont {A.}~\bibnamefont
  {Doostmohammadi}}, \bibinfo {author} {\bibfnamefont {S.~P.}\ \bibnamefont
  {Thampi}}, \ and\ \bibinfo {author} {\bibfnamefont {J.~M.}\ \bibnamefont
  {Yeomans}},\ }\href@noop {} {\bibfield  {journal} {\bibinfo  {journal}
  {Physical review letters}\ }\textbf {\bibinfo {volume} {117}},\ \bibinfo
  {pages} {048102} (\bibinfo {year} {2016})}\BibitemShut {NoStop}%
\bibitem [{\citenamefont {Saw}\ \emph {et~al.}(2017)\citenamefont {Saw},
  \citenamefont {Doostmohammadi}, \citenamefont {Nier}, \citenamefont
  {Kocgozlu}, \citenamefont {Thampi}, \citenamefont {Toyama}, \citenamefont
  {Marcq}, \citenamefont {Lim}, \citenamefont {Yeomans},\ and\ \citenamefont
  {Ladoux}}]{Saw2017}%
  \BibitemOpen
  \bibfield  {author} {\bibinfo {author} {\bibfnamefont {T.~B.}\ \bibnamefont
  {Saw}}, \bibinfo {author} {\bibfnamefont {A.}~\bibnamefont {Doostmohammadi}},
  \bibinfo {author} {\bibfnamefont {V.}~\bibnamefont {Nier}}, \bibinfo {author}
  {\bibfnamefont {L.}~\bibnamefont {Kocgozlu}}, \bibinfo {author}
  {\bibfnamefont {S.}~\bibnamefont {Thampi}}, \bibinfo {author} {\bibfnamefont
  {Y.}~\bibnamefont {Toyama}}, \bibinfo {author} {\bibfnamefont
  {P.}~\bibnamefont {Marcq}}, \bibinfo {author} {\bibfnamefont {C.~T.}\
  \bibnamefont {Lim}}, \bibinfo {author} {\bibfnamefont {J.~M.}\ \bibnamefont
  {Yeomans}}, \ and\ \bibinfo {author} {\bibfnamefont {B.}~\bibnamefont
  {Ladoux}},\ }\href@noop {} {\bibfield  {journal} {\bibinfo  {journal}
  {Nature}\ }\textbf {\bibinfo {volume} {544}},\ \bibinfo {pages} {212}
  (\bibinfo {year} {2017})}\BibitemShut {NoStop}%
\bibitem [{\citenamefont {Trepat}\ \emph {et~al.}(2009)\citenamefont {Trepat},
  \citenamefont {Wasserman}, \citenamefont {Angelini}, \citenamefont {Millet},
  \citenamefont {Weitz}, \citenamefont {Butler},\ and\ \citenamefont
  {Fredberg}}]{Trepat2009}%
  \BibitemOpen
  \bibfield  {author} {\bibinfo {author} {\bibfnamefont {X.}~\bibnamefont
  {Trepat}}, \bibinfo {author} {\bibfnamefont {M.~R.}\ \bibnamefont
  {Wasserman}}, \bibinfo {author} {\bibfnamefont {T.~E.}\ \bibnamefont
  {Angelini}}, \bibinfo {author} {\bibfnamefont {E.}~\bibnamefont {Millet}},
  \bibinfo {author} {\bibfnamefont {D.~A.}\ \bibnamefont {Weitz}}, \bibinfo
  {author} {\bibfnamefont {J.~P.}\ \bibnamefont {Butler}}, \ and\ \bibinfo
  {author} {\bibfnamefont {J.~J.}\ \bibnamefont {Fredberg}},\ }\href@noop {}
  {\bibfield  {journal} {\bibinfo  {journal} {Nature physics}\ }\textbf
  {\bibinfo {volume} {5}},\ \bibinfo {pages} {426} (\bibinfo {year}
  {2009})}\BibitemShut {NoStop}%
\bibitem [{\citenamefont {Trepat}\ and\ \citenamefont
  {Sahai}(2018)}]{Trepat2018}%
  \BibitemOpen
  \bibfield  {author} {\bibinfo {author} {\bibfnamefont {X.}~\bibnamefont
  {Trepat}}\ and\ \bibinfo {author} {\bibfnamefont {E.}~\bibnamefont {Sahai}},\
  }\href@noop {} {\bibfield  {journal} {\bibinfo  {journal} {Nature Physics}\
  }\textbf {\bibinfo {volume} {14}},\ \bibinfo {pages} {671} (\bibinfo {year}
  {2018})}\BibitemShut {NoStop}%
\bibitem [{\citenamefont {Duclos}\ \emph {et~al.}(2014)\citenamefont {Duclos},
  \citenamefont {Garcia}, \citenamefont {Yevick},\ and\ \citenamefont
  {Silberzan}}]{Duclos2014}%
  \BibitemOpen
  \bibfield  {author} {\bibinfo {author} {\bibfnamefont {G.}~\bibnamefont
  {Duclos}}, \bibinfo {author} {\bibfnamefont {S.}~\bibnamefont {Garcia}},
  \bibinfo {author} {\bibfnamefont {H.}~\bibnamefont {Yevick}}, \ and\ \bibinfo
  {author} {\bibfnamefont {P.}~\bibnamefont {Silberzan}},\ }\href@noop {}
  {\bibfield  {journal} {\bibinfo  {journal} {Soft matter}\ }\textbf {\bibinfo
  {volume} {10}},\ \bibinfo {pages} {2346} (\bibinfo {year}
  {2014})}\BibitemShut {NoStop}%
\bibitem [{\citenamefont {Blanch-Mercader}\ \emph {et~al.}(2018)\citenamefont
  {Blanch-Mercader}, \citenamefont {Yashunsky}, \citenamefont {Garcia},
  \citenamefont {Duclos}, \citenamefont {Giomi},\ and\ \citenamefont
  {Silberzan}}]{BlanchMercader2018}%
  \BibitemOpen
  \bibfield  {author} {\bibinfo {author} {\bibfnamefont {C.}~\bibnamefont
  {Blanch-Mercader}}, \bibinfo {author} {\bibfnamefont {V.}~\bibnamefont
  {Yashunsky}}, \bibinfo {author} {\bibfnamefont {S.}~\bibnamefont {Garcia}},
  \bibinfo {author} {\bibfnamefont {G.}~\bibnamefont {Duclos}}, \bibinfo
  {author} {\bibfnamefont {L.}~\bibnamefont {Giomi}}, \ and\ \bibinfo {author}
  {\bibfnamefont {P.}~\bibnamefont {Silberzan}},\ }\href {\doibase
  10.1103/PhysRevLett.120.208101} {\bibfield  {journal} {\bibinfo  {journal}
  {Physical review letters}\ }\textbf {\bibinfo {volume} {120}},\ \bibinfo
  {pages} {208101} (\bibinfo {year} {2018})}\BibitemShut {NoStop}%
\bibitem [{\citenamefont {Garcia}\ \emph {et~al.}(2015)\citenamefont {Garcia},
  \citenamefont {Hannezo}, \citenamefont {Elgeti}, \citenamefont {Joanny},
  \citenamefont {Silberzan},\ and\ \citenamefont {Gov}}]{Garcia2015}%
  \BibitemOpen
  \bibfield  {author} {\bibinfo {author} {\bibfnamefont {S.}~\bibnamefont
  {Garcia}}, \bibinfo {author} {\bibfnamefont {E.}~\bibnamefont {Hannezo}},
  \bibinfo {author} {\bibfnamefont {J.}~\bibnamefont {Elgeti}}, \bibinfo
  {author} {\bibfnamefont {J.-F.}\ \bibnamefont {Joanny}}, \bibinfo {author}
  {\bibfnamefont {P.}~\bibnamefont {Silberzan}}, \ and\ \bibinfo {author}
  {\bibfnamefont {N.~S.}\ \bibnamefont {Gov}},\ }\href@noop {} {\bibfield
  {journal} {\bibinfo  {journal} {Proceedings of the National Academy of
  Sciences}\ }\textbf {\bibinfo {volume} {112}},\ \bibinfo {pages} {15314}
  (\bibinfo {year} {2015})}\BibitemShut {NoStop}%
\bibitem [{\citenamefont {Ghibaudo}\ \emph {et~al.}(2008)\citenamefont
  {Ghibaudo}, \citenamefont {Saez}, \citenamefont {Trichet}, \citenamefont
  {Xayaphoummine}, \citenamefont {Browaeys}, \citenamefont {Silberzan},
  \citenamefont {Buguin},\ and\ \citenamefont {Ladoux}}]{Ghibaudo2008}%
  \BibitemOpen
  \bibfield  {author} {\bibinfo {author} {\bibfnamefont {M.}~\bibnamefont
  {Ghibaudo}}, \bibinfo {author} {\bibfnamefont {A.}~\bibnamefont {Saez}},
  \bibinfo {author} {\bibfnamefont {L.}~\bibnamefont {Trichet}}, \bibinfo
  {author} {\bibfnamefont {A.}~\bibnamefont {Xayaphoummine}}, \bibinfo {author}
  {\bibfnamefont {J.}~\bibnamefont {Browaeys}}, \bibinfo {author}
  {\bibfnamefont {P.}~\bibnamefont {Silberzan}}, \bibinfo {author}
  {\bibfnamefont {A.}~\bibnamefont {Buguin}}, \ and\ \bibinfo {author}
  {\bibfnamefont {B.}~\bibnamefont {Ladoux}},\ }\href@noop {} {\bibfield
  {journal} {\bibinfo  {journal} {Soft Matter}\ }\textbf {\bibinfo {volume}
  {4}},\ \bibinfo {pages} {1836} (\bibinfo {year} {2008})}\BibitemShut
  {NoStop}%
\bibitem [{\citenamefont {Vicsek}\ and\ \citenamefont
  {Zafeiris}(2012)}]{Vicsek2012}%
  \BibitemOpen
  \bibfield  {author} {\bibinfo {author} {\bibfnamefont {T.}~\bibnamefont
  {Vicsek}}\ and\ \bibinfo {author} {\bibfnamefont {A.}~\bibnamefont
  {Zafeiris}},\ }\href {\doibase 10.1016/j.physrep.2012.03.004} {\bibfield
  {journal} {\bibinfo  {journal} {Physics reports}\ }\textbf {\bibinfo {volume}
  {517}},\ \bibinfo {pages} {71} (\bibinfo {year} {2012})}\BibitemShut
  {NoStop}%
\bibitem [{\citenamefont {Chat{\'e}}(2020)}]{Chate2020}%
  \BibitemOpen
  \bibfield  {author} {\bibinfo {author} {\bibfnamefont {H.}~\bibnamefont
  {Chat{\'e}}},\ }\href@noop {} {\bibfield  {journal} {\bibinfo  {journal}
  {Annual Review of Condensed Matter Physics}\ }\textbf {\bibinfo {volume}
  {11}},\ \bibinfo {pages} {189} (\bibinfo {year} {2020})}\BibitemShut
  {NoStop}%
\bibitem [{\citenamefont {Doostmohammadi}\ \emph {et~al.}(2018)\citenamefont
  {Doostmohammadi}, \citenamefont {Ign{\'e}s-Mullol}, \citenamefont {Yeomans},\
  and\ \citenamefont {Sagu{\'e}s}}]{Doostmohammadi2018}%
  \BibitemOpen
  \bibfield  {author} {\bibinfo {author} {\bibfnamefont {A.}~\bibnamefont
  {Doostmohammadi}}, \bibinfo {author} {\bibfnamefont {J.}~\bibnamefont
  {Ign{\'e}s-Mullol}}, \bibinfo {author} {\bibfnamefont {J.~M.}\ \bibnamefont
  {Yeomans}}, \ and\ \bibinfo {author} {\bibfnamefont {F.}~\bibnamefont
  {Sagu{\'e}s}},\ }\href {\doibase 10.1038/s41467-018-05666-8} {\bibfield
  {journal} {\bibinfo  {journal} {Nature communications}\ }\textbf {\bibinfo
  {volume} {9}},\ \bibinfo {pages} {1} (\bibinfo {year} {2018})}\BibitemShut
  {NoStop}%
\bibitem [{\citenamefont {Bechinger}\ \emph {et~al.}(2016)\citenamefont
  {Bechinger}, \citenamefont {Di~Leonardo}, \citenamefont {L{\"o}wen},
  \citenamefont {Reichhardt}, \citenamefont {Volpe},\ and\ \citenamefont
  {Volpe}}]{Bechinger2016}%
  \BibitemOpen
  \bibfield  {author} {\bibinfo {author} {\bibfnamefont {C.}~\bibnamefont
  {Bechinger}}, \bibinfo {author} {\bibfnamefont {R.}~\bibnamefont
  {Di~Leonardo}}, \bibinfo {author} {\bibfnamefont {H.}~\bibnamefont
  {L{\"o}wen}}, \bibinfo {author} {\bibfnamefont {C.}~\bibnamefont
  {Reichhardt}}, \bibinfo {author} {\bibfnamefont {G.}~\bibnamefont {Volpe}}, \
  and\ \bibinfo {author} {\bibfnamefont {G.}~\bibnamefont {Volpe}},\
  }\href@noop {} {\bibfield  {journal} {\bibinfo  {journal} {Reviews of Modern
  Physics}\ }\textbf {\bibinfo {volume} {88}},\ \bibinfo {pages} {045006}
  (\bibinfo {year} {2016})}\BibitemShut {NoStop}%
\bibitem [{\citenamefont {Gompper}\ \emph {et~al.}(2020)\citenamefont
  {Gompper}, \citenamefont {Winkler}, \citenamefont {Speck}, \citenamefont
  {Solon}, \citenamefont {Nardini}, \citenamefont {Peruani}, \citenamefont
  {L{\"o}wen}, \citenamefont {Golestanian}, \citenamefont {Kaupp},
  \citenamefont {Alvarez} \emph {et~al.}}]{Gompper2020}%
  \BibitemOpen
  \bibfield  {author} {\bibinfo {author} {\bibfnamefont {G.}~\bibnamefont
  {Gompper}}, \bibinfo {author} {\bibfnamefont {R.~G.}\ \bibnamefont
  {Winkler}}, \bibinfo {author} {\bibfnamefont {T.}~\bibnamefont {Speck}},
  \bibinfo {author} {\bibfnamefont {A.}~\bibnamefont {Solon}}, \bibinfo
  {author} {\bibfnamefont {C.}~\bibnamefont {Nardini}}, \bibinfo {author}
  {\bibfnamefont {F.}~\bibnamefont {Peruani}}, \bibinfo {author} {\bibfnamefont
  {H.}~\bibnamefont {L{\"o}wen}}, \bibinfo {author} {\bibfnamefont
  {R.}~\bibnamefont {Golestanian}}, \bibinfo {author} {\bibfnamefont {U.~B.}\
  \bibnamefont {Kaupp}}, \bibinfo {author} {\bibfnamefont {L.}~\bibnamefont
  {Alvarez}},  \emph {et~al.},\ }\href {\doibase 10.1088/1361-648X/ab6348}
  {\bibfield  {journal} {\bibinfo  {journal} {Journal of Physics: Condensed
  Matter}\ }\textbf {\bibinfo {volume} {32}},\ \bibinfo {pages} {193001}
  (\bibinfo {year} {2020})}\BibitemShut {NoStop}%
\bibitem [{\citenamefont {Marchetti}\ \emph {et~al.}(2013)\citenamefont
  {Marchetti}, \citenamefont {Joanny}, \citenamefont {Ramaswamy}, \citenamefont
  {Liverpool}, \citenamefont {Prost}, \citenamefont {Rao},\ and\ \citenamefont
  {Simha}}]{Marchetti2013}%
  \BibitemOpen
  \bibfield  {author} {\bibinfo {author} {\bibfnamefont {M.~C.}\ \bibnamefont
  {Marchetti}}, \bibinfo {author} {\bibfnamefont {J.-F.}\ \bibnamefont
  {Joanny}}, \bibinfo {author} {\bibfnamefont {S.}~\bibnamefont {Ramaswamy}},
  \bibinfo {author} {\bibfnamefont {T.~B.}\ \bibnamefont {Liverpool}}, \bibinfo
  {author} {\bibfnamefont {J.}~\bibnamefont {Prost}}, \bibinfo {author}
  {\bibfnamefont {M.}~\bibnamefont {Rao}}, \ and\ \bibinfo {author}
  {\bibfnamefont {R.~A.}\ \bibnamefont {Simha}},\ }\href {\doibase
  10.1103/RevModPhys.85.1143} {\bibfield  {journal} {\bibinfo  {journal}
  {Reviews of Modern Physics}\ }\textbf {\bibinfo {volume} {85}},\ \bibinfo
  {pages} {1143} (\bibinfo {year} {2013})}\BibitemShut {NoStop}%
\bibitem [{\citenamefont {Aubret}\ \emph {et~al.}(2018)\citenamefont {Aubret},
  \citenamefont {Youssef}, \citenamefont {Sacanna},\ and\ \citenamefont
  {Palacci}}]{Aubret2018}%
  \BibitemOpen
  \bibfield  {author} {\bibinfo {author} {\bibfnamefont {A.}~\bibnamefont
  {Aubret}}, \bibinfo {author} {\bibfnamefont {M.}~\bibnamefont {Youssef}},
  \bibinfo {author} {\bibfnamefont {S.}~\bibnamefont {Sacanna}}, \ and\
  \bibinfo {author} {\bibfnamefont {J.}~\bibnamefont {Palacci}},\ }\href@noop
  {} {\bibfield  {journal} {\bibinfo  {journal} {Nature Physics}\ }\textbf
  {\bibinfo {volume} {14}},\ \bibinfo {pages} {1114} (\bibinfo {year}
  {2018})}\BibitemShut {NoStop}%
\bibitem [{\citenamefont {Palacci}\ \emph {et~al.}(2013)\citenamefont
  {Palacci}, \citenamefont {Sacanna}, \citenamefont {Steinberg}, \citenamefont
  {Pine},\ and\ \citenamefont {Chaikin}}]{Palacci2013}%
  \BibitemOpen
  \bibfield  {author} {\bibinfo {author} {\bibfnamefont {J.}~\bibnamefont
  {Palacci}}, \bibinfo {author} {\bibfnamefont {S.}~\bibnamefont {Sacanna}},
  \bibinfo {author} {\bibfnamefont {A.~P.}\ \bibnamefont {Steinberg}}, \bibinfo
  {author} {\bibfnamefont {D.~J.}\ \bibnamefont {Pine}}, \ and\ \bibinfo
  {author} {\bibfnamefont {P.~M.}\ \bibnamefont {Chaikin}},\ }\href {\doibase
  10.1126/science.1230020} {\bibfield  {journal} {\bibinfo  {journal}
  {Science}\ }\textbf {\bibinfo {volume} {339}},\ \bibinfo {pages} {936}
  (\bibinfo {year} {2013})}\BibitemShut {NoStop}%
\bibitem [{\citenamefont {Arlt}\ \emph {et~al.}(2018)\citenamefont {Arlt},
  \citenamefont {Martinez}, \citenamefont {Dawson}, \citenamefont {Pilizota},\
  and\ \citenamefont {Poon}}]{Arlt2018}%
  \BibitemOpen
  \bibfield  {author} {\bibinfo {author} {\bibfnamefont {J.}~\bibnamefont
  {Arlt}}, \bibinfo {author} {\bibfnamefont {V.~A.}\ \bibnamefont {Martinez}},
  \bibinfo {author} {\bibfnamefont {A.}~\bibnamefont {Dawson}}, \bibinfo
  {author} {\bibfnamefont {T.}~\bibnamefont {Pilizota}}, \ and\ \bibinfo
  {author} {\bibfnamefont {W.~C.}\ \bibnamefont {Poon}},\ }\href@noop {}
  {\bibfield  {journal} {\bibinfo  {journal} {Nature communications}\ }\textbf
  {\bibinfo {volume} {9}},\ \bibinfo {pages} {768} (\bibinfo {year}
  {2018})}\BibitemShut {NoStop}%
\bibitem [{\citenamefont {Frangipane}\ \emph {et~al.}(2018)\citenamefont
  {Frangipane}, \citenamefont {Dell'Arciprete}, \citenamefont {Petracchini},
  \citenamefont {Maggi}, \citenamefont {Saglimbeni}, \citenamefont {Bianchi},
  \citenamefont {Vizsnyiczai}, \citenamefont {Bernardini},\ and\ \citenamefont
  {Di~Leonardo}}]{Frangipane2018}%
  \BibitemOpen
  \bibfield  {author} {\bibinfo {author} {\bibfnamefont {G.}~\bibnamefont
  {Frangipane}}, \bibinfo {author} {\bibfnamefont {D.}~\bibnamefont
  {Dell'Arciprete}}, \bibinfo {author} {\bibfnamefont {S.}~\bibnamefont
  {Petracchini}}, \bibinfo {author} {\bibfnamefont {C.}~\bibnamefont {Maggi}},
  \bibinfo {author} {\bibfnamefont {F.}~\bibnamefont {Saglimbeni}}, \bibinfo
  {author} {\bibfnamefont {S.}~\bibnamefont {Bianchi}}, \bibinfo {author}
  {\bibfnamefont {G.}~\bibnamefont {Vizsnyiczai}}, \bibinfo {author}
  {\bibfnamefont {M.~L.}\ \bibnamefont {Bernardini}}, \ and\ \bibinfo {author}
  {\bibfnamefont {R.}~\bibnamefont {Di~Leonardo}},\ }\href@noop {} {\bibfield
  {journal} {\bibinfo  {journal} {Elife}\ }\textbf {\bibinfo {volume} {7}},\
  \bibinfo {pages} {e36608} (\bibinfo {year} {2018})}\BibitemShut {NoStop}%
\bibitem [{\citenamefont {Lemma}\ \emph {et~al.}(2023)\citenamefont {Lemma},
  \citenamefont {Varghese}, \citenamefont {Ross}, \citenamefont {Thomson},
  \citenamefont {Baskaran},\ and\ \citenamefont {Dogic}}]{Lemma2023}%
  \BibitemOpen
  \bibfield  {author} {\bibinfo {author} {\bibfnamefont {L.~M.}\ \bibnamefont
  {Lemma}}, \bibinfo {author} {\bibfnamefont {M.}~\bibnamefont {Varghese}},
  \bibinfo {author} {\bibfnamefont {T.~D.}\ \bibnamefont {Ross}}, \bibinfo
  {author} {\bibfnamefont {M.}~\bibnamefont {Thomson}}, \bibinfo {author}
  {\bibfnamefont {A.}~\bibnamefont {Baskaran}}, \ and\ \bibinfo {author}
  {\bibfnamefont {Z.}~\bibnamefont {Dogic}},\ }\href@noop {} {\bibfield
  {journal} {\bibinfo  {journal} {PNAS nexus}\ }\textbf {\bibinfo {volume}
  {2}},\ \bibinfo {pages} {pgad130} (\bibinfo {year} {2023})}\BibitemShut
  {NoStop}%
\bibitem [{\citenamefont {Zarei}\ \emph {et~al.}(2023)\citenamefont {Zarei},
  \citenamefont {Berezney}, \citenamefont {Hensley}, \citenamefont {Lemma},
  \citenamefont {Senbil}, \citenamefont {Dogic},\ and\ \citenamefont
  {Fraden}}]{Zarei2023}%
  \BibitemOpen
  \bibfield  {author} {\bibinfo {author} {\bibfnamefont {Z.}~\bibnamefont
  {Zarei}}, \bibinfo {author} {\bibfnamefont {J.}~\bibnamefont {Berezney}},
  \bibinfo {author} {\bibfnamefont {A.}~\bibnamefont {Hensley}}, \bibinfo
  {author} {\bibfnamefont {L.}~\bibnamefont {Lemma}}, \bibinfo {author}
  {\bibfnamefont {N.}~\bibnamefont {Senbil}}, \bibinfo {author} {\bibfnamefont
  {Z.}~\bibnamefont {Dogic}}, \ and\ \bibinfo {author} {\bibfnamefont
  {S.}~\bibnamefont {Fraden}},\ }\href@noop {} {\bibfield  {journal} {\bibinfo
  {journal} {Soft matter}\ }\textbf {\bibinfo {volume} {19}},\ \bibinfo {pages}
  {6691} (\bibinfo {year} {2023})}\BibitemShut {NoStop}%
\bibitem [{\citenamefont {Zhang}\ \emph {et~al.}(2021)\citenamefont {Zhang},
  \citenamefont {Redford}, \citenamefont {Ruijgrok}, \citenamefont {Kumar},
  \citenamefont {Mozaffari}, \citenamefont {Zemsky}, \citenamefont {Dinner},
  \citenamefont {Vitelli}, \citenamefont {Bryant}, \citenamefont {Gardel} \emph
  {et~al.}}]{Zhang2021}%
  \BibitemOpen
  \bibfield  {author} {\bibinfo {author} {\bibfnamefont {R.}~\bibnamefont
  {Zhang}}, \bibinfo {author} {\bibfnamefont {S.~A.}\ \bibnamefont {Redford}},
  \bibinfo {author} {\bibfnamefont {P.~V.}\ \bibnamefont {Ruijgrok}}, \bibinfo
  {author} {\bibfnamefont {N.}~\bibnamefont {Kumar}}, \bibinfo {author}
  {\bibfnamefont {A.}~\bibnamefont {Mozaffari}}, \bibinfo {author}
  {\bibfnamefont {S.}~\bibnamefont {Zemsky}}, \bibinfo {author} {\bibfnamefont
  {A.~R.}\ \bibnamefont {Dinner}}, \bibinfo {author} {\bibfnamefont
  {V.}~\bibnamefont {Vitelli}}, \bibinfo {author} {\bibfnamefont
  {Z.}~\bibnamefont {Bryant}}, \bibinfo {author} {\bibfnamefont {M.~L.}\
  \bibnamefont {Gardel}},  \emph {et~al.},\ }\href {\doibase
  10.1038/s41563-020-00901-4} {\bibfield  {journal} {\bibinfo  {journal}
  {Nature Materials}\ }\textbf {\bibinfo {volume} {20}},\ \bibinfo {pages}
  {875} (\bibinfo {year} {2021})}\BibitemShut {NoStop}%
\bibitem [{\citenamefont {Ross}\ \emph {et~al.}(2019)\citenamefont {Ross},
  \citenamefont {Lee}, \citenamefont {Qu}, \citenamefont {Banks}, \citenamefont
  {Phillips},\ and\ \citenamefont {Thomson}}]{Ross2019}%
  \BibitemOpen
  \bibfield  {author} {\bibinfo {author} {\bibfnamefont {T.~D.}\ \bibnamefont
  {Ross}}, \bibinfo {author} {\bibfnamefont {H.~J.}\ \bibnamefont {Lee}},
  \bibinfo {author} {\bibfnamefont {Z.}~\bibnamefont {Qu}}, \bibinfo {author}
  {\bibfnamefont {R.~A.}\ \bibnamefont {Banks}}, \bibinfo {author}
  {\bibfnamefont {R.}~\bibnamefont {Phillips}}, \ and\ \bibinfo {author}
  {\bibfnamefont {M.}~\bibnamefont {Thomson}},\ }\href {\doibase
  10.1038/s41586-019-1447-1} {\bibfield  {journal} {\bibinfo  {journal}
  {Nature}\ }\textbf {\bibinfo {volume} {572}},\ \bibinfo {pages} {224}
  (\bibinfo {year} {2019})}\BibitemShut {NoStop}%
\bibitem [{\citenamefont {Shankar}\ \emph {et~al.}(2022)\citenamefont
  {Shankar}, \citenamefont {Scharrer}, \citenamefont {Bowick},\ and\
  \citenamefont {Marchetti}}]{Shankar2022}%
  \BibitemOpen
  \bibfield  {author} {\bibinfo {author} {\bibfnamefont {S.}~\bibnamefont
  {Shankar}}, \bibinfo {author} {\bibfnamefont {L.~V.}\ \bibnamefont
  {Scharrer}}, \bibinfo {author} {\bibfnamefont {M.~J.}\ \bibnamefont
  {Bowick}}, \ and\ \bibinfo {author} {\bibfnamefont {M.~C.}\ \bibnamefont
  {Marchetti}},\ }\href@noop {} {\bibfield  {journal} {\bibinfo  {journal}
  {arXiv preprint arXiv:2212.00666}\ } (\bibinfo {year} {2022})}\BibitemShut
  {NoStop}%
\bibitem [{\citenamefont {Nasiri}\ \emph {et~al.}(2023)\citenamefont {Nasiri},
  \citenamefont {L{\"o}wen},\ and\ \citenamefont {Liebchen}}]{Nasiri2023}%
  \BibitemOpen
  \bibfield  {author} {\bibinfo {author} {\bibfnamefont {M.}~\bibnamefont
  {Nasiri}}, \bibinfo {author} {\bibfnamefont {H.}~\bibnamefont {L{\"o}wen}}, \
  and\ \bibinfo {author} {\bibfnamefont {B.}~\bibnamefont {Liebchen}},\
  }\href@noop {} {\bibfield  {journal} {\bibinfo  {journal} {Europhysics
  Letters}\ }\textbf {\bibinfo {volume} {142}},\ \bibinfo {pages} {17001}
  (\bibinfo {year} {2023})}\BibitemShut {NoStop}%
\bibitem [{\citenamefont {Kne{\v{z}}evi{\'c}}\ \emph
  {et~al.}(2022)\citenamefont {Kne{\v{z}}evi{\'c}}, \citenamefont {Welker},\
  and\ \citenamefont {Stark}}]{Knezevic2022}%
  \BibitemOpen
  \bibfield  {author} {\bibinfo {author} {\bibfnamefont {M.}~\bibnamefont
  {Kne{\v{z}}evi{\'c}}}, \bibinfo {author} {\bibfnamefont {T.}~\bibnamefont
  {Welker}}, \ and\ \bibinfo {author} {\bibfnamefont {H.}~\bibnamefont
  {Stark}},\ }\href@noop {} {\bibfield  {journal} {\bibinfo  {journal}
  {Scientific Reports}\ }\textbf {\bibinfo {volume} {12}},\ \bibinfo {pages}
  {19437} (\bibinfo {year} {2022})}\BibitemShut {NoStop}%
\bibitem [{\citenamefont {Brunton}\ and\ \citenamefont
  {Kutz}(2022)}]{Brunton2022}%
  \BibitemOpen
  \bibfield  {author} {\bibinfo {author} {\bibfnamefont {S.~L.}\ \bibnamefont
  {Brunton}}\ and\ \bibinfo {author} {\bibfnamefont {J.~N.}\ \bibnamefont
  {Kutz}},\ }\href@noop {} {\emph {\bibinfo {title} {Data-driven science and
  engineering: Machine learning, dynamical systems, and control}}}\ (\bibinfo
  {publisher} {Cambridge University Press},\ \bibinfo {year}
  {2022})\BibitemShut {NoStop}%
\bibitem [{\citenamefont {Dullerud}\ and\ \citenamefont
  {Paganini}(2013)}]{Dullerud2013}%
  \BibitemOpen
  \bibfield  {author} {\bibinfo {author} {\bibfnamefont {G.~E.}\ \bibnamefont
  {Dullerud}}\ and\ \bibinfo {author} {\bibfnamefont {F.}~\bibnamefont
  {Paganini}},\ }\href@noop {} {\emph {\bibinfo {title} {A course in robust
  control theory: a convex approach}}},\ Vol.~\bibinfo {volume} {36}\ (\bibinfo
   {publisher} {Springer Science \& Business Media},\ \bibinfo {year}
  {2013})\BibitemShut {NoStop}%
\bibitem [{\citenamefont {Villani}\ \emph {et~al.}(2009)\citenamefont {Villani}
  \emph {et~al.}}]{Villani2009}%
  \BibitemOpen
  \bibfield  {author} {\bibinfo {author} {\bibfnamefont {C.}~\bibnamefont
  {Villani}} \emph {et~al.},\ }\href@noop {} {\emph {\bibinfo {title} {Optimal
  transport: old and new}}},\ Vol.\ \bibinfo {volume} {338}\ (\bibinfo
  {publisher} {Springer},\ \bibinfo {year} {2009})\BibitemShut {NoStop}%
\bibitem [{\citenamefont {Norton}\ \emph {et~al.}(2020)\citenamefont {Norton},
  \citenamefont {Grover}, \citenamefont {Hagan},\ and\ \citenamefont
  {Fraden}}]{Norton2020}%
  \BibitemOpen
  \bibfield  {author} {\bibinfo {author} {\bibfnamefont {M.~M.}\ \bibnamefont
  {Norton}}, \bibinfo {author} {\bibfnamefont {P.}~\bibnamefont {Grover}},
  \bibinfo {author} {\bibfnamefont {M.~F.}\ \bibnamefont {Hagan}}, \ and\
  \bibinfo {author} {\bibfnamefont {S.}~\bibnamefont {Fraden}},\ }\href
  {\doibase 10.1103/PhysRevLett.125.178005} {\bibfield  {journal} {\bibinfo
  {journal} {Physical review letters}\ }\textbf {\bibinfo {volume} {125}},\
  \bibinfo {pages} {178005} (\bibinfo {year} {2020})}\BibitemShut {NoStop}%
\bibitem [{\citenamefont {Sinigaglia}\ \emph {et~al.}(2023)\citenamefont
  {Sinigaglia}, \citenamefont {Braghin},\ and\ \citenamefont
  {Serra}}]{Sinigaglia2023}%
  \BibitemOpen
  \bibfield  {author} {\bibinfo {author} {\bibfnamefont {C.}~\bibnamefont
  {Sinigaglia}}, \bibinfo {author} {\bibfnamefont {F.}~\bibnamefont {Braghin}},
  \ and\ \bibinfo {author} {\bibfnamefont {M.}~\bibnamefont {Serra}},\
  }\href@noop {} {\bibfield  {journal} {\bibinfo  {journal} {arXiv preprint
  arXiv:2305.00193}\ } (\bibinfo {year} {2023})}\BibitemShut {NoStop}%
\bibitem [{\citenamefont {Toner}\ and\ \citenamefont {Tu}(1995)}]{Toner1995}%
  \BibitemOpen
  \bibfield  {author} {\bibinfo {author} {\bibfnamefont {J.}~\bibnamefont
  {Toner}}\ and\ \bibinfo {author} {\bibfnamefont {Y.}~\bibnamefont {Tu}},\
  }\href {\doibase 10.1103/PhysRevLett.75.4326} {\bibfield  {journal} {\bibinfo
   {journal} {Physical review letters}\ }\textbf {\bibinfo {volume} {75}},\
  \bibinfo {pages} {4326} (\bibinfo {year} {1995})}\BibitemShut {NoStop}%
\bibitem [{\citenamefont {Toner}\ and\ \citenamefont {Tu}(1998)}]{Toner1998}%
  \BibitemOpen
  \bibfield  {author} {\bibinfo {author} {\bibfnamefont {J.}~\bibnamefont
  {Toner}}\ and\ \bibinfo {author} {\bibfnamefont {Y.}~\bibnamefont {Tu}},\
  }\href {\doibase 10.1103/PhysRevE.58.4828} {\bibfield  {journal} {\bibinfo
  {journal} {Physical review E}\ }\textbf {\bibinfo {volume} {58}},\ \bibinfo
  {pages} {4828} (\bibinfo {year} {1998})}\BibitemShut {NoStop}%
\bibitem [{\citenamefont {Lee}\ and\ \citenamefont {Kardar}(2001)}]{Lee2001}%
  \BibitemOpen
  \bibfield  {author} {\bibinfo {author} {\bibfnamefont {H.~Y.}\ \bibnamefont
  {Lee}}\ and\ \bibinfo {author} {\bibfnamefont {M.}~\bibnamefont {Kardar}},\
  }\href {\doibase 10.1103/physreve.64.056113} {\bibfield  {journal} {\bibinfo
  {journal} {Physical Review E}\ }\textbf {\bibinfo {volume} {64}} (\bibinfo
  {year} {2001}),\ 10.1103/physreve.64.056113}\BibitemShut {NoStop}%
\bibitem [{\citenamefont {Husain}\ and\ \citenamefont
  {Rao}(2017)}]{Husain2017}%
  \BibitemOpen
  \bibfield  {author} {\bibinfo {author} {\bibfnamefont {K.}~\bibnamefont
  {Husain}}\ and\ \bibinfo {author} {\bibfnamefont {M.}~\bibnamefont {Rao}},\
  }\href {\doibase 10.1103/physrevlett.118.078104} {\bibfield  {journal}
  {\bibinfo  {journal} {Physical Review Letters}\ }\textbf {\bibinfo {volume}
  {118}} (\bibinfo {year} {2017}),\ 10.1103/physrevlett.118.078104}\BibitemShut
  {NoStop}%
\bibitem [{\citenamefont {Geyer}\ \emph {et~al.}(2018)\citenamefont {Geyer},
  \citenamefont {Morin},\ and\ \citenamefont {Bartolo}}]{Geyer2018}%
  \BibitemOpen
  \bibfield  {author} {\bibinfo {author} {\bibfnamefont {D.}~\bibnamefont
  {Geyer}}, \bibinfo {author} {\bibfnamefont {A.}~\bibnamefont {Morin}}, \ and\
  \bibinfo {author} {\bibfnamefont {D.}~\bibnamefont {Bartolo}},\ }\href
  {\doibase 10.1038/s41563-018-0123-4} {\bibfield  {journal} {\bibinfo
  {journal} {Nature Materials}\ }\textbf {\bibinfo {volume} {17}},\ \bibinfo
  {pages} {789} (\bibinfo {year} {2018})}\BibitemShut {NoStop}%
\bibitem [{\citenamefont {Worlitzer}\ \emph {et~al.}(2021)\citenamefont
  {Worlitzer}, \citenamefont {Ariel}, \citenamefont {Be'er}, \citenamefont
  {Stark}, \citenamefont {B\"{a}r},\ and\ \citenamefont
  {Heidenreich}}]{Worlitzer2021}%
  \BibitemOpen
  \bibfield  {author} {\bibinfo {author} {\bibfnamefont {V.~M.}\ \bibnamefont
  {Worlitzer}}, \bibinfo {author} {\bibfnamefont {G.}~\bibnamefont {Ariel}},
  \bibinfo {author} {\bibfnamefont {A.}~\bibnamefont {Be'er}}, \bibinfo
  {author} {\bibfnamefont {H.}~\bibnamefont {Stark}}, \bibinfo {author}
  {\bibfnamefont {M.}~\bibnamefont {B\"{a}r}}, \ and\ \bibinfo {author}
  {\bibfnamefont {S.}~\bibnamefont {Heidenreich}},\ }\href {\doibase
  10.1088/1367-2630/abe72d} {\bibfield  {journal} {\bibinfo  {journal} {New
  Journal of Physics}\ }\textbf {\bibinfo {volume} {23}},\ \bibinfo {pages}
  {033012} (\bibinfo {year} {2021})}\BibitemShut {NoStop}%
\bibitem [{\citenamefont {Mishra}\ \emph {et~al.}(2010)\citenamefont {Mishra},
  \citenamefont {Baskaran},\ and\ \citenamefont {Marchetti}}]{Mishra2010}%
  \BibitemOpen
  \bibfield  {author} {\bibinfo {author} {\bibfnamefont {S.}~\bibnamefont
  {Mishra}}, \bibinfo {author} {\bibfnamefont {A.}~\bibnamefont {Baskaran}}, \
  and\ \bibinfo {author} {\bibfnamefont {M.~C.}\ \bibnamefont {Marchetti}},\
  }\href {\doibase 10.1103/physreve.81.061916} {\bibfield  {journal} {\bibinfo
  {journal} {Physical Review E}\ }\textbf {\bibinfo {volume} {81}} (\bibinfo
  {year} {2010}),\ 10.1103/physreve.81.061916}\BibitemShut {NoStop}%
\bibitem [{\citenamefont {Gopinath}\ \emph {et~al.}(2012)\citenamefont
  {Gopinath}, \citenamefont {Hagan}, \citenamefont {Marchetti},\ and\
  \citenamefont {Baskaran}}]{Gopinath2012}%
  \BibitemOpen
  \bibfield  {author} {\bibinfo {author} {\bibfnamefont {A.}~\bibnamefont
  {Gopinath}}, \bibinfo {author} {\bibfnamefont {M.~F.}\ \bibnamefont {Hagan}},
  \bibinfo {author} {\bibfnamefont {M.~C.}\ \bibnamefont {Marchetti}}, \ and\
  \bibinfo {author} {\bibfnamefont {A.}~\bibnamefont {Baskaran}},\ }\href
  {\doibase 10.1103/PhysRevE.85.061903} {\bibfield  {journal} {\bibinfo
  {journal} {Physical Review E}\ }\textbf {\bibinfo {volume} {85}},\ \bibinfo
  {pages} {061903} (\bibinfo {year} {2012})}\BibitemShut {NoStop}%
\bibitem [{\citenamefont {Caussin}\ \emph {et~al.}(2014)\citenamefont
  {Caussin}, \citenamefont {Solon}, \citenamefont {Peshkov}, \citenamefont
  {Chat{\'{e}}}, \citenamefont {Dauxois}, \citenamefont {Tailleur},
  \citenamefont {Vitelli},\ and\ \citenamefont {Bartolo}}]{Caussin2014}%
  \BibitemOpen
  \bibfield  {author} {\bibinfo {author} {\bibfnamefont {J.-B.}\ \bibnamefont
  {Caussin}}, \bibinfo {author} {\bibfnamefont {A.}~\bibnamefont {Solon}},
  \bibinfo {author} {\bibfnamefont {A.}~\bibnamefont {Peshkov}}, \bibinfo
  {author} {\bibfnamefont {H.}~\bibnamefont {Chat{\'{e}}}}, \bibinfo {author}
  {\bibfnamefont {T.}~\bibnamefont {Dauxois}}, \bibinfo {author} {\bibfnamefont
  {J.}~\bibnamefont {Tailleur}}, \bibinfo {author} {\bibfnamefont
  {V.}~\bibnamefont {Vitelli}}, \ and\ \bibinfo {author} {\bibfnamefont
  {D.}~\bibnamefont {Bartolo}},\ }\href {\doibase
  10.1103/physrevlett.112.148102} {\bibfield  {journal} {\bibinfo  {journal}
  {Physical Review Letters}\ }\textbf {\bibinfo {volume} {112}} (\bibinfo
  {year} {2014}),\ 10.1103/physrevlett.112.148102}\BibitemShut {NoStop}%
\bibitem [{\citenamefont {Reinken}\ \emph {et~al.}(2018)\citenamefont
  {Reinken}, \citenamefont {Klapp}, \citenamefont {B\"{a}r},\ and\
  \citenamefont {Heidenreich}}]{Reinken2018}%
  \BibitemOpen
  \bibfield  {author} {\bibinfo {author} {\bibfnamefont {H.}~\bibnamefont
  {Reinken}}, \bibinfo {author} {\bibfnamefont {S.~H.~L.}\ \bibnamefont
  {Klapp}}, \bibinfo {author} {\bibfnamefont {M.}~\bibnamefont {B\"{a}r}}, \
  and\ \bibinfo {author} {\bibfnamefont {S.}~\bibnamefont {Heidenreich}},\
  }\href {\doibase 10.1103/physreve.97.022613} {\bibfield  {journal} {\bibinfo
  {journal} {Physical Review E}\ }\textbf {\bibinfo {volume} {97}} (\bibinfo
  {year} {2018}),\ 10.1103/physreve.97.022613}\BibitemShut {NoStop}%
\bibitem [{\citenamefont {Ngamsaad}\ and\ \citenamefont
  {Suantai}(2018)}]{Ngamsaad2018}%
  \BibitemOpen
  \bibfield  {author} {\bibinfo {author} {\bibfnamefont {W.}~\bibnamefont
  {Ngamsaad}}\ and\ \bibinfo {author} {\bibfnamefont {S.}~\bibnamefont
  {Suantai}},\ }\href {\doibase 10.1103/physreve.98.062618} {\bibfield
  {journal} {\bibinfo  {journal} {Physical Review E}\ }\textbf {\bibinfo
  {volume} {98}} (\bibinfo {year} {2018}),\
  10.1103/physreve.98.062618}\BibitemShut {NoStop}%
\bibitem [{\citenamefont {Bertin}\ \emph {et~al.}(2006)\citenamefont {Bertin},
  \citenamefont {Droz},\ and\ \citenamefont {Gr{\'{e}}goire}}]{Bertin2006}%
  \BibitemOpen
  \bibfield  {author} {\bibinfo {author} {\bibfnamefont {E.}~\bibnamefont
  {Bertin}}, \bibinfo {author} {\bibfnamefont {M.}~\bibnamefont {Droz}}, \ and\
  \bibinfo {author} {\bibfnamefont {G.}~\bibnamefont {Gr{\'{e}}goire}},\ }\href
  {\doibase 10.1103/physreve.74.022101} {\bibfield  {journal} {\bibinfo
  {journal} {Physical Review E}\ }\textbf {\bibinfo {volume} {74}} (\bibinfo
  {year} {2006}),\ 10.1103/physreve.74.022101}\BibitemShut {NoStop}%
\bibitem [{\citenamefont {Peshkov}\ \emph {et~al.}(2012)\citenamefont
  {Peshkov}, \citenamefont {Aranson}, \citenamefont {Bertin}, \citenamefont
  {Chat{\'{e}}},\ and\ \citenamefont {Ginelli}}]{Peshkov2012}%
  \BibitemOpen
  \bibfield  {author} {\bibinfo {author} {\bibfnamefont {A.}~\bibnamefont
  {Peshkov}}, \bibinfo {author} {\bibfnamefont {I.~S.}\ \bibnamefont
  {Aranson}}, \bibinfo {author} {\bibfnamefont {E.}~\bibnamefont {Bertin}},
  \bibinfo {author} {\bibfnamefont {H.}~\bibnamefont {Chat{\'{e}}}}, \ and\
  \bibinfo {author} {\bibfnamefont {F.}~\bibnamefont {Ginelli}},\ }\href
  {\doibase 10.1103/physrevlett.109.268701} {\bibfield  {journal} {\bibinfo
  {journal} {Physical Review Letters}\ }\textbf {\bibinfo {volume} {109}}
  (\bibinfo {year} {2012}),\ 10.1103/physrevlett.109.268701}\BibitemShut
  {NoStop}%
\bibitem [{\citenamefont {Solon}\ and\ \citenamefont
  {Tailleur}(2013)}]{Solon2013}%
  \BibitemOpen
  \bibfield  {author} {\bibinfo {author} {\bibfnamefont {A.~P.}\ \bibnamefont
  {Solon}}\ and\ \bibinfo {author} {\bibfnamefont {J.}~\bibnamefont
  {Tailleur}},\ }\href {\doibase 10.1103/physrevlett.111.078101} {\bibfield
  {journal} {\bibinfo  {journal} {Physical Review Letters}\ }\textbf {\bibinfo
  {volume} {111}} (\bibinfo {year} {2013}),\
  10.1103/physrevlett.111.078101}\BibitemShut {NoStop}%
\bibitem [{SIr()}]{SIref}%
  \BibitemOpen
  \href@noop {} {\ }\bibinfo {note} {Supplemental Material: Spatiotemporal
  control of structure in a polar active fluid}\BibitemShut {NoStop}%
\bibitem [{\citenamefont {Kirk}(2004)}]{Kirk2004}%
  \BibitemOpen
  \bibfield  {author} {\bibinfo {author} {\bibfnamefont {D.~E.}\ \bibnamefont
  {Kirk}},\ }\href@noop {} {\emph {\bibinfo {title} {Optimal control theory: an
  introduction}}}\ (\bibinfo  {publisher} {Courier Corporation},\ \bibinfo
  {year} {2004})\BibitemShut {NoStop}%
\bibitem [{\citenamefont {Lenhart}\ and\ \citenamefont
  {Workman}(2007)}]{Lenhart2007}%
  \BibitemOpen
  \bibfield  {author} {\bibinfo {author} {\bibfnamefont {S.}~\bibnamefont
  {Lenhart}}\ and\ \bibinfo {author} {\bibfnamefont {J.~T.}\ \bibnamefont
  {Workman}},\ }\href {\doibase 10.1201/9781420011418} {\emph {\bibinfo {title}
  {Optimal control applied to biological models}}}\ (\bibinfo  {publisher}
  {Chapman and Hall/CRC},\ \bibinfo {year} {2007})\BibitemShut {NoStop}%
\bibitem [{\citenamefont {Kerswell}\ \emph {et~al.}(2014)\citenamefont
  {Kerswell}, \citenamefont {Pringle},\ and\ \citenamefont
  {Willis}}]{Kerswell2014}%
  \BibitemOpen
  \bibfield  {author} {\bibinfo {author} {\bibfnamefont {R.}~\bibnamefont
  {Kerswell}}, \bibinfo {author} {\bibfnamefont {C.~C.}\ \bibnamefont
  {Pringle}}, \ and\ \bibinfo {author} {\bibfnamefont {A.}~\bibnamefont
  {Willis}},\ }\href {\doibase 10.1088/0034-4885/77/8/085901} {\bibfield
  {journal} {\bibinfo  {journal} {Reports on Progress in Physics}\ }\textbf
  {\bibinfo {volume} {77}},\ \bibinfo {pages} {085901} (\bibinfo {year}
  {2014})}\BibitemShut {NoStop}%
\bibitem [{\citenamefont {Borz{\`\i}}\ and\ \citenamefont
  {Schulz}(2011)}]{Borzi2011}%
  \BibitemOpen
  \bibfield  {author} {\bibinfo {author} {\bibfnamefont {A.}~\bibnamefont
  {Borz{\`\i}}}\ and\ \bibinfo {author} {\bibfnamefont {V.}~\bibnamefont
  {Schulz}},\ }\href {\doibase 10.1137/1.9781611972054} {\emph {\bibinfo
  {title} {Computational optimization of systems governed by partial
  differential equations}}}\ (\bibinfo  {publisher} {SIAM},\ \bibinfo {year}
  {2011})\BibitemShut {NoStop}%
\bibitem [{\citenamefont {Dupont}\ \emph {et~al.}(2003)\citenamefont {Dupont},
  \citenamefont {Hoffman}, \citenamefont {Johnson}, \citenamefont {Kirby},
  \citenamefont {Larson}, \citenamefont {Logg},\ and\ \citenamefont
  {Scott}}]{Dupont2003}%
  \BibitemOpen
  \bibfield  {author} {\bibinfo {author} {\bibfnamefont {T.}~\bibnamefont
  {Dupont}}, \bibinfo {author} {\bibfnamefont {J.}~\bibnamefont {Hoffman}},
  \bibinfo {author} {\bibfnamefont {C.}~\bibnamefont {Johnson}}, \bibinfo
  {author} {\bibfnamefont {R.~C.}\ \bibnamefont {Kirby}}, \bibinfo {author}
  {\bibfnamefont {M.~G.}\ \bibnamefont {Larson}}, \bibinfo {author}
  {\bibfnamefont {A.}~\bibnamefont {Logg}}, \ and\ \bibinfo {author}
  {\bibfnamefont {L.~R.}\ \bibnamefont {Scott}},\ }\href {\doibase
  10.11588/ans.2015.100.20553} {\emph {\bibinfo {title} {The fenics project}}}\
  (\bibinfo  {publisher} {Chalmers Finite Element Centre, Chalmers University
  of Technology},\ \bibinfo {year} {2003})\BibitemShut {NoStop}%
\bibitem [{\citenamefont {Bechhoefer}(2021)}]{Bechhoefer2021}%
  \BibitemOpen
  \bibfield  {author} {\bibinfo {author} {\bibfnamefont {J.}~\bibnamefont
  {Bechhoefer}},\ }\href@noop {} {\emph {\bibinfo {title} {Control Theory for
  Physicists}}}\ (\bibinfo  {publisher} {Cambridge University Press},\ \bibinfo
  {year} {2021})\BibitemShut {NoStop}%
\bibitem [{\citenamefont {Schaller}\ \emph {et~al.}(2010)\citenamefont
  {Schaller}, \citenamefont {Weber}, \citenamefont {Semmrich}, \citenamefont
  {Frey},\ and\ \citenamefont {Bausch}}]{Schaller2010}%
  \BibitemOpen
  \bibfield  {author} {\bibinfo {author} {\bibfnamefont {V.}~\bibnamefont
  {Schaller}}, \bibinfo {author} {\bibfnamefont {C.}~\bibnamefont {Weber}},
  \bibinfo {author} {\bibfnamefont {C.}~\bibnamefont {Semmrich}}, \bibinfo
  {author} {\bibfnamefont {E.}~\bibnamefont {Frey}}, \ and\ \bibinfo {author}
  {\bibfnamefont {A.~R.}\ \bibnamefont {Bausch}},\ }\href {\doibase
  10.1038/nature09312} {\bibfield  {journal} {\bibinfo  {journal} {Nature}\
  }\textbf {\bibinfo {volume} {467}},\ \bibinfo {pages} {73} (\bibinfo {year}
  {2010})}\BibitemShut {NoStop}%
\bibitem [{\citenamefont {Huber}\ \emph {et~al.}(2018)\citenamefont {Huber},
  \citenamefont {Suzuki}, \citenamefont {Kr{\"u}ger}, \citenamefont {Frey},\
  and\ \citenamefont {Bausch}}]{Huber2018}%
  \BibitemOpen
  \bibfield  {author} {\bibinfo {author} {\bibfnamefont {L.}~\bibnamefont
  {Huber}}, \bibinfo {author} {\bibfnamefont {R.}~\bibnamefont {Suzuki}},
  \bibinfo {author} {\bibfnamefont {T.}~\bibnamefont {Kr{\"u}ger}}, \bibinfo
  {author} {\bibfnamefont {E.}~\bibnamefont {Frey}}, \ and\ \bibinfo {author}
  {\bibfnamefont {A.}~\bibnamefont {Bausch}},\ }\href@noop {} {\bibfield
  {journal} {\bibinfo  {journal} {Science}\ }\textbf {\bibinfo {volume}
  {361}},\ \bibinfo {pages} {255} (\bibinfo {year} {2018})}\BibitemShut
  {NoStop}%
\bibitem [{\citenamefont {Sciortino}\ and\ \citenamefont
  {Bausch}(2021)}]{Sciortino2021}%
  \BibitemOpen
  \bibfield  {author} {\bibinfo {author} {\bibfnamefont {A.}~\bibnamefont
  {Sciortino}}\ and\ \bibinfo {author} {\bibfnamefont {A.~R.}\ \bibnamefont
  {Bausch}},\ }\href@noop {} {\bibfield  {journal} {\bibinfo  {journal}
  {Proceedings of the National Academy of Sciences}\ }\textbf {\bibinfo
  {volume} {118}},\ \bibinfo {pages} {e2017047118} (\bibinfo {year}
  {2021})}\BibitemShut {NoStop}%
\bibitem [{\citenamefont {Suzuki}\ and\ \citenamefont
  {Bausch}(2017)}]{Suzuki2017}%
  \BibitemOpen
  \bibfield  {author} {\bibinfo {author} {\bibfnamefont {R.}~\bibnamefont
  {Suzuki}}\ and\ \bibinfo {author} {\bibfnamefont {A.~R.}\ \bibnamefont
  {Bausch}},\ }\href@noop {} {\bibfield  {journal} {\bibinfo  {journal} {Nature
  communications}\ }\textbf {\bibinfo {volume} {8}},\ \bibinfo {pages} {41}
  (\bibinfo {year} {2017})}\BibitemShut {NoStop}%
\bibitem [{\citenamefont {Saragosti}\ \emph {et~al.}(2011)\citenamefont
  {Saragosti}, \citenamefont {Calvez}, \citenamefont {Bournaveas},
  \citenamefont {Perthame}, \citenamefont {Buguin},\ and\ \citenamefont
  {Silberzan}}]{Saragosti2011}%
  \BibitemOpen
  \bibfield  {author} {\bibinfo {author} {\bibfnamefont {J.}~\bibnamefont
  {Saragosti}}, \bibinfo {author} {\bibfnamefont {V.}~\bibnamefont {Calvez}},
  \bibinfo {author} {\bibfnamefont {N.}~\bibnamefont {Bournaveas}}, \bibinfo
  {author} {\bibfnamefont {B.}~\bibnamefont {Perthame}}, \bibinfo {author}
  {\bibfnamefont {A.}~\bibnamefont {Buguin}}, \ and\ \bibinfo {author}
  {\bibfnamefont {P.}~\bibnamefont {Silberzan}},\ }\href {\doibase
  10.1073/pnas.1101996108} {\bibfield  {journal} {\bibinfo  {journal}
  {Proceedings of the National Academy of Sciences}\ }\textbf {\bibinfo
  {volume} {108}},\ \bibinfo {pages} {16235} (\bibinfo {year}
  {2011})}\BibitemShut {NoStop}%
\bibitem [{\citenamefont {Pohl}\ and\ \citenamefont {Stark}(2014)}]{Pohl2014}%
  \BibitemOpen
  \bibfield  {author} {\bibinfo {author} {\bibfnamefont {O.}~\bibnamefont
  {Pohl}}\ and\ \bibinfo {author} {\bibfnamefont {H.}~\bibnamefont {Stark}},\
  }\href@noop {} {\bibfield  {journal} {\bibinfo  {journal} {Physical review
  letters}\ }\textbf {\bibinfo {volume} {112}},\ \bibinfo {pages} {238303}
  (\bibinfo {year} {2014})}\BibitemShut {NoStop}%
\bibitem [{\citenamefont {Bhattacharjee}\ \emph {et~al.}(2022)\citenamefont
  {Bhattacharjee}, \citenamefont {Amchin}, \citenamefont {Alert}, \citenamefont
  {Ott},\ and\ \citenamefont {Datta}}]{Bhattacharjee2022}%
  \BibitemOpen
  \bibfield  {author} {\bibinfo {author} {\bibfnamefont {T.}~\bibnamefont
  {Bhattacharjee}}, \bibinfo {author} {\bibfnamefont {D.~B.}\ \bibnamefont
  {Amchin}}, \bibinfo {author} {\bibfnamefont {R.}~\bibnamefont {Alert}},
  \bibinfo {author} {\bibfnamefont {J.~A.}\ \bibnamefont {Ott}}, \ and\
  \bibinfo {author} {\bibfnamefont {S.~S.}\ \bibnamefont {Datta}},\ }\href@noop
  {} {\bibfield  {journal} {\bibinfo  {journal} {Elife}\ }\textbf {\bibinfo
  {volume} {11}},\ \bibinfo {pages} {e71226} (\bibinfo {year}
  {2022})}\BibitemShut {NoStop}%
\bibitem [{\citenamefont {Alert}\ \emph {et~al.}(2022)\citenamefont {Alert},
  \citenamefont {Mart{\'\i}nez-Calvo},\ and\ \citenamefont
  {Datta}}]{Alert2022}%
  \BibitemOpen
  \bibfield  {author} {\bibinfo {author} {\bibfnamefont {R.}~\bibnamefont
  {Alert}}, \bibinfo {author} {\bibfnamefont {A.}~\bibnamefont
  {Mart{\'\i}nez-Calvo}}, \ and\ \bibinfo {author} {\bibfnamefont {S.~S.}\
  \bibnamefont {Datta}},\ }\href@noop {} {\bibfield  {journal} {\bibinfo
  {journal} {Physical review letters}\ }\textbf {\bibinfo {volume} {128}},\
  \bibinfo {pages} {148101} (\bibinfo {year} {2022})}\BibitemShut {NoStop}%
\bibitem [{\citenamefont {Narla}\ \emph {et~al.}(2021)\citenamefont {Narla},
  \citenamefont {Cremer},\ and\ \citenamefont {Hwa}}]{Narla2021}%
  \BibitemOpen
  \bibfield  {author} {\bibinfo {author} {\bibfnamefont {A.~V.}\ \bibnamefont
  {Narla}}, \bibinfo {author} {\bibfnamefont {J.}~\bibnamefont {Cremer}}, \
  and\ \bibinfo {author} {\bibfnamefont {T.}~\bibnamefont {Hwa}},\ }\href@noop
  {} {\bibfield  {journal} {\bibinfo  {journal} {Proceedings of the National
  Academy of Sciences}\ }\textbf {\bibinfo {volume} {118}},\ \bibinfo {pages}
  {e2105138118} (\bibinfo {year} {2021})}\BibitemShut {NoStop}%
\bibitem [{\citenamefont {Cremer}\ \emph {et~al.}(2019)\citenamefont {Cremer},
  \citenamefont {Honda}, \citenamefont {Tang}, \citenamefont {Wong-Ng},
  \citenamefont {Vergassola},\ and\ \citenamefont {Hwa}}]{Cremer2019}%
  \BibitemOpen
  \bibfield  {author} {\bibinfo {author} {\bibfnamefont {J.}~\bibnamefont
  {Cremer}}, \bibinfo {author} {\bibfnamefont {T.}~\bibnamefont {Honda}},
  \bibinfo {author} {\bibfnamefont {Y.}~\bibnamefont {Tang}}, \bibinfo {author}
  {\bibfnamefont {J.}~\bibnamefont {Wong-Ng}}, \bibinfo {author} {\bibfnamefont
  {M.}~\bibnamefont {Vergassola}}, \ and\ \bibinfo {author} {\bibfnamefont
  {T.}~\bibnamefont {Hwa}},\ }\href@noop {} {\bibfield  {journal} {\bibinfo
  {journal} {Nature}\ }\textbf {\bibinfo {volume} {575}},\ \bibinfo {pages}
  {658} (\bibinfo {year} {2019})}\BibitemShut {NoStop}%
\bibitem [{\citenamefont {Bhattacharjee}\ \emph {et~al.}(2021)\citenamefont
  {Bhattacharjee}, \citenamefont {Amchin}, \citenamefont {Ott}, \citenamefont
  {Kratz},\ and\ \citenamefont {Datta}}]{Bhattacharjee2021}%
  \BibitemOpen
  \bibfield  {author} {\bibinfo {author} {\bibfnamefont {T.}~\bibnamefont
  {Bhattacharjee}}, \bibinfo {author} {\bibfnamefont {D.~B.}\ \bibnamefont
  {Amchin}}, \bibinfo {author} {\bibfnamefont {J.~A.}\ \bibnamefont {Ott}},
  \bibinfo {author} {\bibfnamefont {F.}~\bibnamefont {Kratz}}, \ and\ \bibinfo
  {author} {\bibfnamefont {S.~S.}\ \bibnamefont {Datta}},\ }\href@noop {}
  {\bibfield  {journal} {\bibinfo  {journal} {Biophysical Journal}\ }\textbf
  {\bibinfo {volume} {120}},\ \bibinfo {pages} {3483} (\bibinfo {year}
  {2021})}\BibitemShut {NoStop}%
\bibitem [{\citenamefont {Brenner}\ \emph {et~al.}(1998)\citenamefont
  {Brenner}, \citenamefont {Levitov},\ and\ \citenamefont
  {Budrene}}]{Brenner1998}%
  \BibitemOpen
  \bibfield  {author} {\bibinfo {author} {\bibfnamefont {M.~P.}\ \bibnamefont
  {Brenner}}, \bibinfo {author} {\bibfnamefont {L.~S.}\ \bibnamefont
  {Levitov}}, \ and\ \bibinfo {author} {\bibfnamefont {E.~O.}\ \bibnamefont
  {Budrene}},\ }\href@noop {} {\bibfield  {journal} {\bibinfo  {journal}
  {Biophysical journal}\ }\textbf {\bibinfo {volume} {74}},\ \bibinfo {pages}
  {1677} (\bibinfo {year} {1998})}\BibitemShut {NoStop}%
\bibitem [{\citenamefont {Fei}\ \emph {et~al.}(2020)\citenamefont {Fei},
  \citenamefont {Mao}, \citenamefont {Yan}, \citenamefont {Alert},
  \citenamefont {Stone}, \citenamefont {Bassler}, \citenamefont {Wingreen},\
  and\ \citenamefont {Ko{\v{s}}mrlj}}]{Fei2020}%
  \BibitemOpen
  \bibfield  {author} {\bibinfo {author} {\bibfnamefont {C.}~\bibnamefont
  {Fei}}, \bibinfo {author} {\bibfnamefont {S.}~\bibnamefont {Mao}}, \bibinfo
  {author} {\bibfnamefont {J.}~\bibnamefont {Yan}}, \bibinfo {author}
  {\bibfnamefont {R.}~\bibnamefont {Alert}}, \bibinfo {author} {\bibfnamefont
  {H.~A.}\ \bibnamefont {Stone}}, \bibinfo {author} {\bibfnamefont {B.~L.}\
  \bibnamefont {Bassler}}, \bibinfo {author} {\bibfnamefont {N.~S.}\
  \bibnamefont {Wingreen}}, \ and\ \bibinfo {author} {\bibfnamefont
  {A.}~\bibnamefont {Ko{\v{s}}mrlj}},\ }\href@noop {} {\bibfield  {journal}
  {\bibinfo  {journal} {Proceedings of the National Academy of Sciences}\
  }\textbf {\bibinfo {volume} {117}},\ \bibinfo {pages} {7622} (\bibinfo {year}
  {2020})}\BibitemShut {NoStop}%
\bibitem [{\citenamefont {Mayer}\ \emph {et~al.}(1999)\citenamefont {Mayer},
  \citenamefont {Kapoor}, \citenamefont {Haggarty}, \citenamefont {King},
  \citenamefont {Schreiber},\ and\ \citenamefont {Mitchison}}]{Mayer1999}%
  \BibitemOpen
  \bibfield  {author} {\bibinfo {author} {\bibfnamefont {T.~U.}\ \bibnamefont
  {Mayer}}, \bibinfo {author} {\bibfnamefont {T.~M.}\ \bibnamefont {Kapoor}},
  \bibinfo {author} {\bibfnamefont {S.~J.}\ \bibnamefont {Haggarty}}, \bibinfo
  {author} {\bibfnamefont {R.~W.}\ \bibnamefont {King}}, \bibinfo {author}
  {\bibfnamefont {S.~L.}\ \bibnamefont {Schreiber}}, \ and\ \bibinfo {author}
  {\bibfnamefont {T.~J.}\ \bibnamefont {Mitchison}},\ }\href {\doibase
  10.1126/science.286.5441.971} {\bibfield  {journal} {\bibinfo  {journal}
  {Science}\ }\textbf {\bibinfo {volume} {286}},\ \bibinfo {pages} {971}
  (\bibinfo {year} {1999})}\BibitemShut {NoStop}%
\bibitem [{\citenamefont {Schaffner}\ and\ \citenamefont
  {Jos{\'e}}(2006)}]{Schaffner2006}%
  \BibitemOpen
  \bibfield  {author} {\bibinfo {author} {\bibfnamefont {S.~C.}\ \bibnamefont
  {Schaffner}}\ and\ \bibinfo {author} {\bibfnamefont {J.~V.}\ \bibnamefont
  {Jos{\'e}}},\ }\href@noop {} {\bibfield  {journal} {\bibinfo  {journal}
  {Proceedings of the National Academy of Sciences}\ }\textbf {\bibinfo
  {volume} {103}},\ \bibinfo {pages} {11166} (\bibinfo {year}
  {2006})}\BibitemShut {NoStop}%
\bibitem [{\citenamefont {Loughlin}\ \emph {et~al.}(2010)\citenamefont
  {Loughlin}, \citenamefont {Heald},\ and\ \citenamefont
  {N{\'e}d{\'e}lec}}]{Loughlin2010}%
  \BibitemOpen
  \bibfield  {author} {\bibinfo {author} {\bibfnamefont {R.}~\bibnamefont
  {Loughlin}}, \bibinfo {author} {\bibfnamefont {R.}~\bibnamefont {Heald}}, \
  and\ \bibinfo {author} {\bibfnamefont {F.}~\bibnamefont {N{\'e}d{\'e}lec}},\
  }\href@noop {} {\bibfield  {journal} {\bibinfo  {journal} {Journal of cell
  biology}\ }\textbf {\bibinfo {volume} {191}},\ \bibinfo {pages} {1239}
  (\bibinfo {year} {2010})}\BibitemShut {NoStop}%
\bibitem [{\citenamefont {Shirasu-Hiza}\ \emph {et~al.}(2004)\citenamefont
  {Shirasu-Hiza}, \citenamefont {Perlman}, \citenamefont {Wittmann},
  \citenamefont {Karsenti},\ and\ \citenamefont {Mitchison}}]{ShirasuHiza2004}%
  \BibitemOpen
  \bibfield  {author} {\bibinfo {author} {\bibfnamefont {M.}~\bibnamefont
  {Shirasu-Hiza}}, \bibinfo {author} {\bibfnamefont {Z.~E.}\ \bibnamefont
  {Perlman}}, \bibinfo {author} {\bibfnamefont {T.}~\bibnamefont {Wittmann}},
  \bibinfo {author} {\bibfnamefont {E.}~\bibnamefont {Karsenti}}, \ and\
  \bibinfo {author} {\bibfnamefont {T.~J.}\ \bibnamefont {Mitchison}},\ }\href
  {\doibase 10.1016/j.cub.2004.10.029} {\bibfield  {journal} {\bibinfo
  {journal} {Current biology}\ }\textbf {\bibinfo {volume} {14}},\ \bibinfo
  {pages} {1941} (\bibinfo {year} {2004})}\BibitemShut {NoStop}%
\bibitem [{\citenamefont {Heidemann}\ and\ \citenamefont
  {Kirschner}(1975)}]{Heidemann1975}%
  \BibitemOpen
  \bibfield  {author} {\bibinfo {author} {\bibfnamefont {S.~R.}\ \bibnamefont
  {Heidemann}}\ and\ \bibinfo {author} {\bibfnamefont {M.~W.}\ \bibnamefont
  {Kirschner}},\ }\href@noop {} {\bibfield  {journal} {\bibinfo  {journal} {The
  Journal of cell biology}\ }\textbf {\bibinfo {volume} {67}},\ \bibinfo
  {pages} {105} (\bibinfo {year} {1975})}\BibitemShut {NoStop}%
\bibitem [{\citenamefont {Heidemann}\ and\ \citenamefont
  {Kirschner}(1978)}]{Heidemann1978}%
  \BibitemOpen
  \bibfield  {author} {\bibinfo {author} {\bibfnamefont {S.~R.}\ \bibnamefont
  {Heidemann}}\ and\ \bibinfo {author} {\bibfnamefont {M.~W.}\ \bibnamefont
  {Kirschner}},\ }\href@noop {} {\bibfield  {journal} {\bibinfo  {journal}
  {Journal of Experimental Zoology}\ }\textbf {\bibinfo {volume} {204}},\
  \bibinfo {pages} {431} (\bibinfo {year} {1978})}\BibitemShut {NoStop}%
\bibitem [{\citenamefont {Schmit}\ \emph {et~al.}(1983)\citenamefont {Schmit},
  \citenamefont {Vantard}, \citenamefont {De~Mey},\ and\ \citenamefont
  {Lambert}}]{Schmit1983}%
  \BibitemOpen
  \bibfield  {author} {\bibinfo {author} {\bibfnamefont {A.-C.}\ \bibnamefont
  {Schmit}}, \bibinfo {author} {\bibfnamefont {M.}~\bibnamefont {Vantard}},
  \bibinfo {author} {\bibfnamefont {J.}~\bibnamefont {De~Mey}}, \ and\ \bibinfo
  {author} {\bibfnamefont {A.-M.}\ \bibnamefont {Lambert}},\ }\href@noop {}
  {\bibfield  {journal} {\bibinfo  {journal} {Plant cell reports}\ }\textbf
  {\bibinfo {volume} {2}},\ \bibinfo {pages} {285} (\bibinfo {year}
  {1983})}\BibitemShut {NoStop}%
\bibitem [{\citenamefont {Kumagai}\ and\ \citenamefont
  {Hasezawa}(2001)}]{Kumagai2001}%
  \BibitemOpen
  \bibfield  {author} {\bibinfo {author} {\bibfnamefont {F.}~\bibnamefont
  {Kumagai}}\ and\ \bibinfo {author} {\bibfnamefont {S.}~\bibnamefont
  {Hasezawa}},\ }\href@noop {} {\bibfield  {journal} {\bibinfo  {journal}
  {Plant biology}\ }\textbf {\bibinfo {volume} {3}},\ \bibinfo {pages} {4}
  (\bibinfo {year} {2001})}\BibitemShut {NoStop}%
\bibitem [{\citenamefont {Najma}\ \emph {et~al.}(2024)\citenamefont {Najma},
  \citenamefont {Wei}, \citenamefont {Baskaran}, \citenamefont {Foster},\ and\
  \citenamefont {Duclos}}]{Najma2024}%
  \BibitemOpen
  \bibfield  {author} {\bibinfo {author} {\bibfnamefont {B.}~\bibnamefont
  {Najma}}, \bibinfo {author} {\bibfnamefont {W.-S.}\ \bibnamefont {Wei}},
  \bibinfo {author} {\bibfnamefont {A.}~\bibnamefont {Baskaran}}, \bibinfo
  {author} {\bibfnamefont {P.~J.}\ \bibnamefont {Foster}}, \ and\ \bibinfo
  {author} {\bibfnamefont {G.}~\bibnamefont {Duclos}},\ }\href@noop {}
  {\bibfield  {journal} {\bibinfo  {journal} {Proceedings of the National
  Academy of Sciences}\ }\textbf {\bibinfo {volume} {121}},\ \bibinfo {pages}
  {e2300174121} (\bibinfo {year} {2024})}\BibitemShut {NoStop}%
\bibitem [{\citenamefont {Wollrab}\ \emph {et~al.}(2019)\citenamefont
  {Wollrab}, \citenamefont {Belmonte}, \citenamefont {Baldauf}, \citenamefont
  {Leptin}, \citenamefont {N{\'e}del{\'e}c},\ and\ \citenamefont
  {Koenderink}}]{Wollrab2019}%
  \BibitemOpen
  \bibfield  {author} {\bibinfo {author} {\bibfnamefont {V.}~\bibnamefont
  {Wollrab}}, \bibinfo {author} {\bibfnamefont {J.~M.}\ \bibnamefont
  {Belmonte}}, \bibinfo {author} {\bibfnamefont {L.}~\bibnamefont {Baldauf}},
  \bibinfo {author} {\bibfnamefont {M.}~\bibnamefont {Leptin}}, \bibinfo
  {author} {\bibfnamefont {F.}~\bibnamefont {N{\'e}del{\'e}c}}, \ and\ \bibinfo
  {author} {\bibfnamefont {G.~H.}\ \bibnamefont {Koenderink}},\ }\href@noop {}
  {\bibfield  {journal} {\bibinfo  {journal} {Journal of cell science}\
  }\textbf {\bibinfo {volume} {132}},\ \bibinfo {pages} {jcs219717} (\bibinfo
  {year} {2019})}\BibitemShut {NoStop}%
\bibitem [{\citenamefont {Foster}\ \emph {et~al.}(2015)\citenamefont {Foster},
  \citenamefont {F{\"u}rthauer}, \citenamefont {Shelley},\ and\ \citenamefont
  {Needleman}}]{Foster2015}%
  \BibitemOpen
  \bibfield  {author} {\bibinfo {author} {\bibfnamefont {P.~J.}\ \bibnamefont
  {Foster}}, \bibinfo {author} {\bibfnamefont {S.}~\bibnamefont
  {F{\"u}rthauer}}, \bibinfo {author} {\bibfnamefont {M.~J.}\ \bibnamefont
  {Shelley}}, \ and\ \bibinfo {author} {\bibfnamefont {D.~J.}\ \bibnamefont
  {Needleman}},\ }\href@noop {} {\bibfield  {journal} {\bibinfo  {journal}
  {Elife}\ }\textbf {\bibinfo {volume} {4}},\ \bibinfo {pages} {e10837}
  (\bibinfo {year} {2015})}\BibitemShut {NoStop}%
\bibitem [{\citenamefont {Colin}\ \emph {et~al.}(2018)\citenamefont {Colin},
  \citenamefont {Singaravelu}, \citenamefont {Th{\'e}ry}, \citenamefont
  {Blanchoin},\ and\ \citenamefont {Gueroui}}]{Colin2018}%
  \BibitemOpen
  \bibfield  {author} {\bibinfo {author} {\bibfnamefont {A.}~\bibnamefont
  {Colin}}, \bibinfo {author} {\bibfnamefont {P.}~\bibnamefont {Singaravelu}},
  \bibinfo {author} {\bibfnamefont {M.}~\bibnamefont {Th{\'e}ry}}, \bibinfo
  {author} {\bibfnamefont {L.}~\bibnamefont {Blanchoin}}, \ and\ \bibinfo
  {author} {\bibfnamefont {Z.}~\bibnamefont {Gueroui}},\ }\href@noop {}
  {\bibfield  {journal} {\bibinfo  {journal} {Current Biology}\ }\textbf
  {\bibinfo {volume} {28}},\ \bibinfo {pages} {2647} (\bibinfo {year}
  {2018})}\BibitemShut {NoStop}%
\bibitem [{\citenamefont {Luo}\ \emph {et~al.}(2013)\citenamefont {Luo},
  \citenamefont {Yu}, \citenamefont {Lieu}, \citenamefont {Allard},
  \citenamefont {Mogilner}, \citenamefont {Sheetz},\ and\ \citenamefont
  {Bershadsky}}]{Luo2013}%
  \BibitemOpen
  \bibfield  {author} {\bibinfo {author} {\bibfnamefont {W.}~\bibnamefont
  {Luo}}, \bibinfo {author} {\bibfnamefont {C.-h.}\ \bibnamefont {Yu}},
  \bibinfo {author} {\bibfnamefont {Z.~Z.}\ \bibnamefont {Lieu}}, \bibinfo
  {author} {\bibfnamefont {J.}~\bibnamefont {Allard}}, \bibinfo {author}
  {\bibfnamefont {A.}~\bibnamefont {Mogilner}}, \bibinfo {author}
  {\bibfnamefont {M.~P.}\ \bibnamefont {Sheetz}}, \ and\ \bibinfo {author}
  {\bibfnamefont {A.~D.}\ \bibnamefont {Bershadsky}},\ }\href@noop {}
  {\bibfield  {journal} {\bibinfo  {journal} {Journal of Cell Biology}\
  }\textbf {\bibinfo {volume} {202}},\ \bibinfo {pages} {1057} (\bibinfo {year}
  {2013})}\BibitemShut {NoStop}%
\bibitem [{\citenamefont {Stam}\ \emph {et~al.}(2017)\citenamefont {Stam},
  \citenamefont {Freedman}, \citenamefont {Banerjee}, \citenamefont {Weirich},
  \citenamefont {Dinner},\ and\ \citenamefont {Gardel}}]{Stam2017}%
  \BibitemOpen
  \bibfield  {author} {\bibinfo {author} {\bibfnamefont {S.}~\bibnamefont
  {Stam}}, \bibinfo {author} {\bibfnamefont {S.~L.}\ \bibnamefont {Freedman}},
  \bibinfo {author} {\bibfnamefont {S.}~\bibnamefont {Banerjee}}, \bibinfo
  {author} {\bibfnamefont {K.~L.}\ \bibnamefont {Weirich}}, \bibinfo {author}
  {\bibfnamefont {A.~R.}\ \bibnamefont {Dinner}}, \ and\ \bibinfo {author}
  {\bibfnamefont {M.~L.}\ \bibnamefont {Gardel}},\ }\href@noop {} {\bibfield
  {journal} {\bibinfo  {journal} {Proceedings of the National Academy of
  Sciences}\ }\textbf {\bibinfo {volume} {114}},\ \bibinfo {pages} {E10037}
  (\bibinfo {year} {2017})}\BibitemShut {NoStop}%
\bibitem [{\citenamefont {Berezney}\ \emph {et~al.}(2022)\citenamefont
  {Berezney}, \citenamefont {Goode}, \citenamefont {Fraden},\ and\
  \citenamefont {Dogic}}]{Berezney2022}%
  \BibitemOpen
  \bibfield  {author} {\bibinfo {author} {\bibfnamefont {J.}~\bibnamefont
  {Berezney}}, \bibinfo {author} {\bibfnamefont {B.~L.}\ \bibnamefont {Goode}},
  \bibinfo {author} {\bibfnamefont {S.}~\bibnamefont {Fraden}}, \ and\ \bibinfo
  {author} {\bibfnamefont {Z.}~\bibnamefont {Dogic}},\ }\href@noop {}
  {\bibfield  {journal} {\bibinfo  {journal} {Proceedings of the National
  Academy of Sciences}\ }\textbf {\bibinfo {volume} {119}},\ \bibinfo {pages}
  {e2115895119} (\bibinfo {year} {2022})}\BibitemShut {NoStop}%
\bibitem [{\citenamefont {Thoresen}\ \emph {et~al.}(2011)\citenamefont
  {Thoresen}, \citenamefont {Lenz},\ and\ \citenamefont
  {Gardel}}]{Thoresen2011}%
  \BibitemOpen
  \bibfield  {author} {\bibinfo {author} {\bibfnamefont {T.}~\bibnamefont
  {Thoresen}}, \bibinfo {author} {\bibfnamefont {M.}~\bibnamefont {Lenz}}, \
  and\ \bibinfo {author} {\bibfnamefont {M.~L.}\ \bibnamefont {Gardel}},\
  }\href@noop {} {\bibfield  {journal} {\bibinfo  {journal} {Biophysical
  journal}\ }\textbf {\bibinfo {volume} {100}},\ \bibinfo {pages} {2698}
  (\bibinfo {year} {2011})}\BibitemShut {NoStop}%
\bibitem [{\citenamefont {K{\"o}ster}\ \emph {et~al.}(2016)\citenamefont
  {K{\"o}ster}, \citenamefont {Husain}, \citenamefont {Iljazi}, \citenamefont
  {Bhat}, \citenamefont {Bieling}, \citenamefont {Mullins}, \citenamefont
  {Rao},\ and\ \citenamefont {Mayor}}]{Koster2016}%
  \BibitemOpen
  \bibfield  {author} {\bibinfo {author} {\bibfnamefont {D.~V.}\ \bibnamefont
  {K{\"o}ster}}, \bibinfo {author} {\bibfnamefont {K.}~\bibnamefont {Husain}},
  \bibinfo {author} {\bibfnamefont {E.}~\bibnamefont {Iljazi}}, \bibinfo
  {author} {\bibfnamefont {A.}~\bibnamefont {Bhat}}, \bibinfo {author}
  {\bibfnamefont {P.}~\bibnamefont {Bieling}}, \bibinfo {author} {\bibfnamefont
  {R.~D.}\ \bibnamefont {Mullins}}, \bibinfo {author} {\bibfnamefont
  {M.}~\bibnamefont {Rao}}, \ and\ \bibinfo {author} {\bibfnamefont
  {S.}~\bibnamefont {Mayor}},\ }\href@noop {} {\bibfield  {journal} {\bibinfo
  {journal} {Proceedings of the National Academy of Sciences}\ }\textbf
  {\bibinfo {volume} {113}},\ \bibinfo {pages} {E1645} (\bibinfo {year}
  {2016})}\BibitemShut {NoStop}%
\bibitem [{\citenamefont {Glaser}\ \emph {et~al.}(2016)\citenamefont {Glaser},
  \citenamefont {Schnau{\ss}}, \citenamefont {Tschirner}, \citenamefont
  {Schmidt}, \citenamefont {Moebius-Winkler}, \citenamefont {K{\"a}s},\ and\
  \citenamefont {Smith}}]{Glaser2016}%
  \BibitemOpen
  \bibfield  {author} {\bibinfo {author} {\bibfnamefont {M.}~\bibnamefont
  {Glaser}}, \bibinfo {author} {\bibfnamefont {J.}~\bibnamefont {Schnau{\ss}}},
  \bibinfo {author} {\bibfnamefont {T.}~\bibnamefont {Tschirner}}, \bibinfo
  {author} {\bibfnamefont {B.~S.}\ \bibnamefont {Schmidt}}, \bibinfo {author}
  {\bibfnamefont {M.}~\bibnamefont {Moebius-Winkler}}, \bibinfo {author}
  {\bibfnamefont {J.~A.}\ \bibnamefont {K{\"a}s}}, \ and\ \bibinfo {author}
  {\bibfnamefont {D.~M.}\ \bibnamefont {Smith}},\ }\href@noop {} {\bibfield
  {journal} {\bibinfo  {journal} {New Journal of Physics}\ }\textbf {\bibinfo
  {volume} {18}},\ \bibinfo {pages} {055001} (\bibinfo {year}
  {2016})}\BibitemShut {NoStop}%
\bibitem [{\citenamefont {N{\'e}d{\'e}lec}\ \emph {et~al.}(2003)\citenamefont
  {N{\'e}d{\'e}lec}, \citenamefont {Surrey},\ and\ \citenamefont
  {Karsenti}}]{Nedelec2003}%
  \BibitemOpen
  \bibfield  {author} {\bibinfo {author} {\bibfnamefont {F.}~\bibnamefont
  {N{\'e}d{\'e}lec}}, \bibinfo {author} {\bibfnamefont {T.}~\bibnamefont
  {Surrey}}, \ and\ \bibinfo {author} {\bibfnamefont {E.}~\bibnamefont
  {Karsenti}},\ }\href@noop {} {\bibfield  {journal} {\bibinfo  {journal}
  {Current opinion in cell biology}\ }\textbf {\bibinfo {volume} {15}},\
  \bibinfo {pages} {118} (\bibinfo {year} {2003})}\BibitemShut {NoStop}%
\bibitem [{\citenamefont {Joshi}\ \emph {et~al.}(2022)\citenamefont {Joshi},
  \citenamefont {Ray}, \citenamefont {Lemma}, \citenamefont {Varghese},
  \citenamefont {Sharp}, \citenamefont {Dogic}, \citenamefont {Baskaran},\ and\
  \citenamefont {Hagan}}]{Joshi2022}%
  \BibitemOpen
  \bibfield  {author} {\bibinfo {author} {\bibfnamefont {C.}~\bibnamefont
  {Joshi}}, \bibinfo {author} {\bibfnamefont {S.}~\bibnamefont {Ray}}, \bibinfo
  {author} {\bibfnamefont {L.~M.}\ \bibnamefont {Lemma}}, \bibinfo {author}
  {\bibfnamefont {M.}~\bibnamefont {Varghese}}, \bibinfo {author}
  {\bibfnamefont {G.}~\bibnamefont {Sharp}}, \bibinfo {author} {\bibfnamefont
  {Z.}~\bibnamefont {Dogic}}, \bibinfo {author} {\bibfnamefont
  {A.}~\bibnamefont {Baskaran}}, \ and\ \bibinfo {author} {\bibfnamefont
  {M.~F.}\ \bibnamefont {Hagan}},\ }\href@noop {} {\bibfield  {journal}
  {\bibinfo  {journal} {Physical review letters}\ }\textbf {\bibinfo {volume}
  {129}},\ \bibinfo {pages} {258001} (\bibinfo {year} {2022})}\BibitemShut
  {NoStop}%
\bibitem [{\citenamefont {Supekar}\ \emph {et~al.}(2023)\citenamefont
  {Supekar}, \citenamefont {Song}, \citenamefont {Hastewell}, \citenamefont
  {Choi}, \citenamefont {Mietke},\ and\ \citenamefont {Dunkel}}]{Dunkel2023}%
  \BibitemOpen
  \bibfield  {author} {\bibinfo {author} {\bibfnamefont {R.}~\bibnamefont
  {Supekar}}, \bibinfo {author} {\bibfnamefont {B.}~\bibnamefont {Song}},
  \bibinfo {author} {\bibfnamefont {A.}~\bibnamefont {Hastewell}}, \bibinfo
  {author} {\bibfnamefont {G.~P.}\ \bibnamefont {Choi}}, \bibinfo {author}
  {\bibfnamefont {A.}~\bibnamefont {Mietke}}, \ and\ \bibinfo {author}
  {\bibfnamefont {J.}~\bibnamefont {Dunkel}},\ }\href@noop {} {\bibfield
  {journal} {\bibinfo  {journal} {Proceedings of the National Academy of
  Sciences}\ }\textbf {\bibinfo {volume} {120}},\ \bibinfo {pages}
  {e2206994120} (\bibinfo {year} {2023})}\BibitemShut {NoStop}%
\bibitem [{\citenamefont {Golden}\ \emph {et~al.}(2023)\citenamefont {Golden},
  \citenamefont {Grigoriev}, \citenamefont {Nambisan},\ and\ \citenamefont
  {Fernandez-Nieves}}]{Golden2023}%
  \BibitemOpen
  \bibfield  {author} {\bibinfo {author} {\bibfnamefont {M.}~\bibnamefont
  {Golden}}, \bibinfo {author} {\bibfnamefont {R.~O.}\ \bibnamefont
  {Grigoriev}}, \bibinfo {author} {\bibfnamefont {J.}~\bibnamefont {Nambisan}},
  \ and\ \bibinfo {author} {\bibfnamefont {A.}~\bibnamefont
  {Fernandez-Nieves}},\ }\href@noop {} {\bibfield  {journal} {\bibinfo
  {journal} {Science Advances}\ }\textbf {\bibinfo {volume} {9}},\ \bibinfo
  {pages} {eabq6120} (\bibinfo {year} {2023})}\BibitemShut {NoStop}%
\end{thebibliography}
\end{document}